\documentclass[12pt,preprint]{aastex}
\usepackage[colorlinks,citecolor=blue]{hyperref}
\def\coi    {$^{12}$CO}
\def\coii   {$^{13}$CO}
\def\coiii  {C$^{18}$O}
\def\kms    {km s$^{-1}$}
\def\Hii    {H{\scriptsize~II}}
\usepackage{subfigure}

\begin{document}

\title{MOLECULAR GAS toward THE GEMINI OB1 MOLECULAR CLOUD COMPLEX I: OBSERVATION DATA}

\author{Chen Wang\altaffilmark{1,2,3}, Ji Yang\altaffilmark{1,2,4}, Ye Xu\altaffilmark{1,2}, Facheng Li\altaffilmark{1,2}, Yang Su\altaffilmark{1,2} and Shaobo Zhang\altaffilmark{1,2}}
\affil{Purple Mountain Observatory, Chinese Academy of Sciences,
    Nanjing, Jiangsu 210008}

\altaffiltext{1}{Purple Mountain Observatory, Chinese Academy of Sciences, Nanjing 210008, China}
\altaffiltext{2}{Key Laboratory of Radio Astronomy, Chinese Academy of Sciences, Nanjing 210008, China}
\altaffiltext{3}{University of Chinese Academy of Sciences, 19A Yuquan Road, Shijingshan District, Beijing 100049, China}
\altaffiltext{4}{Corresponding author: jiyang@pmo.ac.cn}

\begin{abstract}
We present a large-scale mapping toward the GEM OB1 association in the galactic anti-center direction. The $9^{\circ} \times 6^{\circ}.5$ area was mapped in {$^{12}$CO}, {$^{13}$CO}, and {C$^{18}$O} with $\sim$50$''$ angular resolution at 30$''$ sampling. The region was divided into four main components based on spatial distribution and velocity: the Gemini OB1 Giant Molecular Cloud (GGMC) Complex, the Lynds Dark Clouds and the West Front Clouds, the Swallow and Horn, and the Remote Clouds. The GGMC Complex is located in the Perseus arm, while the Lynds Dark Clouds and the West Front Clouds are located in the Local arm. Swallow and Horn are revealed for the first time in this paper. The two clouds have a similar velocity interval ([11, 21] km s$^{-1}$) and have similar sizes (0.6 and 0.8 deg$^{2}$). We analyzed the structure of these clouds in detail and calculated their parameters (mass, temperature, etc.). Two elongated structures were discovered in a longitude-velocity map in the velocity interval [11, 30] km s$^{-1}$. We also found an interesting filament that shows a 0.8 km s$^{-1}$ pc$^{-1}$ gradient perpendicular to the direction of the long axis.

\end{abstract}





\section{Introduction}
CO emission plays an essential role in the study of molecular clouds (MCs), because it is readily detected even in diffuse molecular gas (\citealt{1999ARA&A..37..311E}). This tracer reveals the distributions of molecular gas and the kinematic information of detected clouds (\citealt{1987ARA&A..25...23S}). MCs are thought to be the place where stars form, so studying them helps us to understand the star formation process and even galactic structures (\citealt{1987ARA&A..25...23S}, \citealt{1999ARA&A..37..311E}, \citealt{2007ARA&A..45..565M}).

Over the past several decades, many large MC surveys have been performed. The first galactic plane ($ [b] < 10-20^{\circ}$) survey was conducted by \citet{1987ApJ...322..706D, 2001ApJ...547..792D} using the CO molecular spectral line (J = 1-0) with an angular resolution of $\sim8'$. Thanks to this survey, we gained insight into the large-scale structures of MCs in our galaxy.

With the degeneracy of the radial velocity in the Galactic anti-center, the kinematic distances in this region have large uncertainties. Hence, it is difficult to distinguish different MCs and to investigate their properties.

\citet{1998ApJS..117..387K} performed a large-scale MC survey in {$^{13}$CO} (J = 1-0) of the Gemini region using the two 4-m millimeter-wavelength telescopes at Nagoya University. They showed that the {$^{13}$CO} clouds associated with IRAS point sources, which are regarded as star-forming sites, have higher column densities N(H$_{2}$), sizes (R), and cloud masses (M$_\mathrm{cloud}$) but lower M$_\mathrm{vir}$$/$M$_\mathrm{cloud}$ than the other clouds. \citet{1995ApJ...445..246C, 1995ApJ...450..201C} investigated the morphology and physical properties of the Gem OB1 cloud complex by mapping more than 32 deg$^{2}$ of the complex in {$^{12}$CO} and {$^{13}$CO} ( J = 1-0) with QUARRY on the FCRAO 14 m telescope, and also conducted a study of global star formation activities along with a combination of CS (J = 2-1) and infrared data. They analyzed the spatial structure of the MC in the region and revealed that the arc-like molecular gas and dense cores might be related to star formation. They concluded that the star formation activity in such regions is mainly triggered.

Although \citet{1998ApJS..117..387K} and \citet{1995ApJ...445..246C, 1995ApJ...450..201C} achieved fruitful results for this region, there are some interesting structures and clouds that still need to be studied in detail. In particular, unbiased {$^{13}$CO} and {C$^{18}$O} emission will help reveal denser areas, and a study of the CO isotopic abundance ratio and outflow in this region is also needed. Such studies will be discussed in a forthcoming papers (C.Wang et al. in preparation, Y. Li et al. submitted). The current paper instead focuses on the distribution of molecular gas and the properties of each cloud.

\section{Observations}

We performed observations of the J = 1-0 transitions of {$^{12}$CO}, {$^{13}$CO}, and {C$^{18}$O} of the GEM OB1 complex with the Purple Mountain Observatory (PMO) 13.7~m telescope at Delingha from 2012 September to 2013 May\footnote{http://www.radioast.csdb.cn/, one can submit an application to download the data.}. These observations were obtained as part of the Milky Way Imaging Scroll Painting (MWISP) project, which is a Multi-Line Galactic Plane Survey of CO and its Isotopic Transitions. This area was mapped using a $3\times3$ pixel Superconducting Spectroscopic Array Receiver (SSAR), which was made with Supercondcutor-Insulator-Superconductor (SIS) mixers. The SSAR works in sideband separation mode and employs a fast Fourier transform spectrometer (FFTS) (\citealt{2012ITTST...2..593S}, \citealt{2011AcASn..52..152Z}).

Our observations were made in 234 cells of dimension $30'\times30'$, which covered an area of total 58.5 deg$^2$. The cells were mapped using the on-the-fly (OTF) observation mode with a scan speed of 50$''$ s$^{-1}$ and a step of $15 ''$ along the Galactic longitude and latitude, while the standard chopper wheel method was used for calibration (\citealt{1973ARA&A..11...51P}). Observations of {$^{12}$CO}, {$^{13}$CO}, and {C$^{18}$O} in the $J = 1-0$ transition were obtained simultaneously, as the instantaneous bandwidth of 1 GHz was arranged for the backends. Each spectrometer provides 16,384 channels, resulting in a spectral resolution of 61 kHz, equivalent to a velocity resolution of about 0.16 km s$^{-1}$ for {$^{12}$CO} and 0.17 km s$^{-1}$ for {$^{13}$CO} and {C$^{18}$O}. The half-power beamwidth of the telescope was about $49''$ for {$^{12}$CO} and $51''$ for {$^{12}$CO} and {C$^{18}$O}, while the pointing accuracy of the telescope was better than $4''$ for each observing epoch. The typical system temperatures during the observations were $\sim$280 K for {$^{12}$CO} and $\sim$185 K for {$^{13}$CO} and {C$^{18}$O}.

All of the CO data used in this study are expressed in brightness temperature, which is the antenna temperature, $T_\mathrm{A}$, divided by the main beam efficiency ($T_\mathrm{R} = T_\mathrm{A}/(f_\mathrm{b} \times \eta_\mathrm{mb}$, $\eta_{mb} = 0.46$, assuming a beam filling factor of $f_\mathrm{b}$ $\sim$1). The calibrated OTF data were then re-gridded to $30''$ pixels and mosaicked to a FITS cube using the GILDAS software package (\citealt{2000ASPC..217..299G}). A first-order baselining was applied to the spectra. The resulting mean rms noise was $\sim$0.45~K for {$^{12}$CO} and $\sim$0.25~K for {$^{13}$CO} and {C$^{18}$O}. These data represent the largest {$^{12}$CO}, {$^{13}$CO}, and {C$^{18}$O} maps to date for such a moderately fine resolution, with fully sampled grids and low noise levels, which are essential for understanding the large-scale structure of the GEM OB1 complex.

\section{Overview}

Our observations cover not only the GEM OB1 complex, but also other MCs that are close to the field-of-view. IC 443 is a supernova remnant (SNR) associated with the GEM OB1 association, and as \citet{2014ApJ...788..122S} have already presented an up-to-date analysis of its properties, it will not be investigated in this paper.

\begin{figure}
   \centering
   \includegraphics[width=\textwidth, angle=0,clip=true,keepaspectratio=true,trim=0 0 0 0mm]{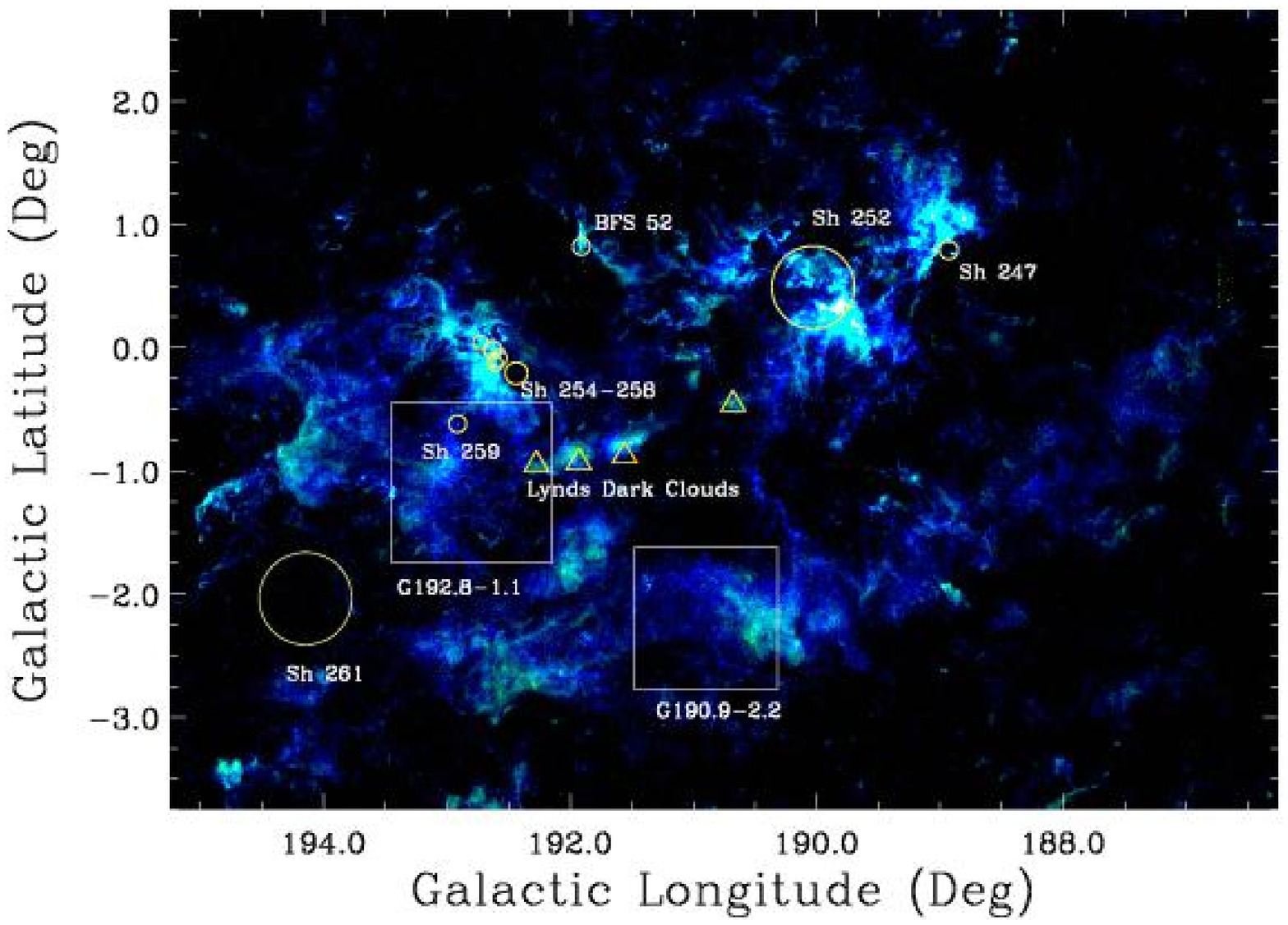}
   \caption{False color map of the GEM OB1 region. Blue, green and red represent the integrated intensities of {$^{12}$CO}, {$^{13}$CO} and {C$^{18}$O}, respectively. The integrated velocity ranges are [-14, 30], [-11, 28] and [-3, 10] {km s$^{-1}$} for {$^{12}$CO}, {$^{13}$CO} and {C$^{18}$O}, respectively. The squares symbolize supernova remnants, the yellow circles indicate \Hii\ regions, and the triangles indicate Lynds Dark Clouds.}
    \label{Fig:figure-1}
\end{figure}

Figure~\ref{Fig:figure-1} presents an integrated map of CO and its other two isotope molecules. Various interesting objects are marked in the figure, including 10 {H{\scriptsize~II}} regions (\citealt{1959ApJS....4..257S}, \citealt{1982ApJS...49..183B}), 4 Lynds dark clouds (\citealt{1962ApJS....7....1L}), and 2 SNRs (\citealt{1969AuJPh..22..211M}, \citealt{1970AuJPh..23..105C}). The {H{\scriptsize~II}} regions Sh 247, Sh 254-258, Sh 252, and BFS 52 are associated with the GEM OB1 association, and are located in the Perseus Arm at a distance of 2 kpc from Earth (\citealt{1995ApJ...445..246C}, \citealt{2009ApJ...693..397R}, \citealt{2010A&A...511A...2R}, \citealt{2011PASJ...63....9N}). The molecular gas with the strongest CO emission is spatially associated with these {H{\scriptsize~II}} regions. The {H{\scriptsize~II}} region Sh 259 is at a distance of 8.71 kpc (\citealt{2015AJ....150..147F}), which we will discuss in more detail in Sect~\ref{Sec:discussion}. We note that there is no CO emission along the line-of-sight (LoS) to Sh 261. The Lynds dark clouds are located in the middle of Figure~\ref{Fig:figure-1} and they form a straight chain.

\begin{figure}
   \centering
   \includegraphics[width=\textwidth, angle=0,clip=true,keepaspectratio=true,trim=0 0 0 0mm]{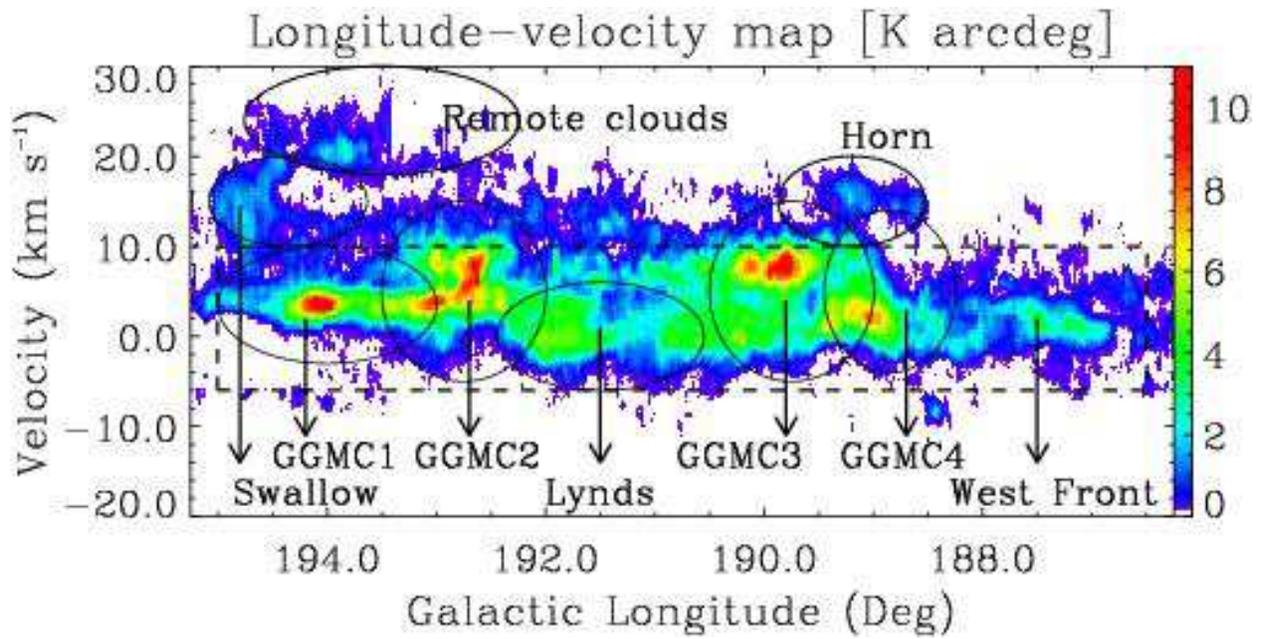}
   \caption{Longitude-velocity map of {$^{12}$CO} emission integrated over the latitude of $6.5^{\circ}$. The rectangle symbolizes the West Front, and the ellipses present the other clouds.}
    \label{Fig:LV}
\end{figure}

The velocity distribution of the clouds in this area is presented in the longitude-velocity ($L-V$) map shown in Figure~\ref{Fig:LV}. The local standard of rest (LSR) velocity range of all the molecular gas is confined to [-14, 30] {km s$^{-1}$}, and most of the gas is concentrated to the range [-3, 10] {km s$^{-1}$}.

\begin{figure}
   \centering
   \subfigure[]{
   \label{Fig:figure-3-A}
   \includegraphics[width=0.48\textwidth, angle=0,clip=true,keepaspectratio=true,trim=0 0 0 0mm]{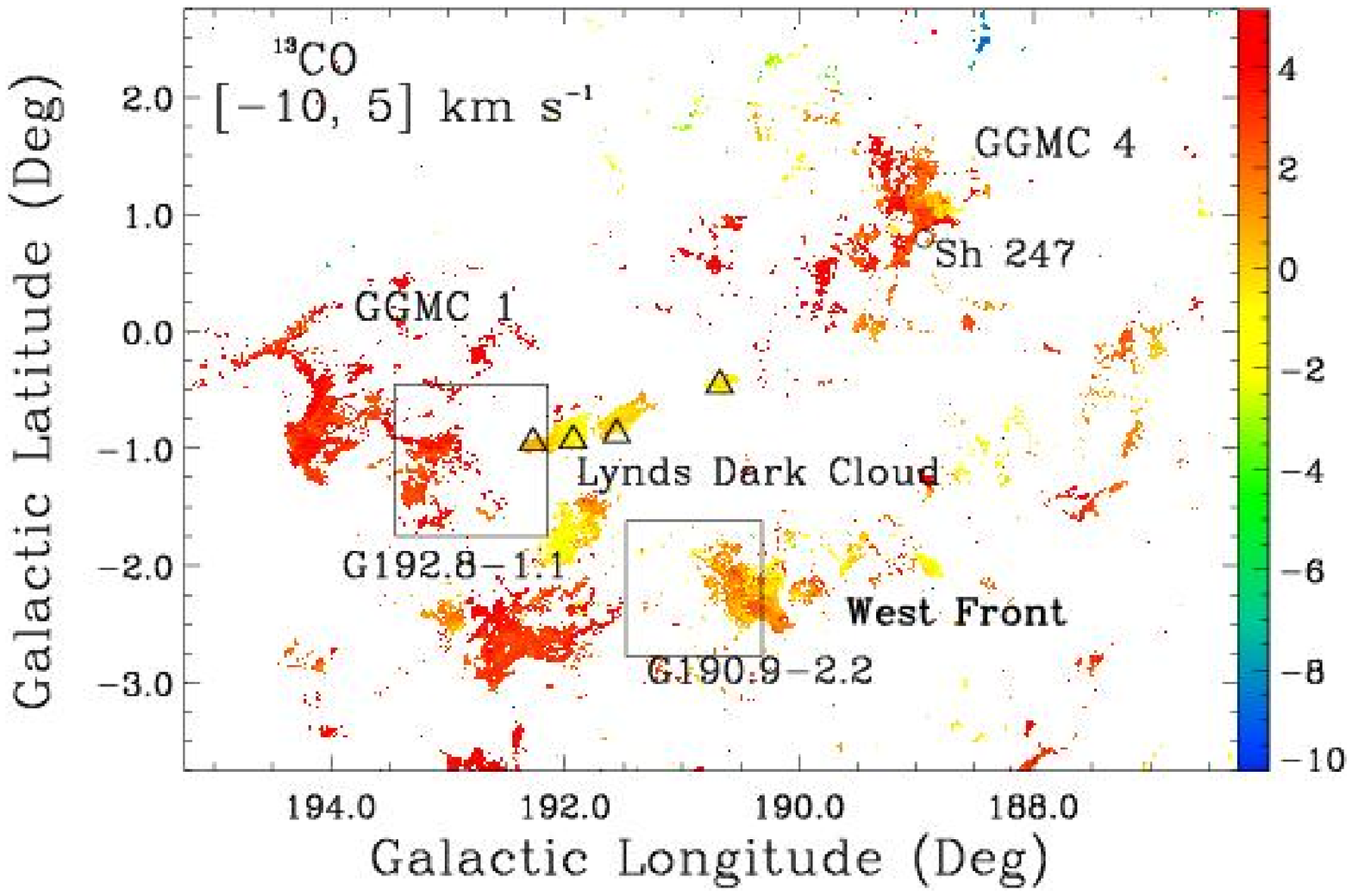}}
   \subfigure[]{
   \label{Fig:figure-3-B}
   \includegraphics[width=0.46\textwidth, angle=0,clip=true,keepaspectratio=true,trim=0 0 0 0mm]{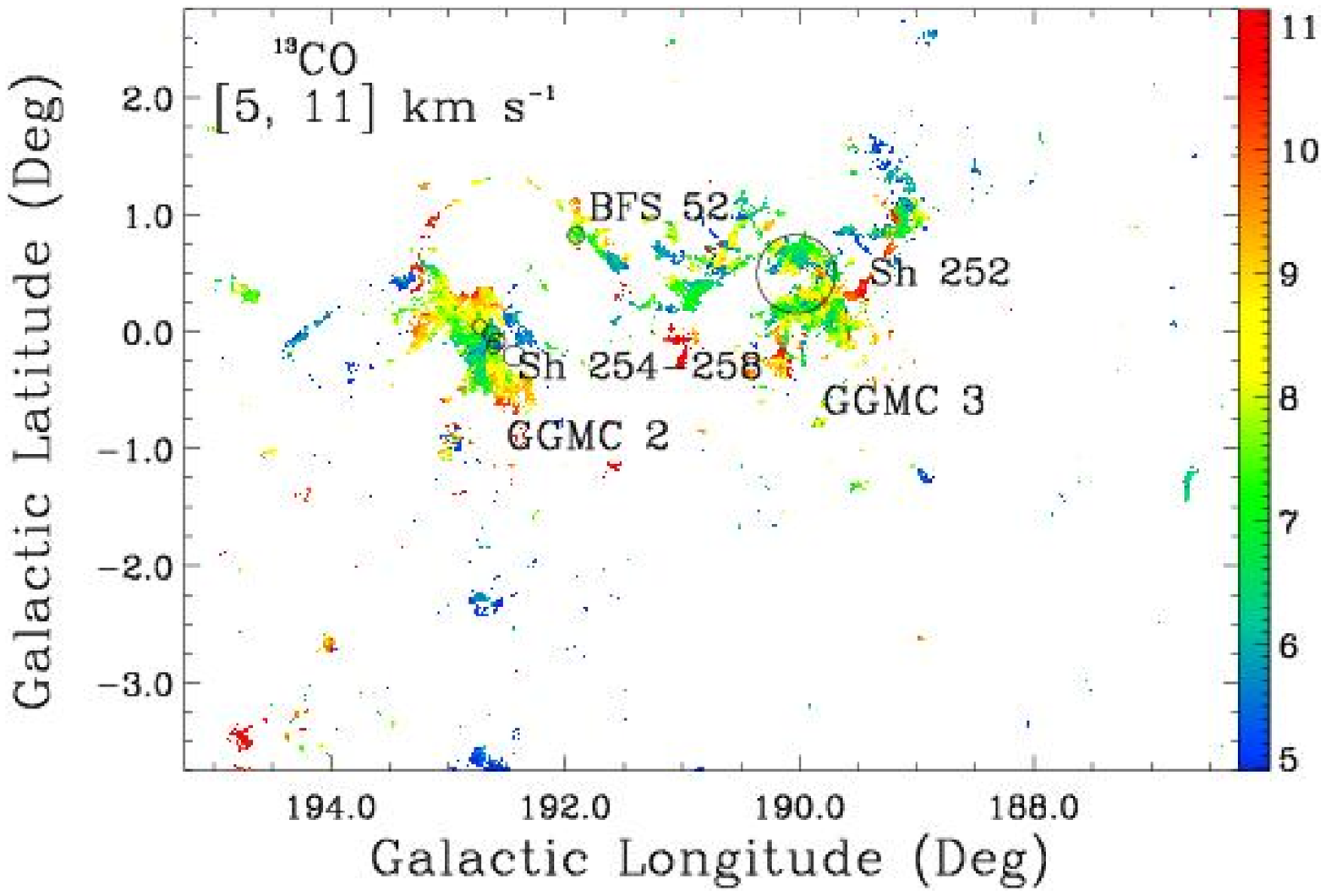}}
   \subfigure[]{
   \label{Fig:figure-3-C}
   \includegraphics[width=0.48\textwidth, angle=0,clip=true,keepaspectratio=true,trim=0 0 0 0mm]{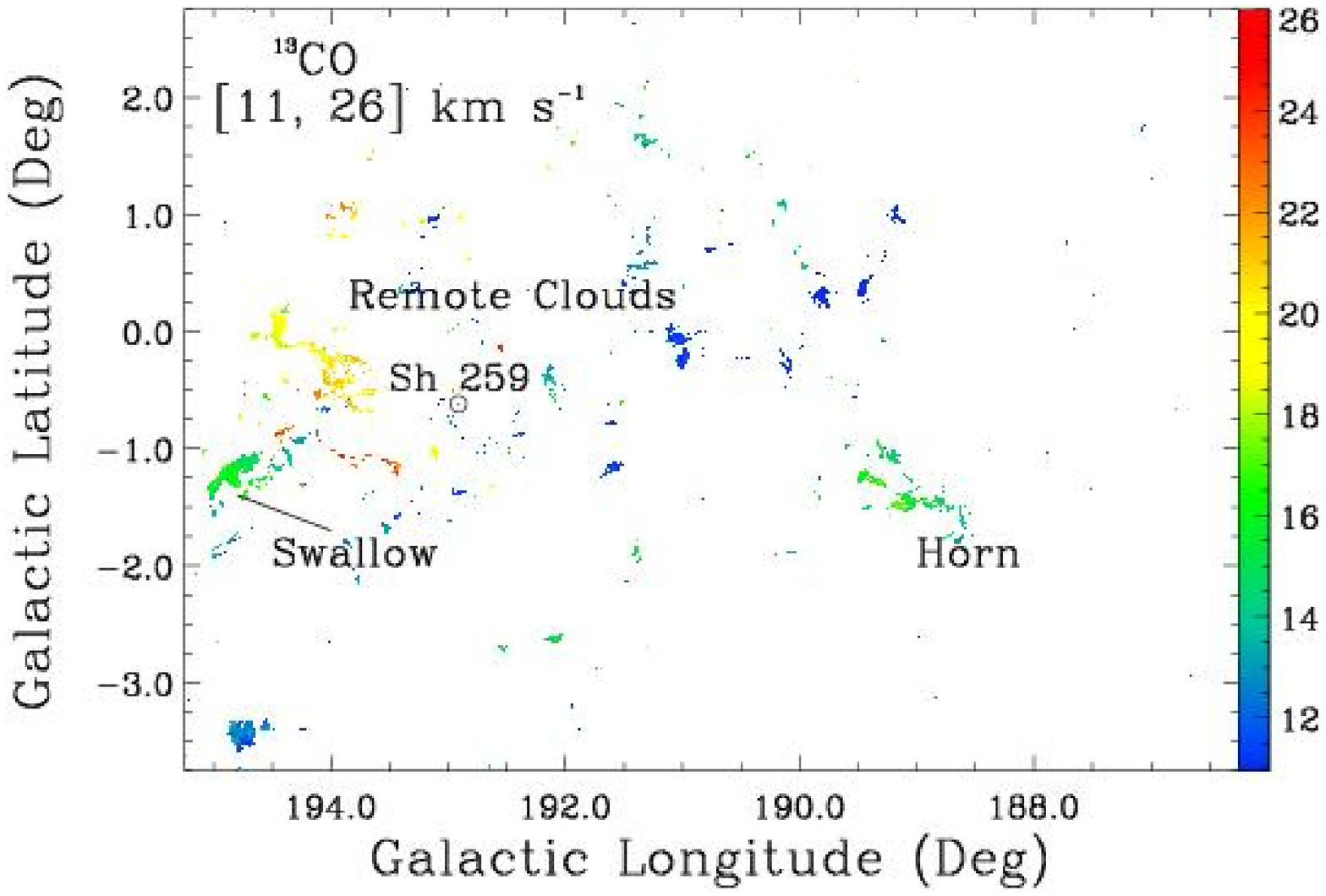}}

   \caption{Peak-velocity map of {$^{13}$CO}. The color indicates the velocity of the peak spectrum. Panel (a) shows a peak-velocity map of {$^{13}$CO} in the velocity range of [-10, 5] {km s$^{-1}$}, while (b) and (c) show the map in the velocity ranges of [5, 11] {km s$^{-1}$} and [11, 26] {km s$^{-1}$}. The meanings of the symbols are the same as in Fig.1.}
    \label{Fig:figure-3}
\end{figure}

\begin{figure}
   \centering
   \includegraphics[width=0.8\textwidth, angle=0,clip=true,keepaspectratio=true,trim=0 0 0 0mm]{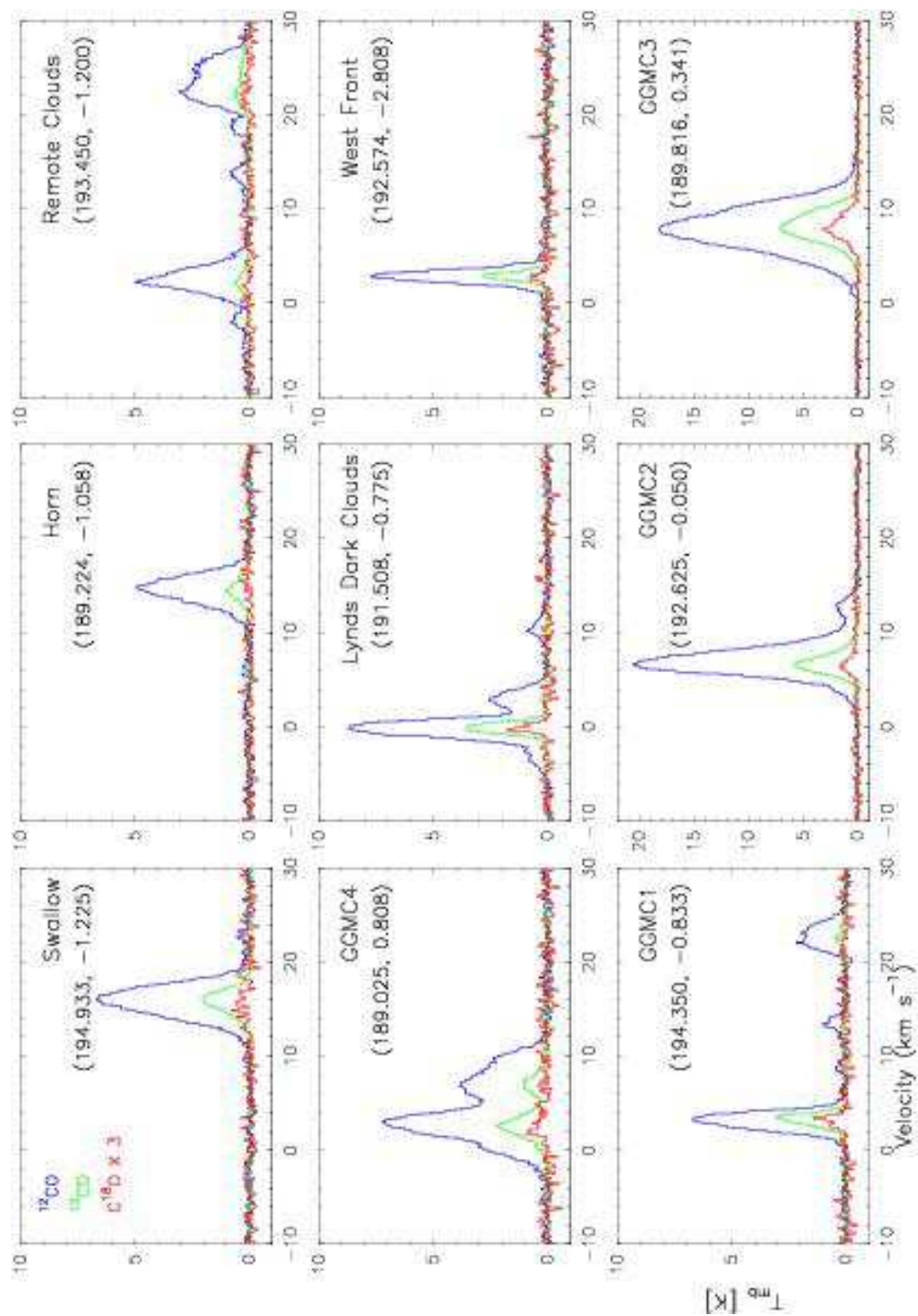}
   \caption{Single-point spectra of {$^{12}$CO} (blue), {$^{13}$CO} (green), and {C$^{18}$O} (red) of the clouds. The top right corners of the panels show the names of clouds and positions of spectra using Galactic coordinates. Note that the vertical coordinate scales of spectra from GGMC 2 and GGMC 3 are different from others.}
    \label{Fig:typicalspec}
\end{figure}

In order to distinguish clouds with different velocity ranges, we separated the velocity into three bins according to the peak positions of the spectra. The three velocity ranges are [-10, 5], [5, 11], [11, 26] {km s$^{-1}$}, as shown in Figure~\ref{Fig:figure-3}. Based on the observations and the discussions of \citet{1995ApJ...445..246C, 1995ApJ...450..201C}, \citet{1998ApJS..117..387K}, \citet{2009ApJ...693..397R}, \citet{2010A&A...511A...2R}, and \citet{2011PASJ...63....9N}, we suggest that the area is mainly divided into the 12 MCs that are shown in Figure~\ref{Fig:figure-3}. The details of the MCs are listed in Table~\ref{tbl-1}. Two new clouds have been discovered in this survey: ``Swallow'' and ``Horn.'' Figure~\ref{Fig:typicalspec} shows individual {$^{12}$CO}, {$^{13}$CO}, and {C$^{18}$O} spectra of the clouds.

In addition to the 12 MCs, there is some gas in the range of [10, 30] {km s$^{-1}$} (Figure~\ref{Fig:LV}). The gas shows two elongated structures in the $L-V$ map, one connecting the Remote Clouds and Horn, and another connecting Swallow and Horn. We will study these elongated structures in detail in future works.

The Gemini OB1 giant molecular cloud (GGMC) complex is the largest MC complex in this area, which includes GGMC~1, GGMC~2, GGMC~3, and GGMC~4. Detailed information about these clouds is presented in Sect~\ref{Sec:ggmc}. Sect~\ref{Sec:lynds} studies the Lynds dark clouds, which are located 400 pc away, while the West Front clouds are introduced in Sect~\ref{Sec:west}. Sect~\ref{Sec:others} presents the two newly discovered Swallow and Horn MCs, while the Remote Clouds are also introduced in this section.

\section{The Gemini OB1 molecular cloud complex}
\label{Sec:ggmc}

The GGMC complex is mainly located in the upper part of Figure~\ref{Fig:figure-1}, where four independent giant molecular clouds (GMCs) are shown in the upper parts of Figure~\ref{Fig:figure-3}. The size of the complex is 6.6$^{\circ}$ $\times$ 2.8$^{\circ}$ along the direction of the GGMC arrangement. The complex is located in the Perseus arm, and it is associated with Sh~247, Sh~252, Sh~254-258, BFS~52, and the GEM~OB1 association, and its distance is suggested to be 2 kpc (\citealt{1995ApJ...445..246C}, \citealt{2009ApJ...693..397R}, \citealt{2010A&A...511A...2R}, \citealt{2011PASJ...63....9N}).

\subsection{GGMC~1}

\begin{figure}
   \centering
   \subfigure[]{
   \label{Fig:figure-GGMC1-A}
   \includegraphics[width=0.44\textwidth, angle=0,clip=true,keepaspectratio=true,trim=0 0 0 0mm]{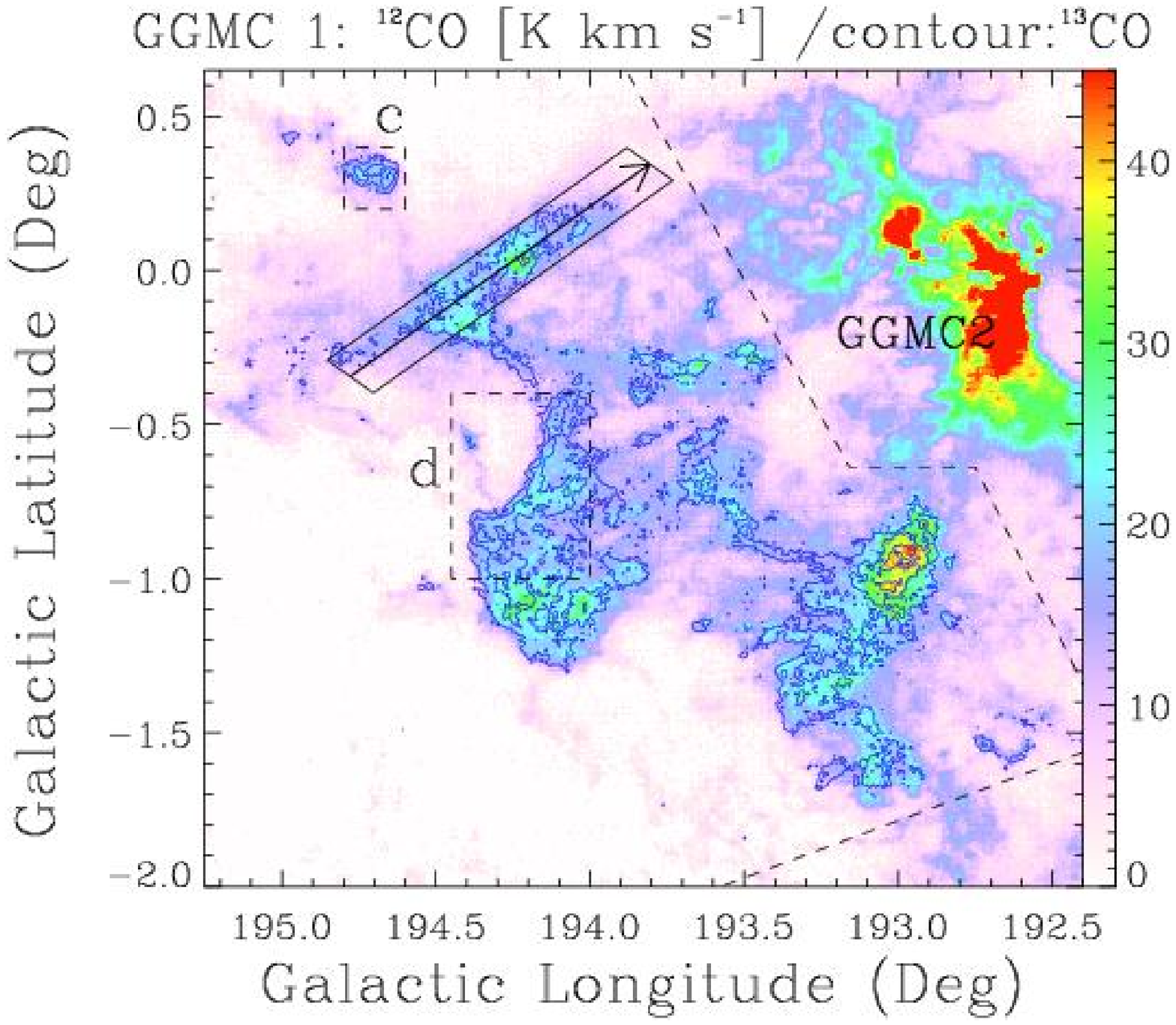}}
   \subfigure[]{
   \label{Fig:figure-GGMC1-B}
   \includegraphics[width=0.44\textwidth, angle=0,clip=true,keepaspectratio=true,trim=0 0 0 0mm]{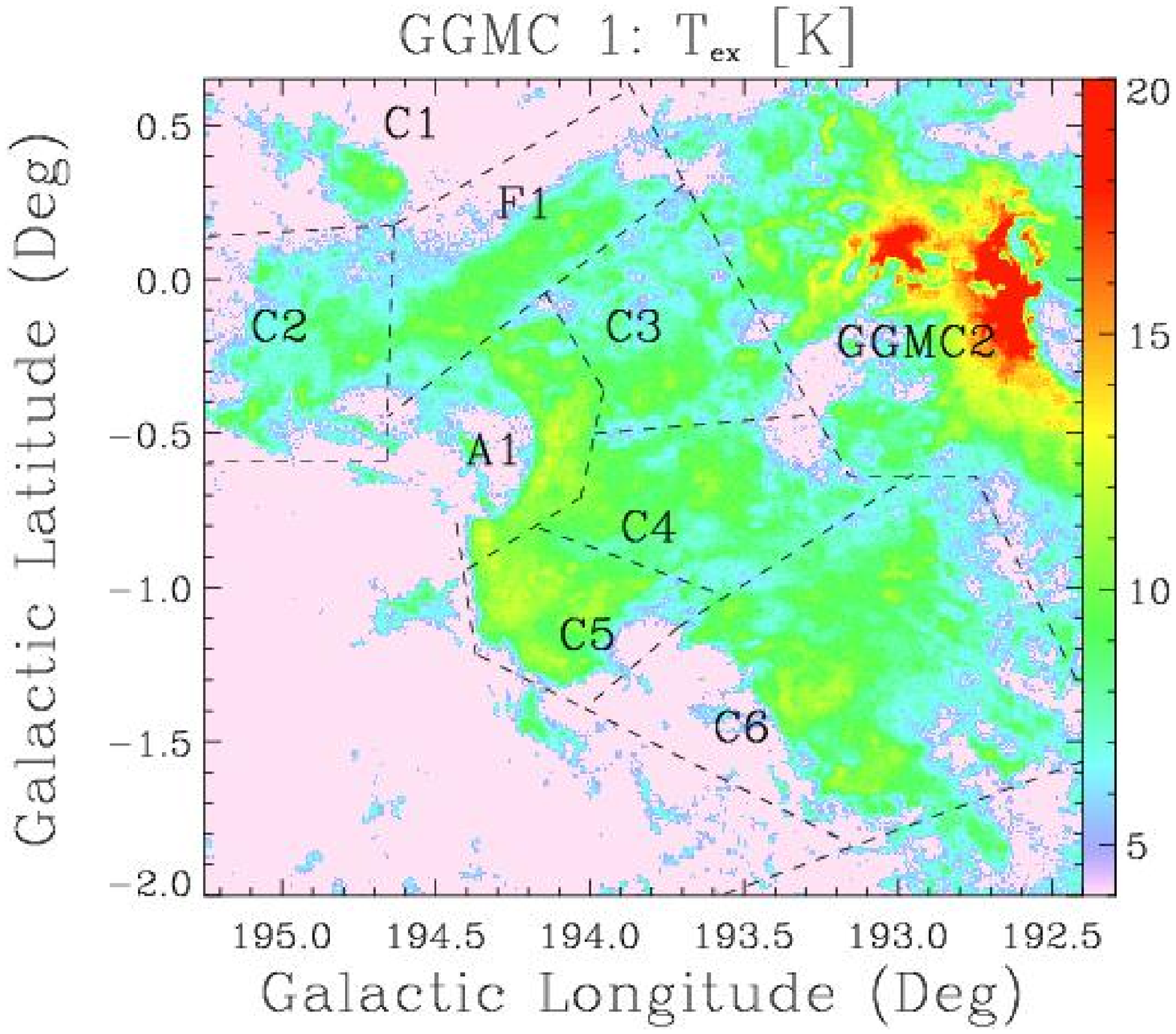}}
   \subfigure[]{
   \label{Fig:figure-GGMC1-C2}
   \includegraphics[width=0.3\textwidth, angle=0,clip=true,keepaspectratio=true,trim=0 0 0 0mm]{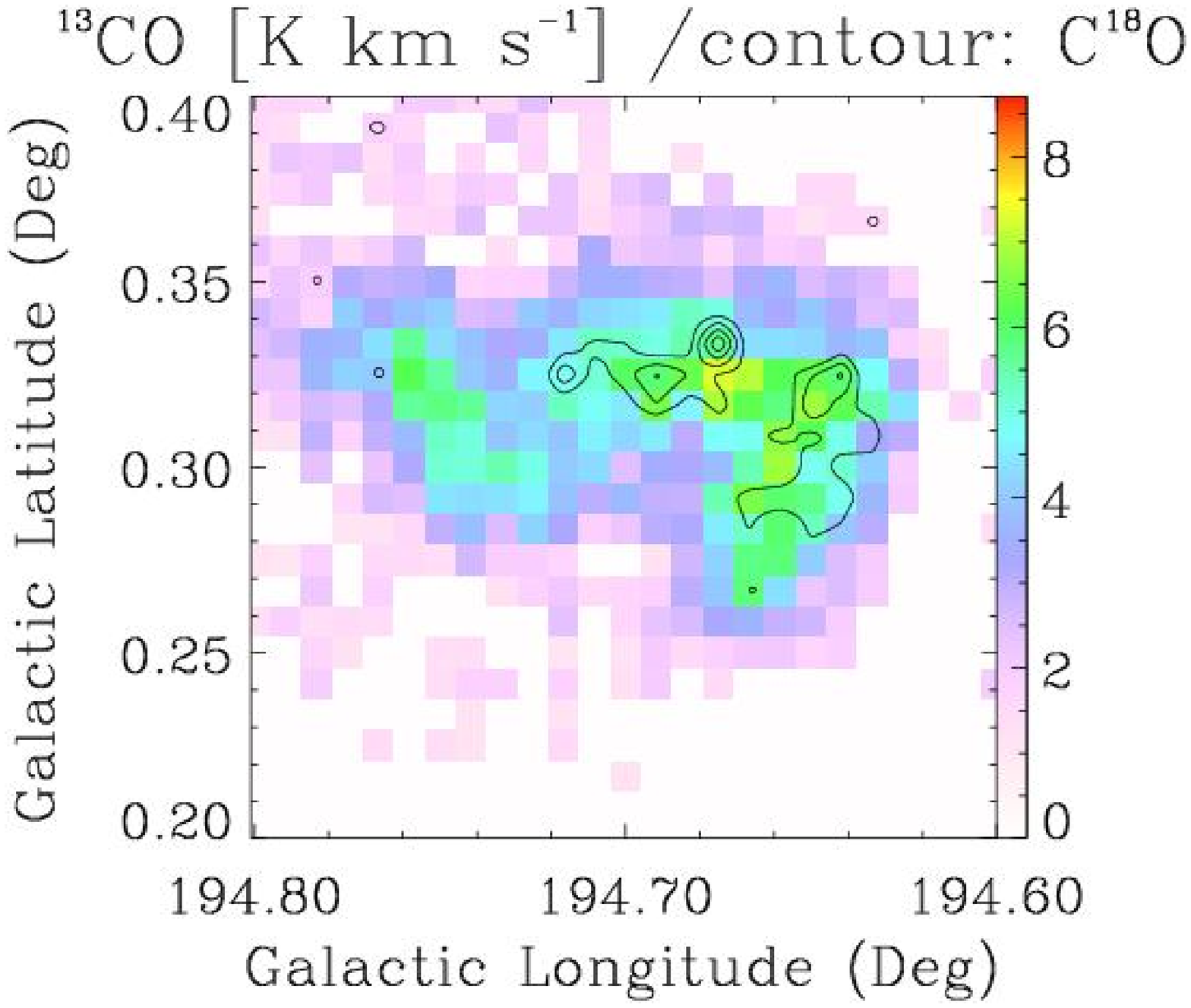}}
   \subfigure[]{
   \label{Fig:figure-GGMC1-C3}
   \includegraphics[width=0.3\textwidth, angle=0,clip=true,keepaspectratio=true,trim=0 0 0 0mm]{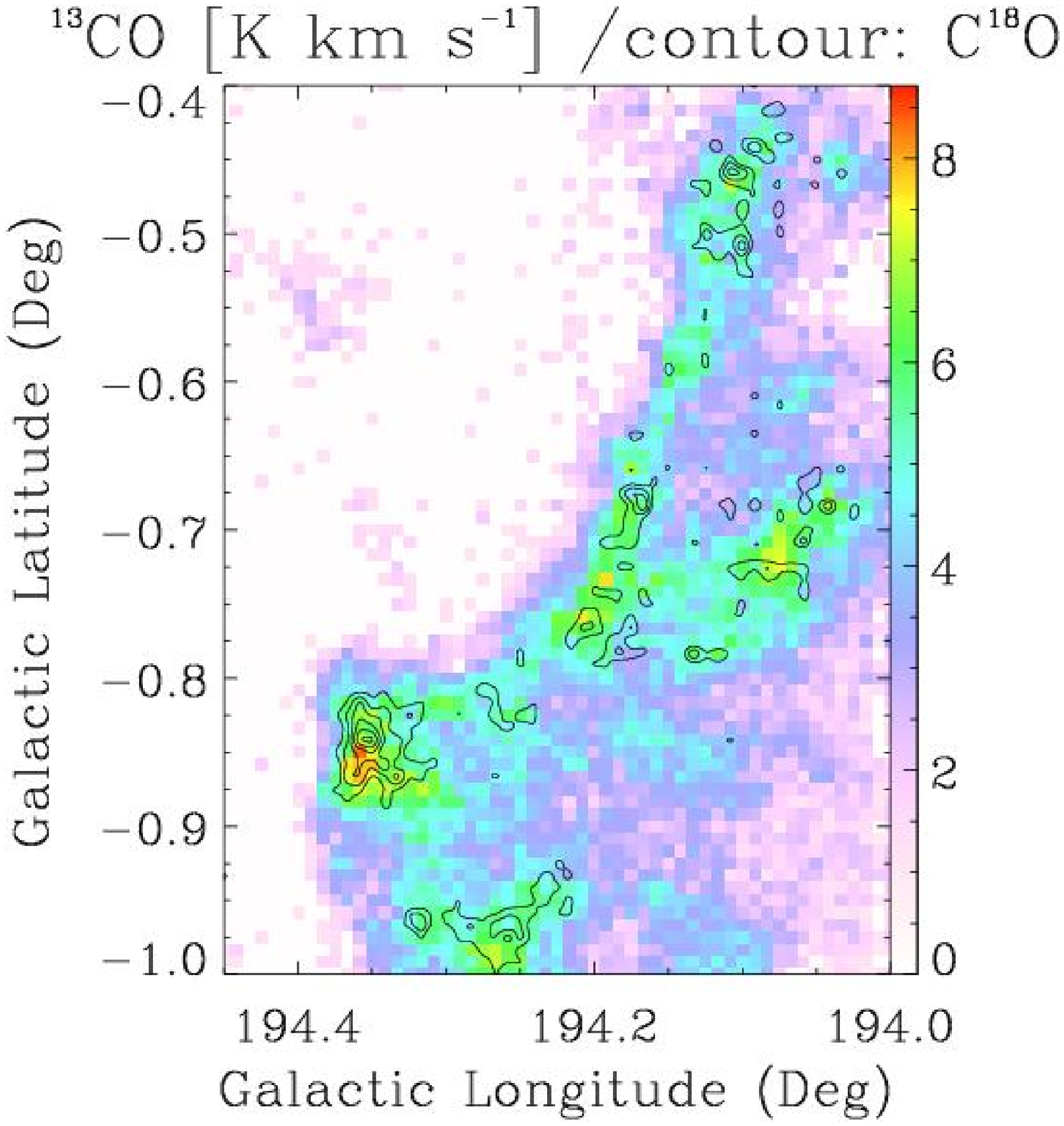}}

   \caption{Gas distribution of GGMC1. (a) The background shows {$^{12}$CO} integrated intensity in the range of [-3,~10] {km s$^{-1}$}. The contours indicate the {$^{13}$CO} integrated intensity in the range of [-0.3,~10] {km s$^{-1}$}, and the steps are 6, 12, 18 $\times$ 0.36 K \kms(1$\sigma$). The dashed rectangles show boundaries of Figure~\ref{Fig:figure-GGMC1-C2} and Figure~\ref{Fig:figure-GGMC1-C3}. The solid arrow and rectangle show the direction and width of the P-V diagram of Figure~\ref{Fig:figure-ray1pv}. (b) The excitation temperature map of GGMC1. The dashed lines show the boundaries of the subregions. (c) and (d) The background shows {$^{13}$CO} integrated intensity. The contours indicate the {C$^{18}$O} integrated intensity in the range of [2, 9] {km s$^{-1}$}. }
    \label{Fig:figure-GGMC1}
\end{figure}

GGMC~1 is located to the south of the complex (in this paper, we define north as the top and east as the left, in Galactic Coordinates) and has a velocity interval of [-3, 10] {km s$^{-1}$} (Figure~\ref{Fig:figure-3-A}). The {$^{12}$CO} emission shows the overall distribution of the clouds (Figure~\ref{Fig:figure-GGMC1-A}). Two distinct structures can be identified: one is a filamentary structure near the top of the cloud, and the other is a semi-circular structure in the middle of the cloud. These structures are more clearly shown in the {$^{13}$CO} emission contours in Figure~\ref{Fig:figure-GGMC1-A}. We note that very little CO emission is located at the center of the semi-circular structure, which will be discussed in future papers. The {C$^{18}$O} emission in this cloud is relatively weak, and it is mainly located to the west of the semi-circular structure (Figure~\ref{Fig:figure-GGMC1-C3}) and northeast of the cloud (Figure~\ref{Fig:figure-GGMC1-C2}). We divided the cloud into eight subregions (Figure~\ref{Fig:figure-GGMC1-B}), and their detailed information is shown in Table~\ref{tbl-GGMC1}.

Assuming {$^{12}$CO} is optically thick, the excitation temperature can be calculated (\citealt{1997ApJ...476..781B}) as
\begin{equation}
T_{\mathrm{ex}}=\frac{h\nu_{12}/k}{\ln(1+\frac{h\nu_{12}/k}{T_{\mathrm{MB,^{12}CO}}+\frac{h\nu_{12}/k}{\exp(h\nu_{12}/T_{\mathrm{bg}}-1)}})}
\end{equation}
where $h$ and $k$ are the Planck constant and Boltzmann constant, respectively. $\nu_{12}$ and $T_{\mathrm{MB,^{12}CO}}$ are the frequency and peak main beam temperature of {$^{12}$CO}, respectively. $T_{\mathrm{bg}}=2.7$~K is the temperature of the cosmic microwave background. When the constants are substituted for their numerical values, the formula can be simplified to
\begin{equation}\label{eq:Tex}
T_{\mathrm{ex}}=\frac{5.532}{\ln(1+\frac{5.532}{T_{\mathrm{MB,^{12}CO}}+0.819})} \quad \mathrm{K}
\end{equation}

Figure~\ref{Fig:figure-GGMC1-B} shows the distribution of the excitation temperatures of GGMC~1, which are within the range of 4.5-13 K. The molecular gas near the semi-circular structure has higher temperatures than the other regions.

\begin{figure}
   \centering
   \subfigure[]{
   \label{Fig:figure-ray1pv-A}
   \includegraphics[width=0.465\textwidth, angle=0,clip=true,keepaspectratio=true,trim=0 0 0 0mm]{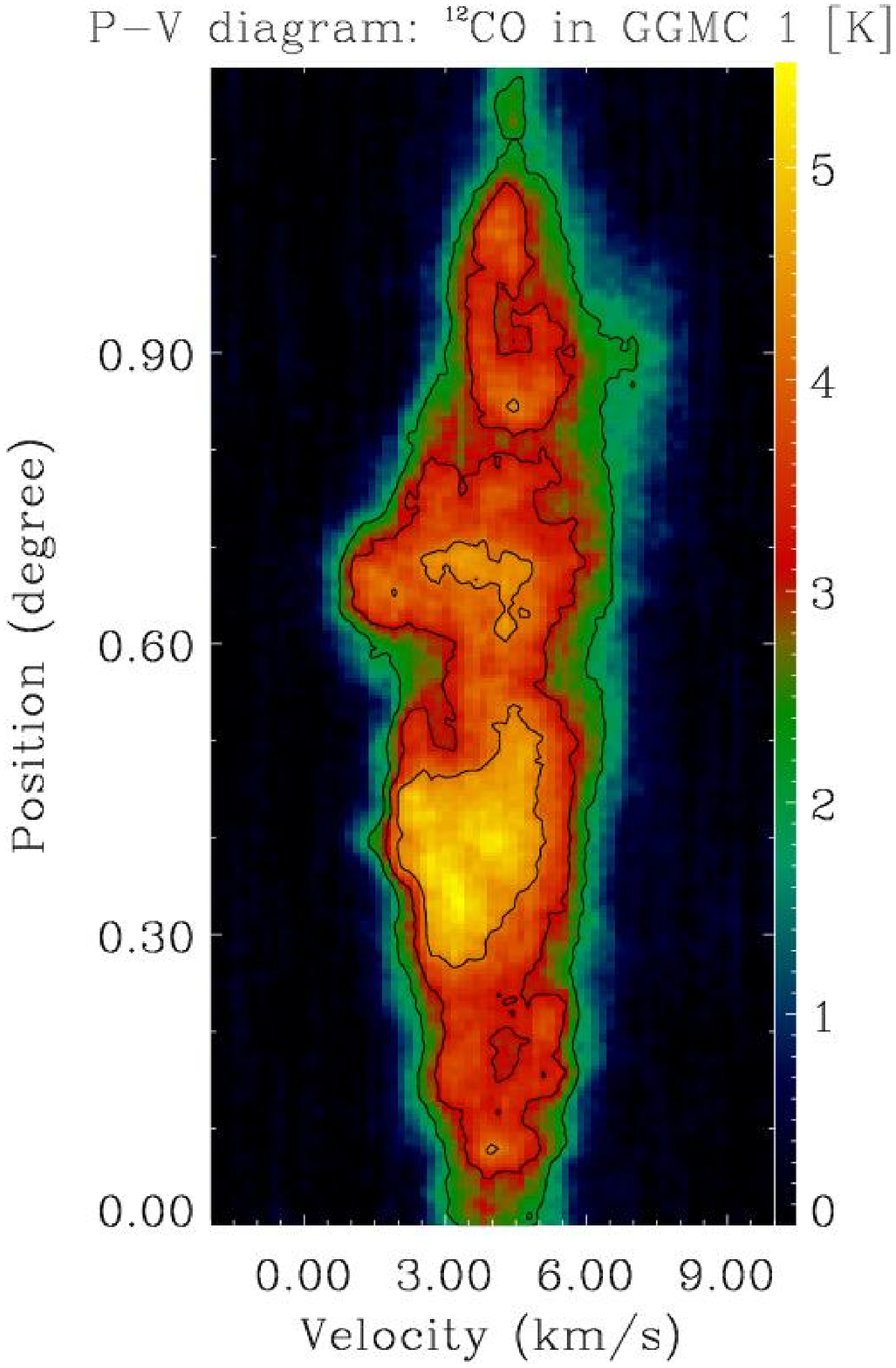}}
   \subfigure[]{
   \label{Fig:figure-ray1pv-B}
   \includegraphics[width=0.46\textwidth, angle=0,clip=true,keepaspectratio=true,trim=0 0 0 0mm]{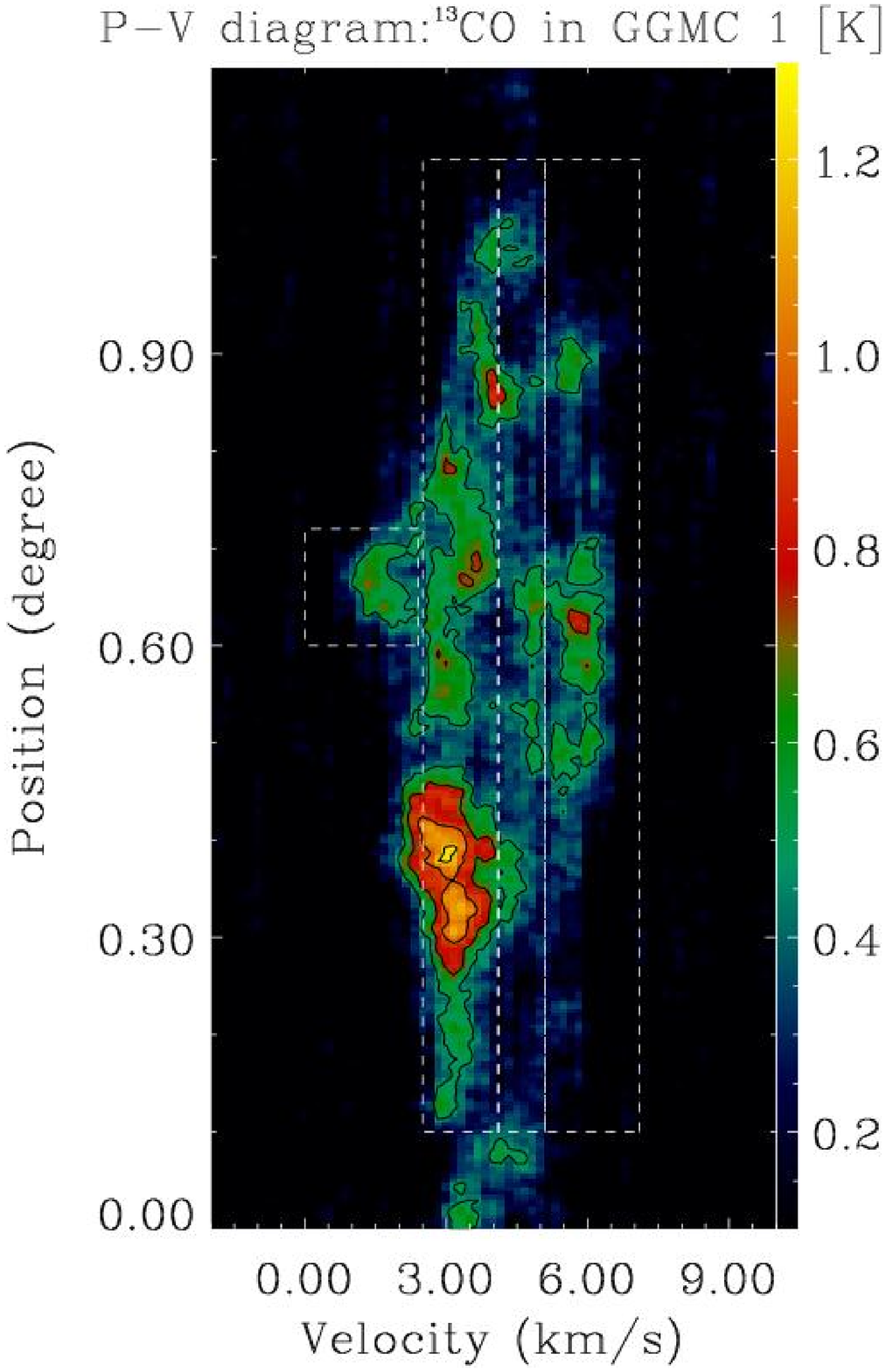}}

   \caption{(a) P-V diagram of {$^{12}$CO} along the arrows in Figure~\ref{Fig:figure-GGMC1-A}; the width of the belt is 22 pixel ($11'$, solid rectangle in Figure~\ref{Fig:figure-GGMC1-A}). (b) P-V diagram of {$^{13}$CO} along the arrows in Figure~\ref{Fig:figure-GGMC1-A}; the width of the belt is the same as in (a). Four velocity components can be identified in {$^{13}$CO} (dashed lines in the figure). Figure~\ref{Fig:figure-ray1} shows the spatial distribution of those components.}
    \label{Fig:figure-ray1pv}
\end{figure}

\begin{figure}
   \centering
   \includegraphics[width=\textwidth, angle=0,clip=true,keepaspectratio=true,trim=0 0 0 0mm]{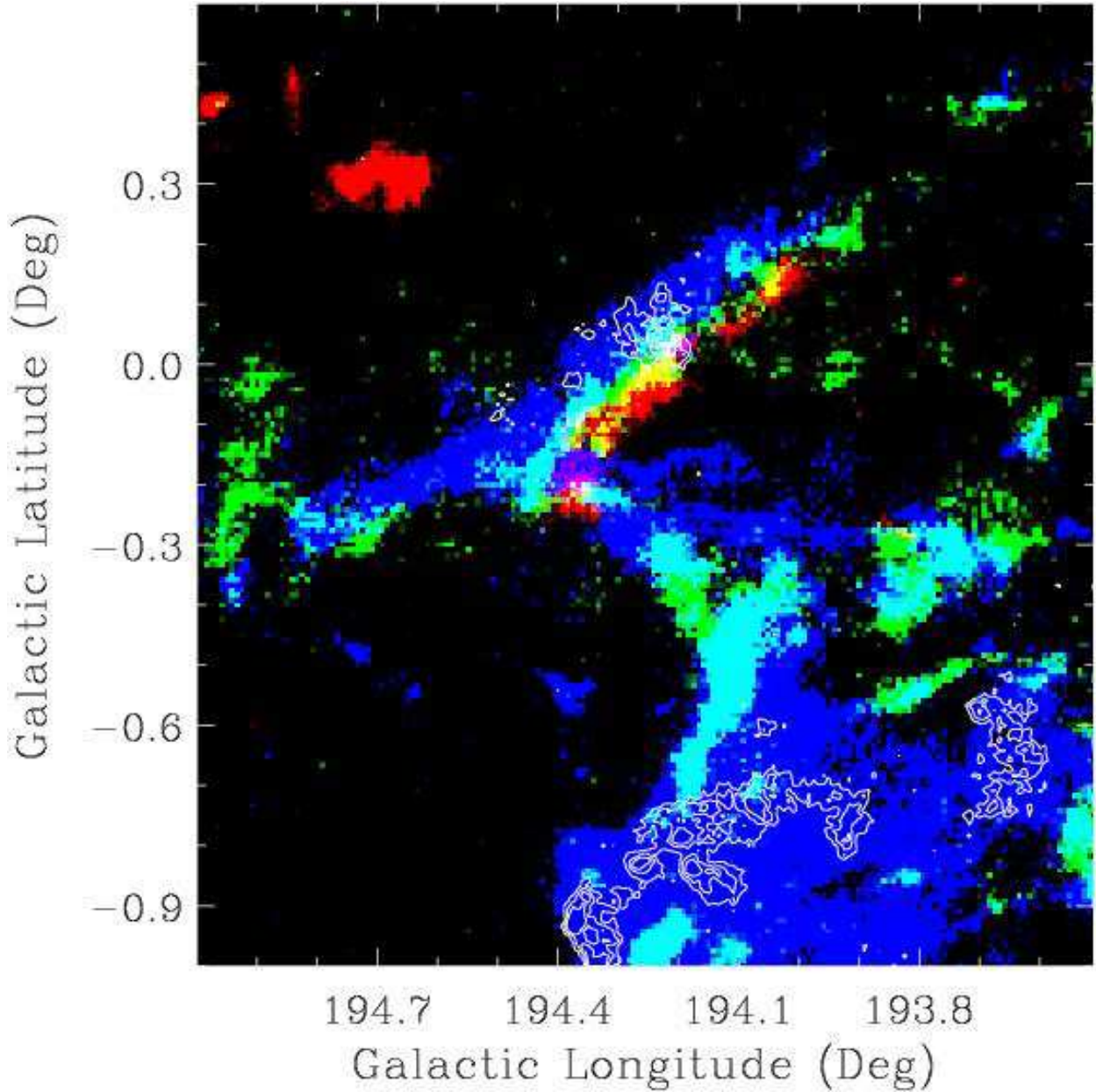}
   \caption{False color map of {$^{13}$CO} in GGMC1. Blue, green and red represent the integrated intensity of {$^{13}$CO} within [2.5,~4.1], [4.2,~5.1], and [5.2,~7.2] {km s$^{-1}$} respectively. The white contours indicate the integrated intensity of {$^{13}$CO} within [0,~2.4] {km s$^{-1}$}, and the contour levels begin at 0.8 K {km s$^{-1}$} with increments of 0.48 K {km s$^{-1}$} (3$\sigma$).}
    \label{Fig:figure-ray1}
\end{figure}

By analyzing the kinematic information of GGMC~1, we find that subregion F1 has a special velocity structure. Figure~\ref{Fig:figure-ray1pv} shows the position-velocity (P-V) diagram of F1 (the black arrow has a width of 11 arcmin in Figure~\ref{Fig:figure-GGMC1-A}). There is no obvious velocity structure from the {$^{12}$CO} data, but four velocity components can be clearly seen from the {$^{13}$CO} data. The velocity ranges of these components, respectively, are [0, 2.4], [2.4, 4.1], [4.1, 5.1], and [5.1, 7.2] {km s$^{-1}$}. Figure~\ref{Fig:figure-ray1} shows the distribution of the integrated intensity of these components. Component 1 ([0, 2.4] {km s$^{-1}$}) is located in the middle of the filament, while the others orderly line up along the axis of the filament. The interesting velocity structure of filament F1 will be discussed in the future. Subregion C6, which is in the southwest of Figure~\ref{Fig:figure-GGMC1-B}, overlaps with SNR G~192.8-1.1 (\citealt{1969AuJPh..22..211M}, \citealt{1970AuJPh..23..105C}) along the LoS. We will discuss this association in our follow-up paper.

The mass of GGMC~1 is 2 $\times$ $10^{5}$ M$_{\odot}$, according to the {$^{12}$CO} integral intensity and X-factor (1.8 $\times$ $10^{20}$ cm$^{-2}$ (K km$^{-1}$)$^{-1}$, \citealt{2001ApJ...547..792D}). The LTE molecular gas mass derived from {$^{13}$CO} is 3.6 $\times$ $10^{4}$ M$_{\odot}$ (\citealt{1997ApJ...476..781B}, N(H$_{2}$)/N(\coii) = 7 $\times$ $10^{5}$), while the LTE molecular gas mass derived from \coiii\ is 2.0 $\times$ $10^{3}$ M$_{\odot}$ (\citealt{1997ApJ...476..781B}, N(H$_{2}$)/N(\coiii) = 7 $\times$ $10^{6}$).

\subsection{GGMC~2}

\begin{figure}
   \centering
   \subfigure[]{
   \label{Fig:figure-GGMC2-A}
   \includegraphics[width=0.45\textwidth, angle=0,clip=true,keepaspectratio=true,trim=0 0 0 0mm]{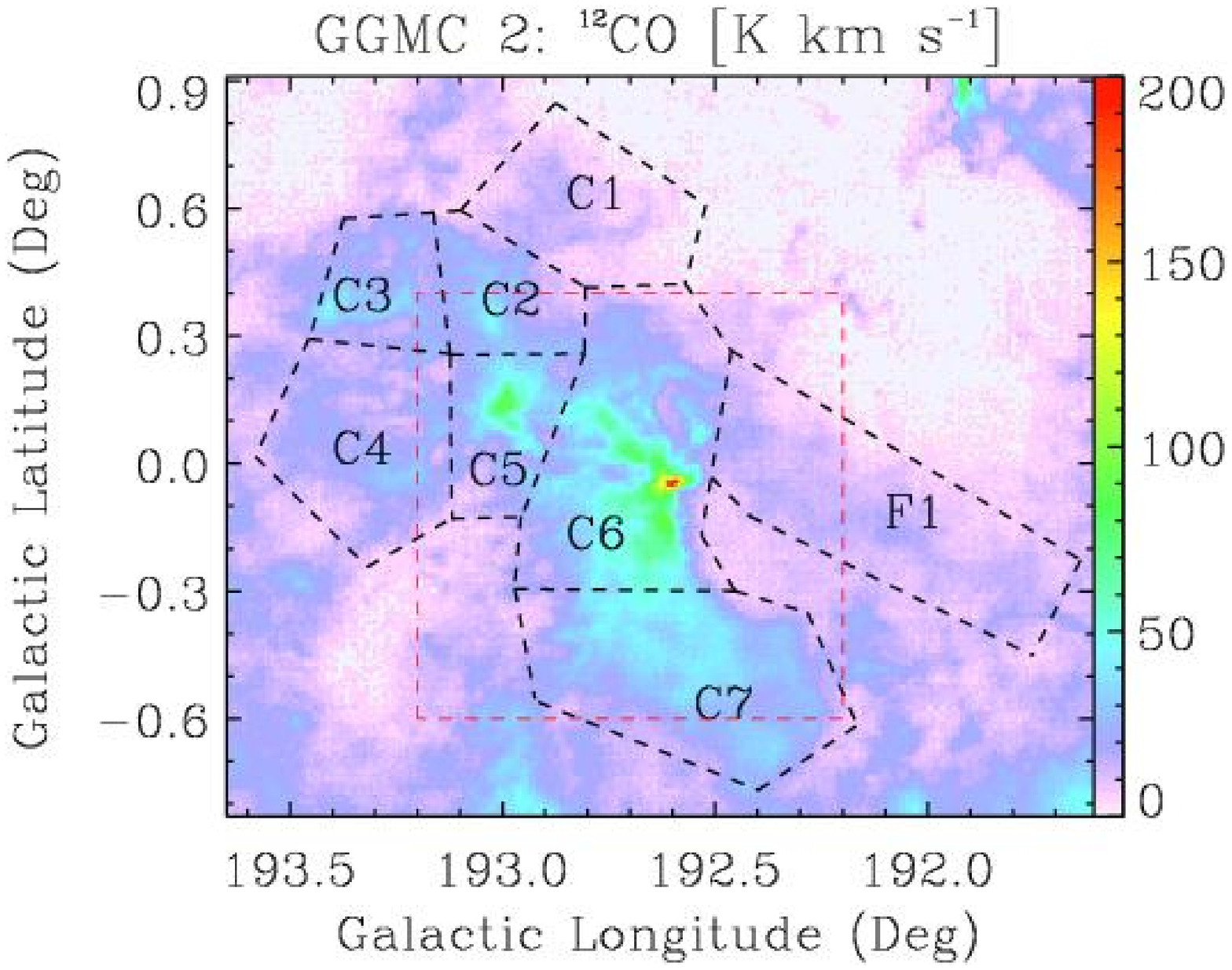}}
   \subfigure[]{
   \label{Fig:figure-GGMC2-B}
   \includegraphics[width=0.425\textwidth, angle=0,clip=true,keepaspectratio=true,trim=0 0 0 0mm]{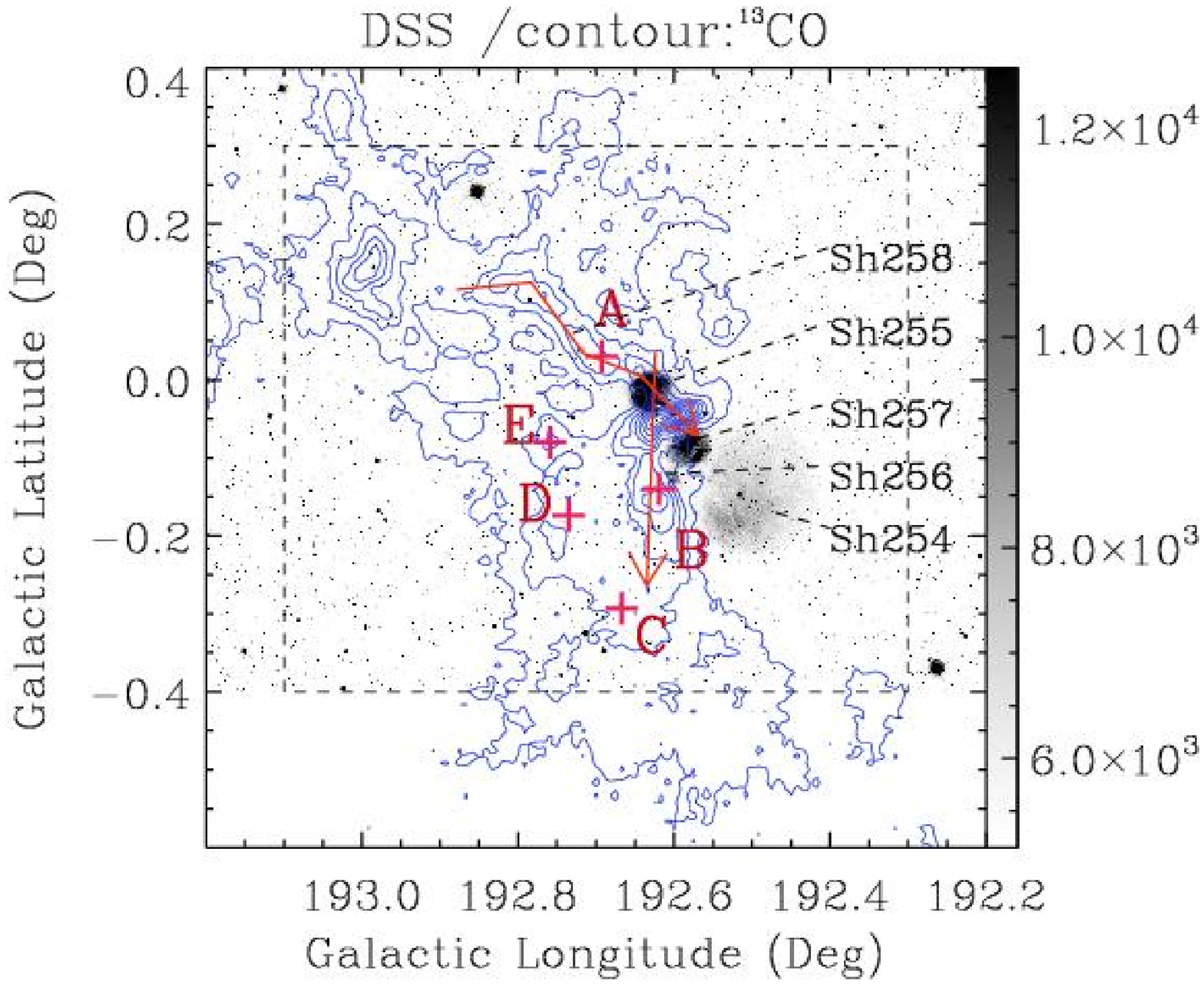}}
   \subfigure[]{
   \label{Fig:figure-GGMC2-C}
   \includegraphics[width=0.45\textwidth, angle=0,clip=true,keepaspectratio=true,trim=0 0 0 0mm]{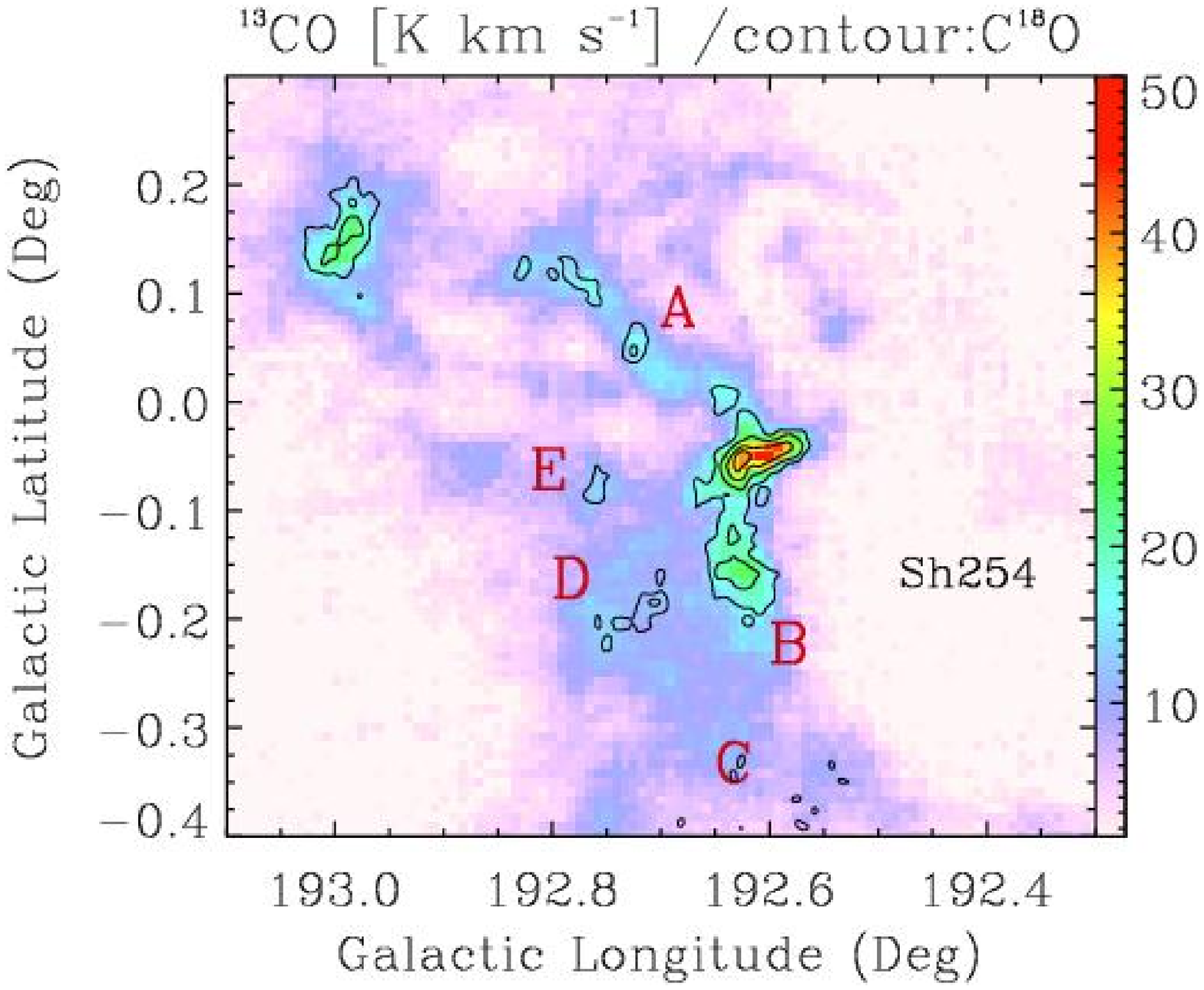}}
   \subfigure[]{
   \label{Fig:figure-GGMC2-D}
   \includegraphics[width=0.45\textwidth, angle=0,clip=true,keepaspectratio=true,trim=0 0 0 0mm]{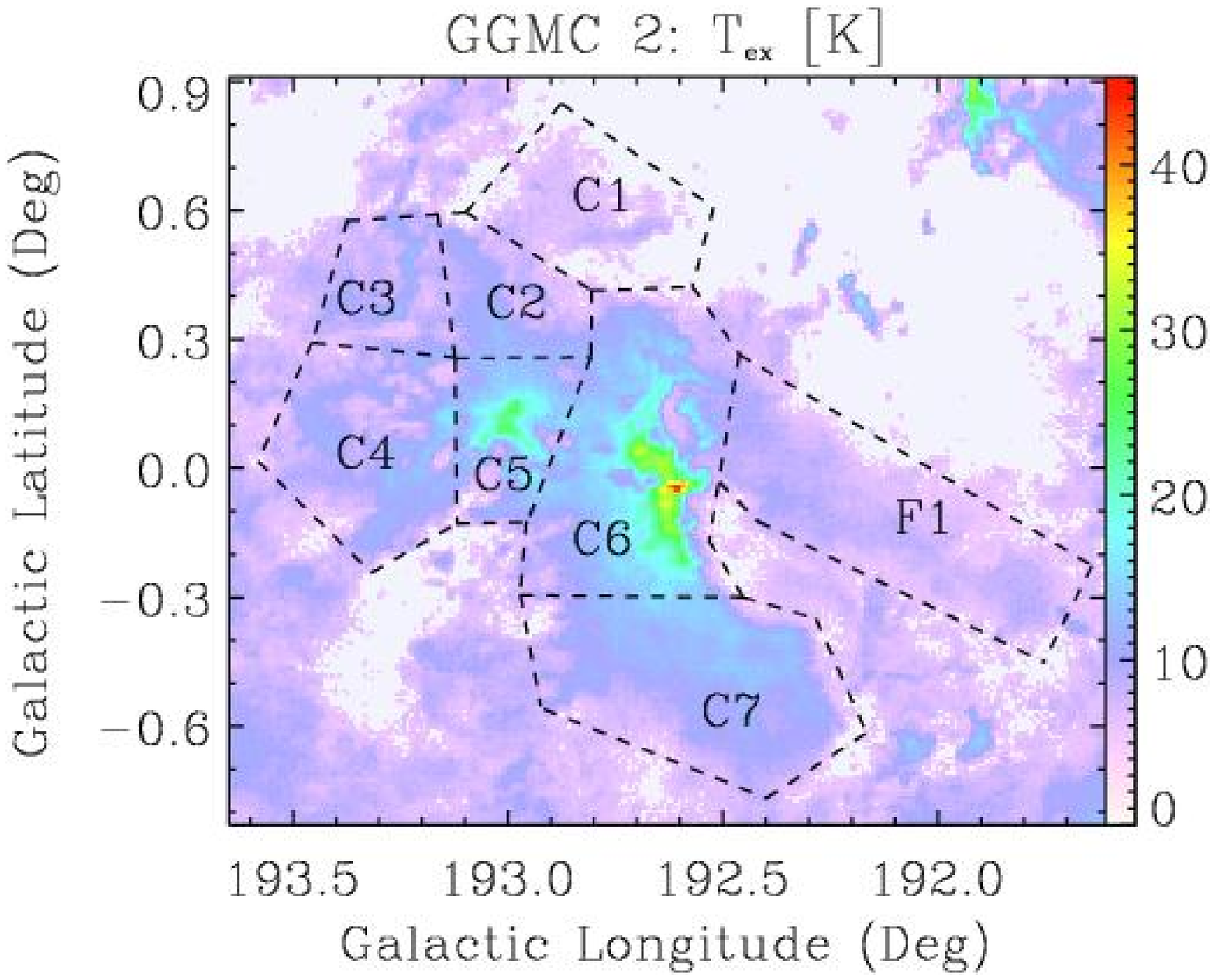}}

   \caption{Gas distribution of the GGMC2 main body. (a) {$^{12}$CO} integrated intensity map in the range of [-5, 16] {km s$^{-1}$}. The dashed lines show the boundaries of the subregions. The red rectangle shows area of Figure~\ref{Fig:figure-GGMC2-B}. (b) The background is from the Palomar Digital Sky Survey. The contours indicate the integrated intensity of {$^{13}$CO} in GGMC~2 with the range of [-1.5, 16] \kms; the steps are 10, 20, 25, 30, 35, 40, 50, 70, 100, 120, 130 $\times$ 0.42 K \kms(1$\sigma$). The box represents the special region (\citealt{2009AJ....138..975B}). The crosses represent five substructures. The two broken lines show the position and direction of the P-V diagrams of Figure~\ref{Fig:figure-GGMC2pv}. The rectangle shows the boundaries of Figure~\ref{Fig:figure-GGMC2-C}. (c) Integrated intensity map of \coii. The contours indicate the integrated intensity of \coiii\ in GGMC~2 with the range of [4,~11] \kms; the steps are 5, 8, 12, 15 $\times$ 0.26 K \kms(1$\sigma$). (d) Excitation temperature map of GGMC2. The dashed lines show the boundaries of the subregions. }
    \label{Fig:figure-GGMC2}
\end{figure}

GGMC~2 is located in the middle of the complex with a velocity interval of [-1.5, 16] {km s$^{-1}$}, as shown in Figure~\ref{Fig:figure-3-B}. Figure~\ref{Fig:figure-GGMC2-A} shows the molecular gas distribution of the GGMC 2 main body. The {$^{12}$CO} emission of the main body of GGMC~2 is concentrated in subregions C5 and C6, which is consistent with the work of \citet{2009AJ....138..975B} who used CO (J = 2-1 and J = 3-2) mapping. There are two shell structures to the west of subregion C6, while C6 itself is associated with five optical \Hii\ regions: Sh 254, Sh 255, Sh 256, Sh 257, and Sh 258. There is no {$^{13}$CO} emission along the LoS to Sh 254 (Figure~\ref{Fig:figure-GGMC2-B}), while the molecular gas is distributed to the east of the Sh 254, which is the largest optical \Hii\ region in GGMC~2. Subregion F1 is a diffuse and elongated cloud, which was identified via its {$^{12}$CO} emission (Figure~\ref{Fig:figure-GGMC2-A}); it is seen, however, that its {$^{13}$CO} emission is very weak (Figure~\ref{Fig:figure-GGMC2-B}). Two distinct filamentary structures, labeled Ridge A and Ridge B, are seen in the {$^{13}$CO} emission map (red lines in Figure~\ref{Fig:figure-GGMC2-B}).

\begin{figure}
   \centering
   \subfigure[]{
   \label{Fig:figure-GGMC2pv-A}
   \includegraphics[width=0.48\textwidth, angle=0,clip=true,keepaspectratio=true,trim=0 0 0 0mm]{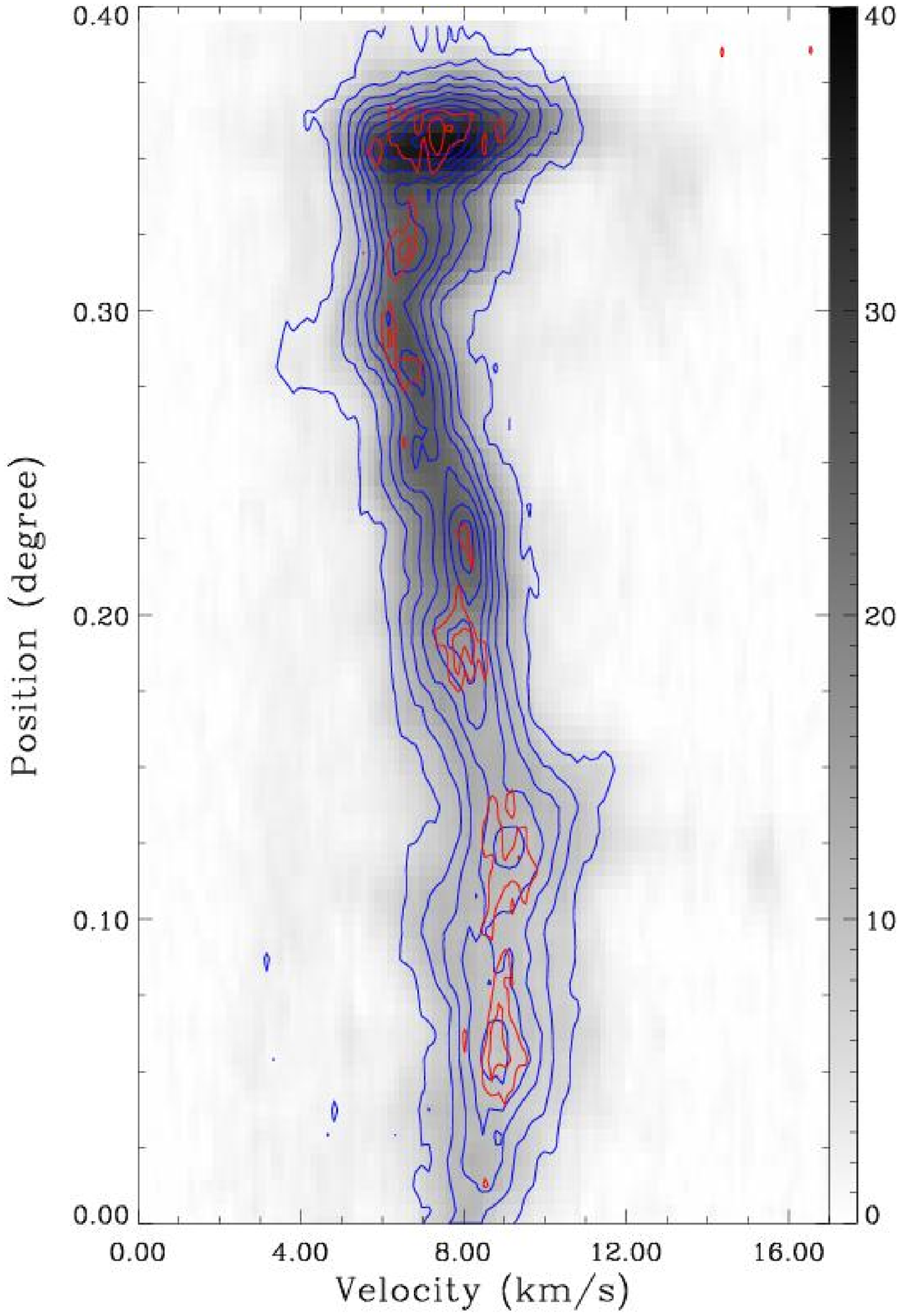}}
   \subfigure[]{
   \label{Fig:figure-GGMC2pv-B}
   \includegraphics[width=0.48\textwidth, angle=0,clip=true,keepaspectratio=true,trim=0 0 0 0mm]{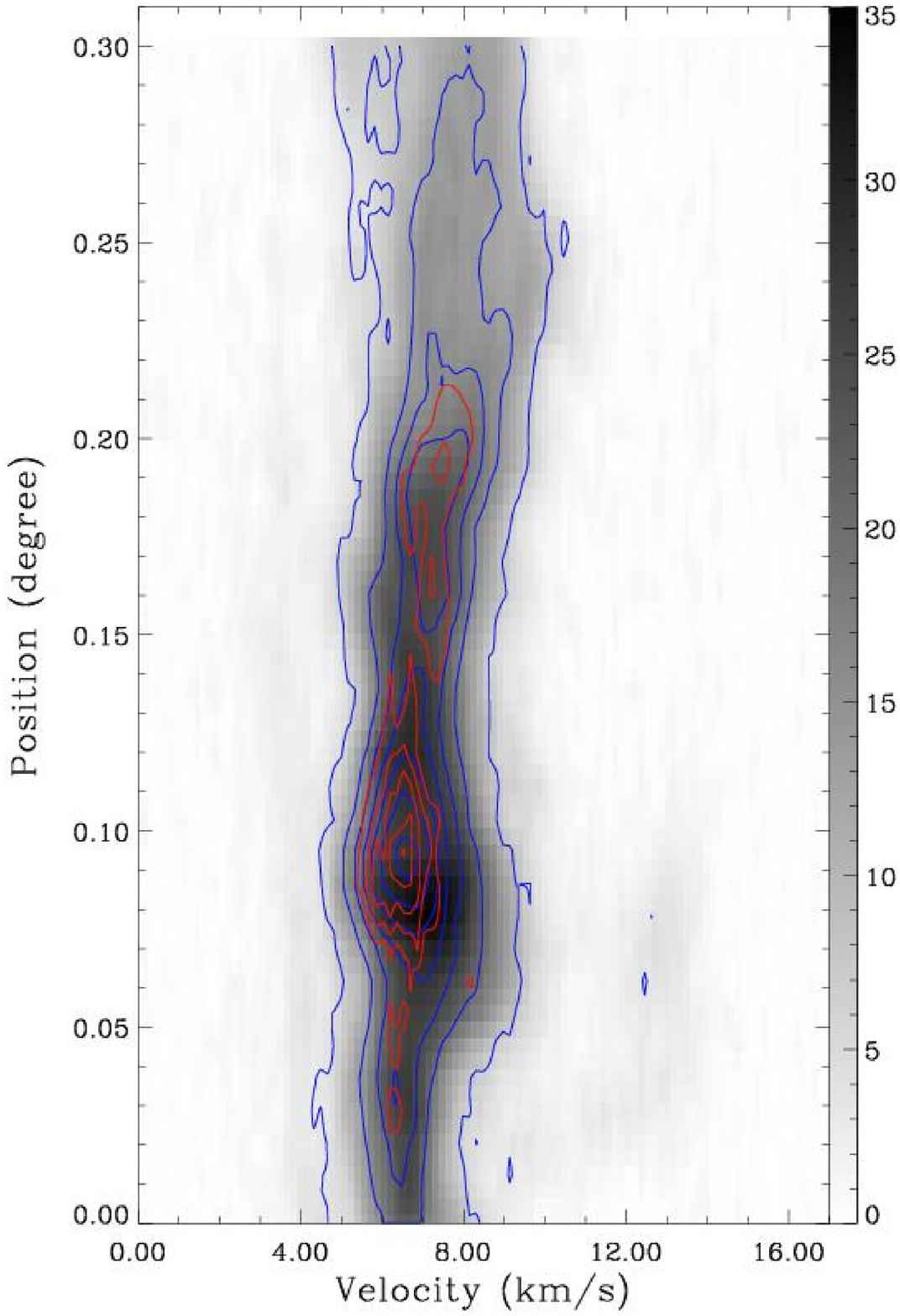}}

   \caption{(a) P-V diagram of \coi, {$^{13}$CO}, and \coiii\ along the red broken line in ``ridge A" (Figure~\ref{Fig:figure-GGMC2-B}). The background shows the emission of \coi. The blue contours indicate {$^{13}$CO} emission and the contour levels begin at 0.75 K with increments of 1.25 K. The red contours indicate \coiii\ emission and the steps are 0.75, 1 K.
   (b) P-V diagram of \coi, {$^{13}$CO}, and \coiii\ along the red broken line in ``ridge B" (Figure~\ref{Fig:figure-GGMC2-B}). The background shows the data of \coi. The blue contours indicate {$^{13}$CO} and the contour levels begin at 0.75 K with increments of 2.5 K. The red contours indicate \coiii\ emission and the steps are 0.75, 1.25, 1.75, 2.75 K.}
    \label{Fig:figure-GGMC2pv}
\end{figure}

{C$^{18}$O} emission in the main body of GGMC~2 is stronger than that in GGMC~1 (Figure~\ref{Fig:figure-GGMC2-C}), and it is concentrated into subregions C5 , Ridge A, and Ridge B. \citet{2009AJ....138..975B} identified three structures: Ridge A, Ridge B, and Cloud C. These structures are also identified in our map (red letters and a plus ¡®+¡¯ symbol in Figure~\ref{Fig:figure-GGMC2-B}), and we newly identify two structures, Cloud D and Cloud E, based on our {$^{13}$CO} and {C$^{18}$O} emission. Ridge~A and Ridge~B are both filamentary structures. Their P-V diagrams are shown in Figure~\ref{Fig:figure-GGMC2pv-A} and Figure~\ref{Fig:figure-GGMC2pv-B}, respectively. According to their distance of 2~kpc, the lengths of Ridge~A and ridge~B are 14 pc and 10.5 pc, respectively.

\begin{figure}
   \centering
   \subfigure[]{
   \label{Fig:figure-BFS-A}
   \includegraphics[width=0.55\textwidth, angle=0,clip=true,keepaspectratio=true,trim=0 0 0 0mm]{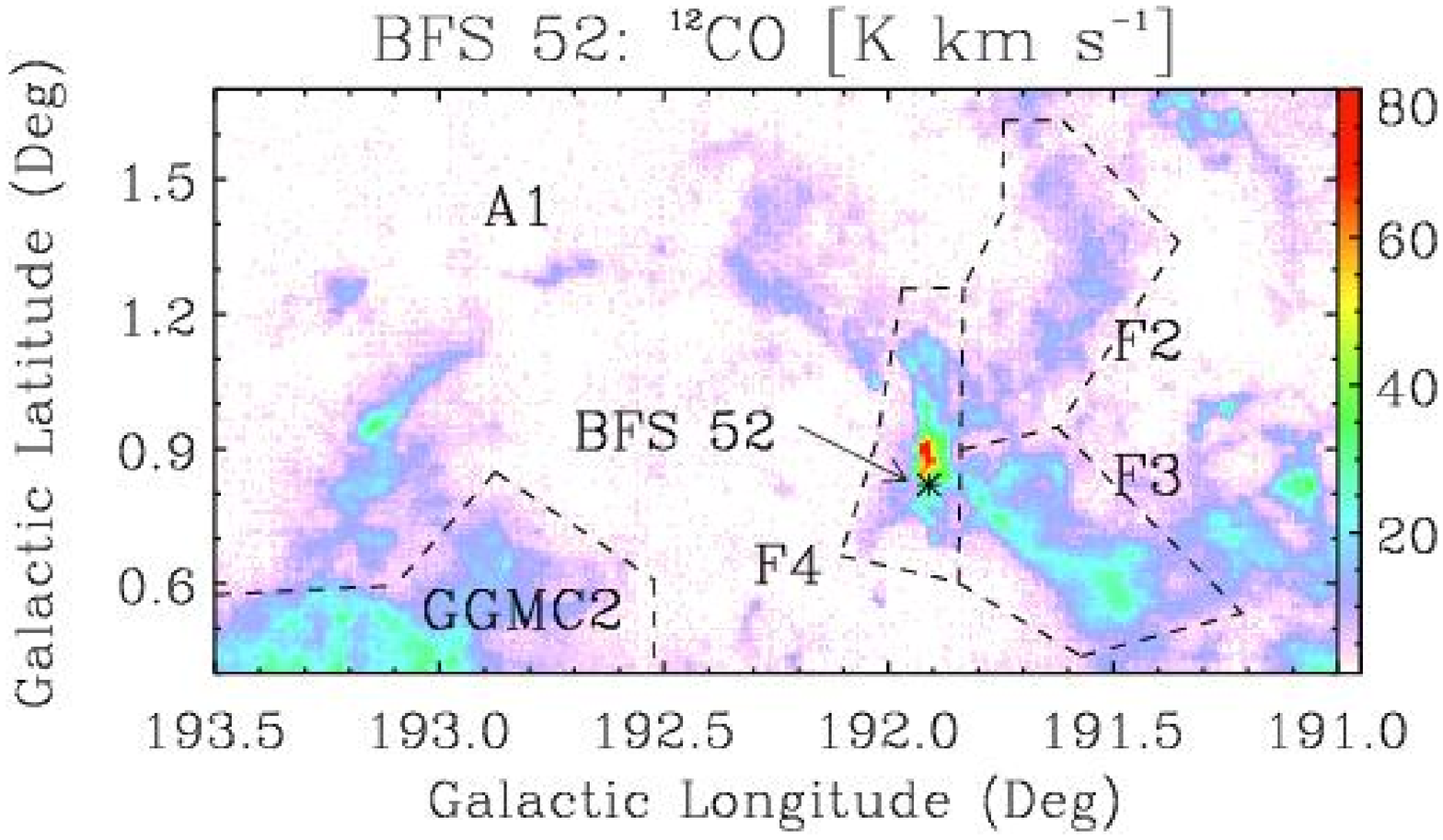}}
   \subfigure[]{
   \label{Fig:figure-BFS-B}
   \includegraphics[width=0.35\textwidth, angle=0,clip=true,keepaspectratio=true,trim=0 0 0 0mm]{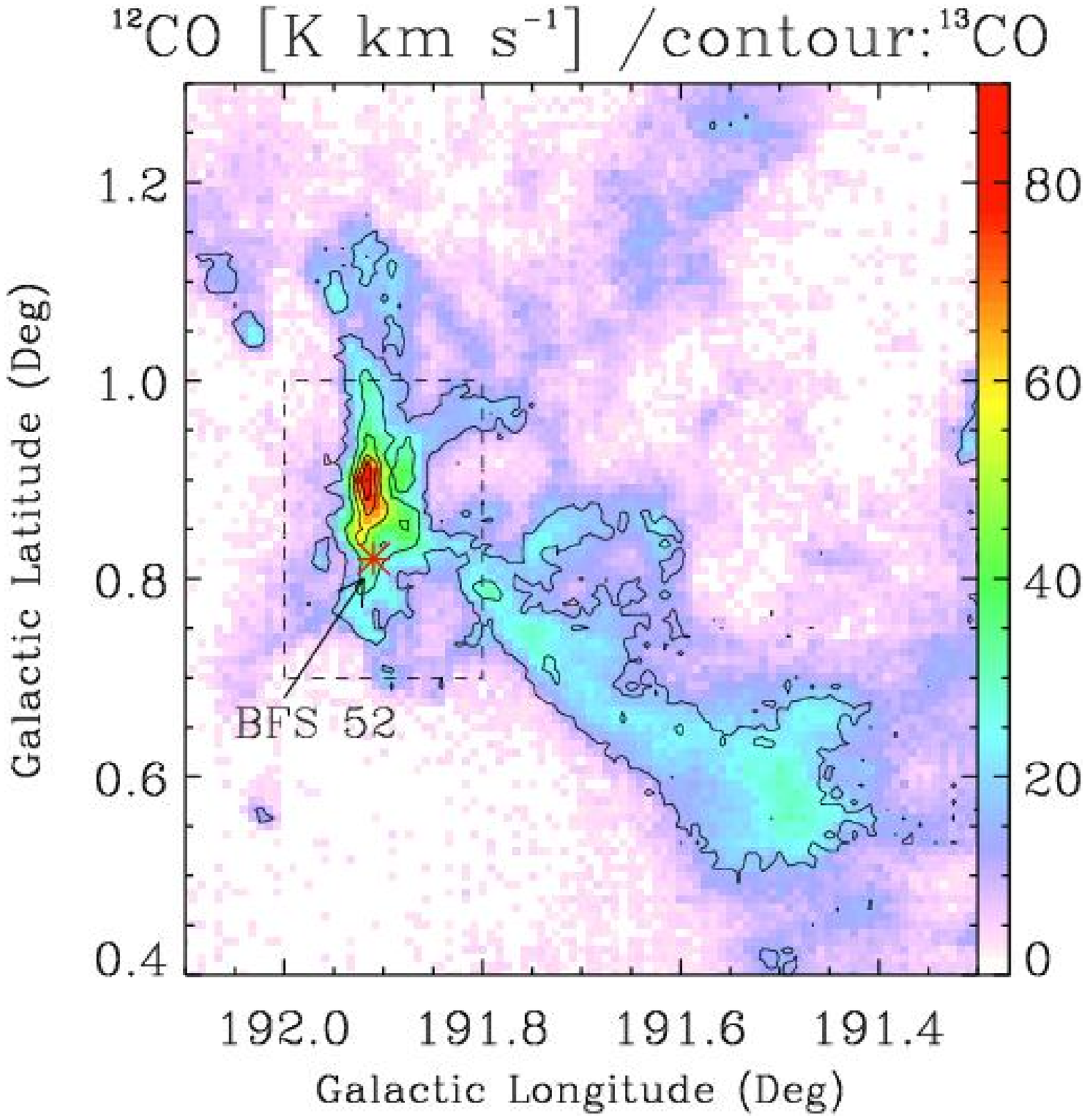}}
   \subfigure[]{
   \label{Fig:figure-BFS-C}
   \includegraphics[width=0.35\textwidth, angle=0,clip=true,keepaspectratio=true,trim=0 0 0 0mm]{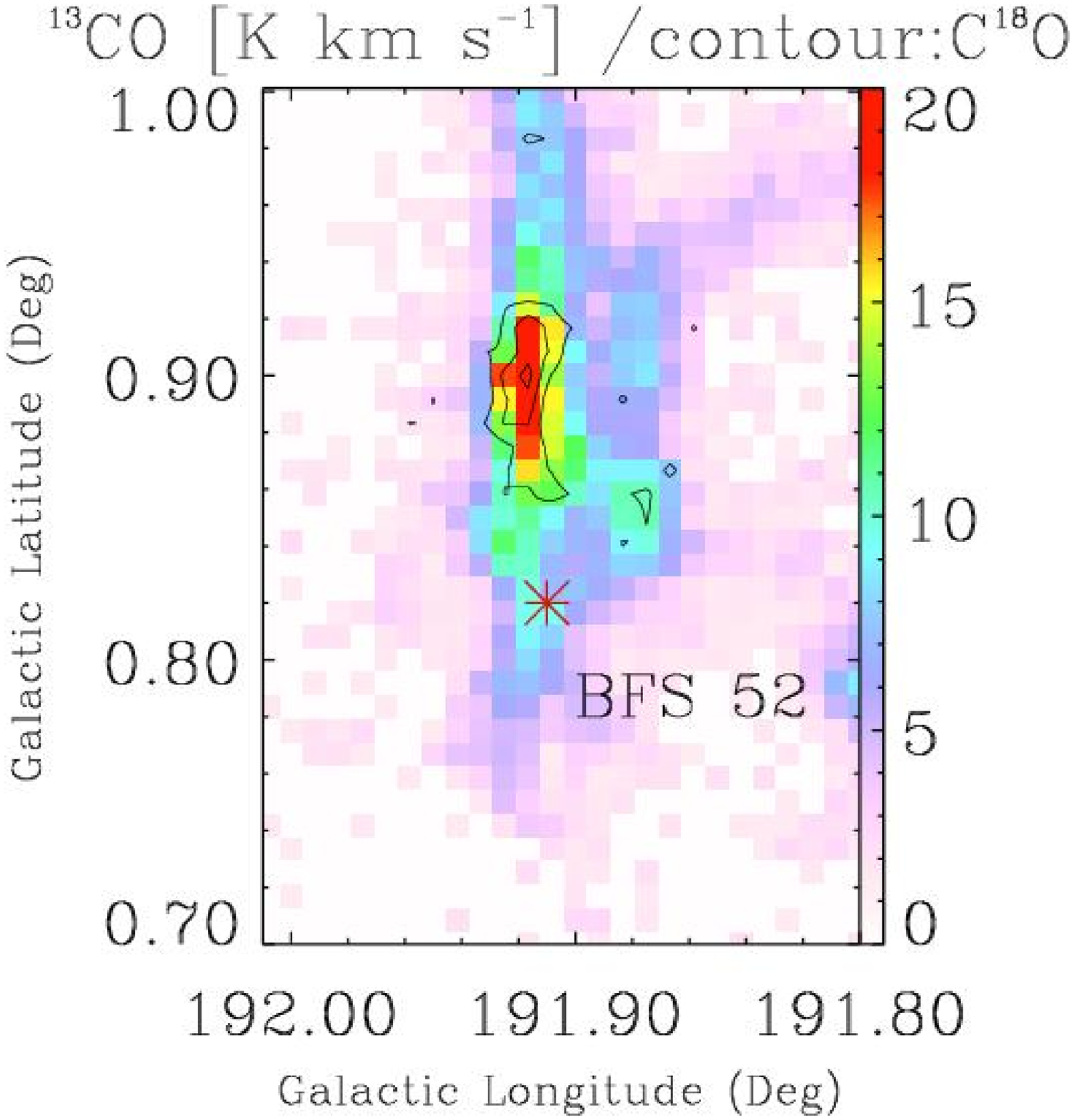}}
   \subfigure[]{
   \label{Fig:figure-BFS-D}
   \includegraphics[width=0.55\textwidth, angle=0,clip=true,keepaspectratio=true,trim=0 0 0 0mm]{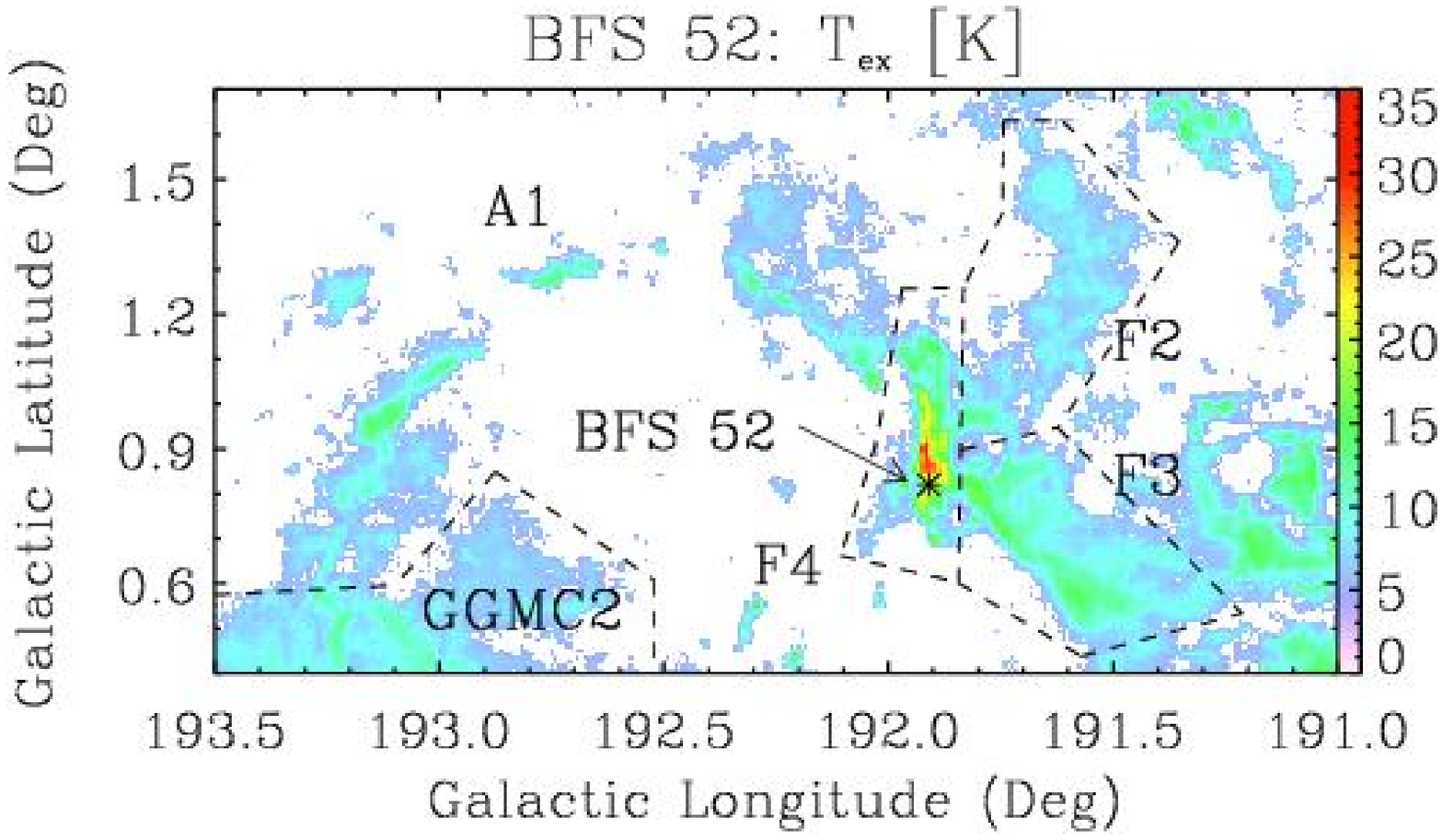}}

   \caption{Gas distribution of the BFS 52 cloud. (a) {$^{12}$CO} integrated intensity map in the range of [-5,~16] {km s$^{-1}$}. The dashed lines show the boundaries of the subregions. (b) Background: {$^{12}$CO} integrated intensity map in the range of [-5, 16] \kms. The contours indicate the {$^{13}$CO} integrated intensity in the range of [-1.5,~16] \kms. The contour levels begin at 2.09 K \kms\ (4$\sigma$) with increments of 4.17 K \kms. The rectangle shows area of Figure~\ref{Fig:figure-BFS-C}. (c) {$^{13}$CO} integrated intensity map. The contours indicate \coiii\ integrated intensity in the range of [4,~11] \kms. The contour levels begin at 1 K \kms\ (3$\sigma$) with increments of 1 K \kms. (d) The excitation temperature map of the BFS 52 cloud.}
    \label{Fig:figure-BFS}
\end{figure}

There are some intriguing clouds located to the northwest of the main body of GGMC~2, which are named BFS~52 cloud (Figure~\ref{Fig:figure-BFS-A}). These clouds are a part of GGMC~2, and are associated with {H{\scriptsize~II}} region BFS~52 (\citealt{1982ApJS...49..183B}, \citealt{1995ApJ...445..246C}). There is a shell structure that connects the GGMC~2 main body and BFS~52 cloud to the east of the BFS~52 cloud (Figure~\ref{Fig:figure-BFS-A}). To the west of the BFS~52 cloud, three filamentary structures are seen. F4 and F3 have strong {$^{13}$CO} emission, but F2 is invisible in {$^{13}$CO} (Figure~\ref{Fig:figure-BFS-B}). The {H{\scriptsize~II}} region BFS 52 is located to the south of F4. {C$^{18}$O} emission in the BFS~52 cloud is concentrated in the region near BFS 52 (Figure~\ref{Fig:figure-BFS-C}).

The excitation temperature of GGMC~2 is higher than that of GGMC~1 (Figure~\ref{Fig:figure-GGMC2-D} and Figure~\ref{Fig:figure-BFS-D}). The maximum excitation temperature of GGMC~2 is 45 K, while for BFS 52 cloud it is 31~K (Table~\ref{tbl-GGMC2}). Ridge A and Ridge B have higher excitation temperatures than the other subregions in GGMC~2, and the temperatures of the molecular gas in Ridge~A and Ridge~B are above 20~K. The excitation temperature of F1 is higher than that in other subregions in the BFS~52 cloud.

Using the adopted X-factor, the mass traced by {$^{12}$CO} emitted by GGMC~2 is 1.8 $\times$ $10^{5}$ M$_{\odot}$, where the main body of GGMC~2 accounts for eighty percent of the total mass (Table~\ref{tbl-1} and Table~\ref{tbl-GGMC2}). The molecular masses derived from {$^{13}$CO} and {C$^{18}$O} are 7.4 $\times$ $10^{4}$ M$_{\odot}$ and 1.0 $\times$ $10^{4}$ M$_{\odot}$, respectively (Table~\ref{tbl-1}).

\subsection{GGMC~3}

\begin{figure}
   \centering
   \subfigure[]{
   \label{Fig:figure-GGMC3-A}
   \includegraphics[width=0.5\textwidth, angle=0,clip=true,keepaspectratio=true,trim=0 0 0 0mm]{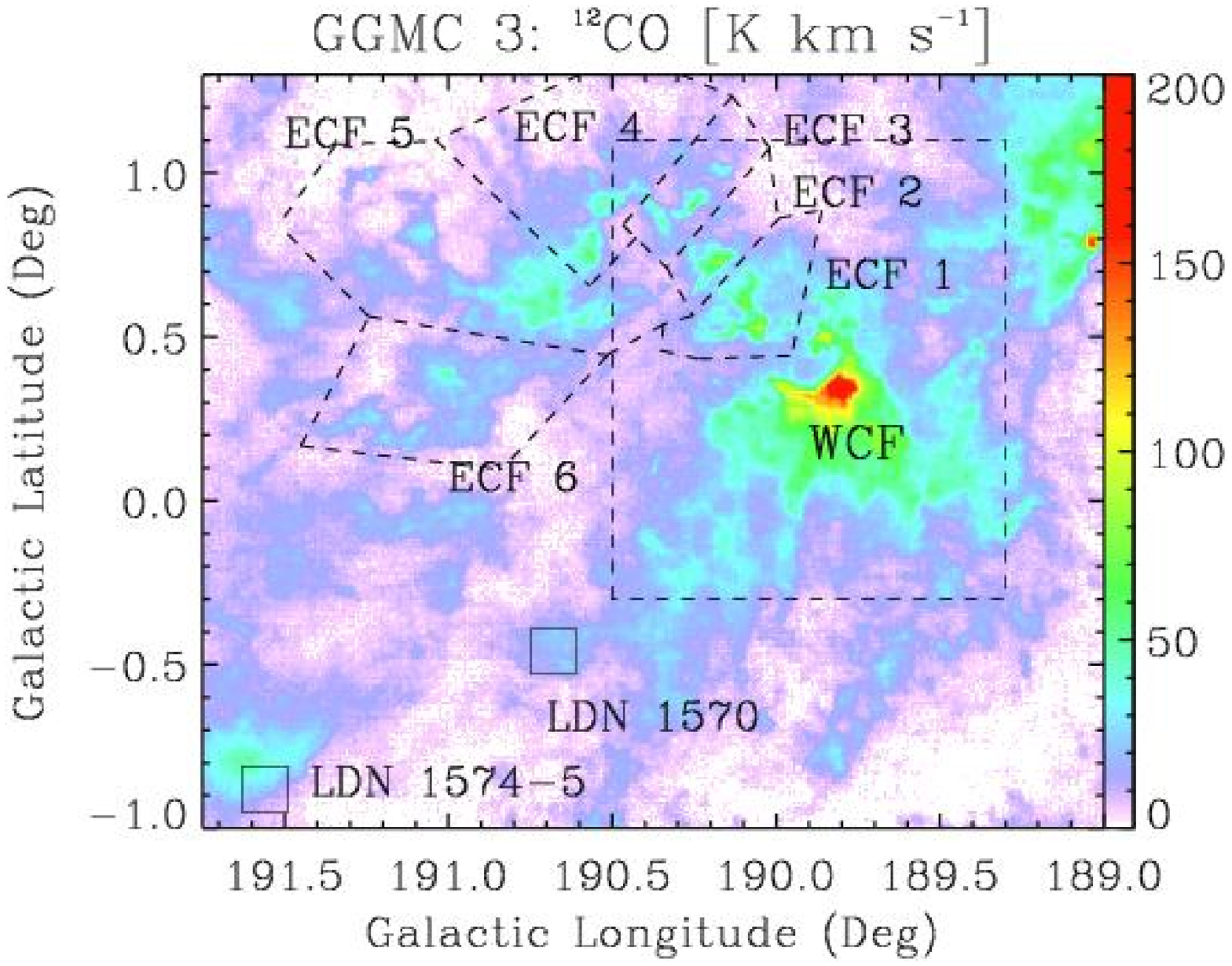}}
   \subfigure[]{
   \label{Fig:figure-GGMC3-B}
   \includegraphics[width=0.47\textwidth, angle=0,clip=true,keepaspectratio=true,trim=0 0 0 0mm]{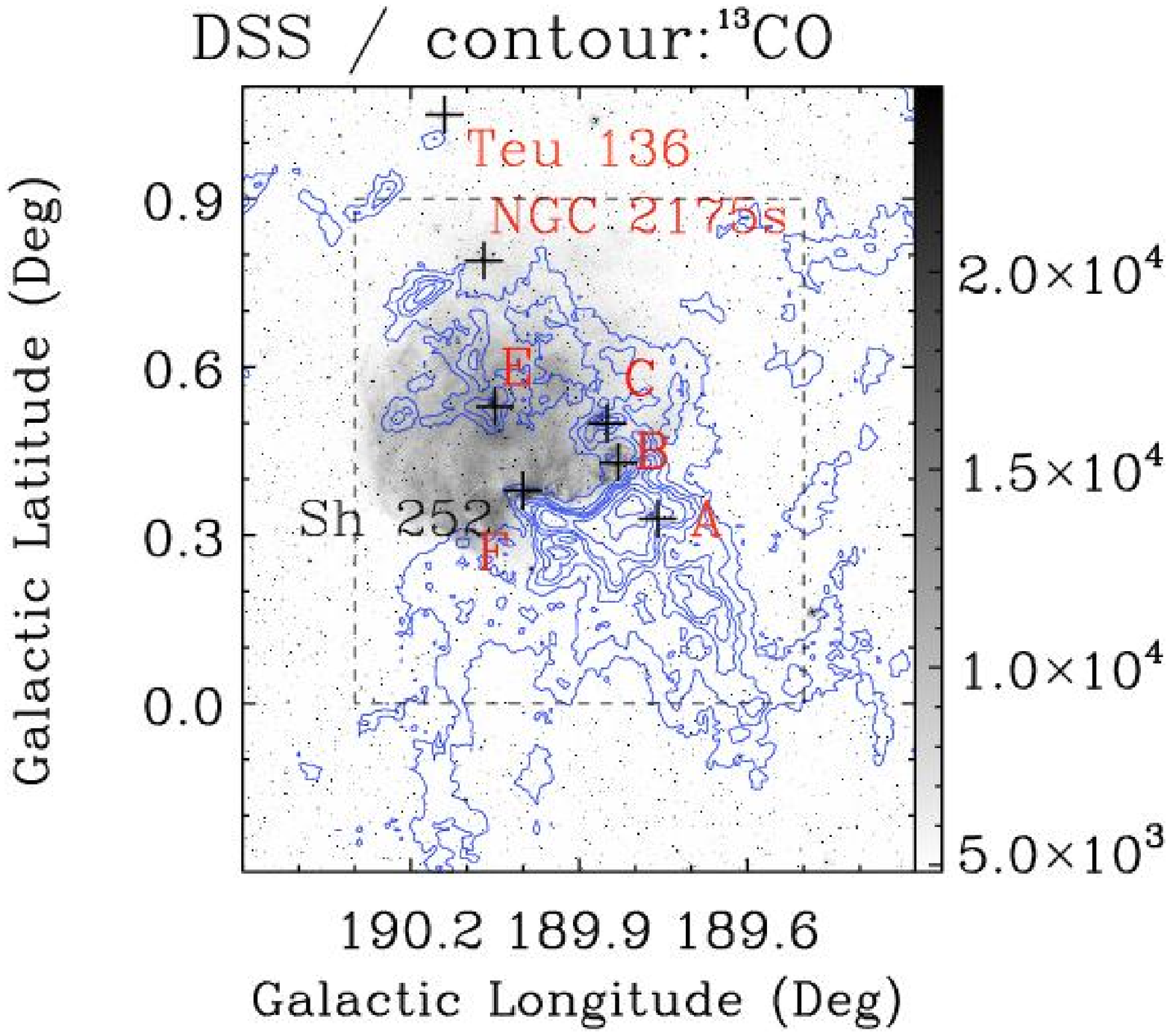}}
   \subfigure[]{
   \label{Fig:figure-GGMC3-C}
   \includegraphics[width=0.4\textwidth, angle=0,clip=true,keepaspectratio=true,trim=0 0 0 0mm]{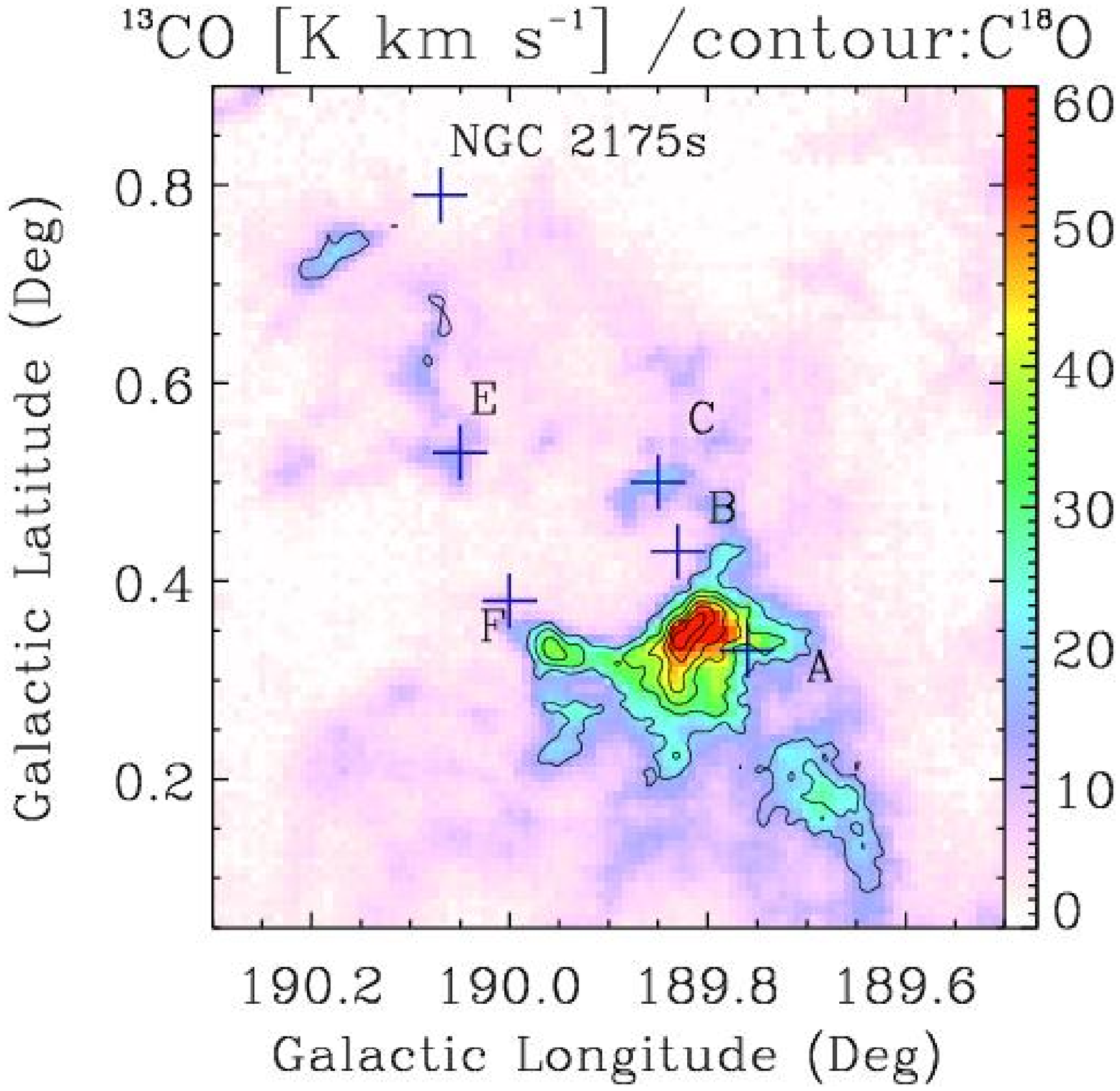}}
   \subfigure[]{
   \label{Fig:figure-GGMC3-D}
   \includegraphics[width=0.5\textwidth, angle=0,clip=true,keepaspectratio=true,trim=0 0 0 0mm]{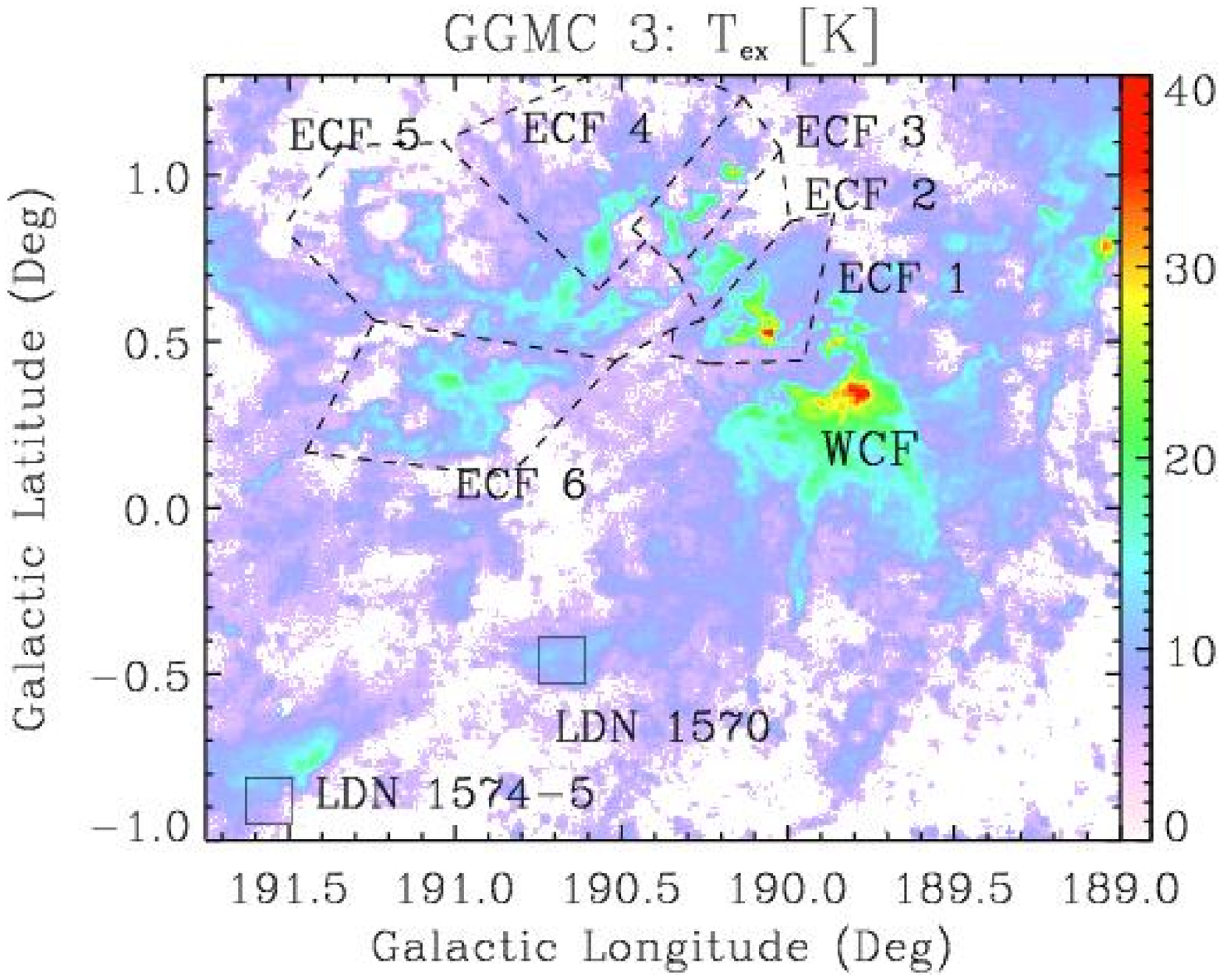}}

   \caption{Gas distribution of GGMC3. (a) {$^{12}$CO} integrated intensity map in the range of [-5, 16] {km s$^{-1}$}. The dashed lines show the boundaries of the subregions and the dashed rectangle shows the area of Figure~\ref{Fig:figure-GGMC3-B}. (b) The background shows the Palomar Digital Sky Survey. The contours indicate the integrated intensity of {$^{13}$CO} with the range of [-1.5, 16] \kms; the steps are 10, 20, 30, 40, 50, 70, 100, 130 $\times$ 0.42 K \kms(1$\sigma$). The pluses represent compact \Hii\ regions (A, B, C, E. \citet{1977A&A....59...43F}) and embedded clusters (A, C, E NGC2175s, Teu136. \citet{2011MNRAS.414.3769B}). (c) The background shows the {$^{13}$CO} integrated intensity. The contours indicate the integrated intensity of \coiii\ with the range of [4, 11] \kms, and the steps are 5, 10, 15, 20, 25, 30 $\times$ 0.25 K \kms(1$\sigma$). The pluses have the same meaning as those in (b). (d) The excitation temperature map of GGMC 3.}
    \label{Fig:figure-GGMC3}
\end{figure}

GGMC~3 is located to the right of the BFS~52 cloud and has the same velocity range (Figure~\ref{Fig:figure-3}). It includes two fragments, which are, respectively, the western cloud fragment (WCF) and the eastern cloud fragment (ECF; \citet{1979ApJ...232..158L}), and are located at the eastern and western sides of Sh~252 (Figure~\ref{Fig:figure-GGMC3-A}). ECF includes six subregions, all of which are filamentary structures. The {$^{12}$CO} emission intensity of the WCF is much stronger than that of the ECF. The {$^{12}$CO} molecular gas is concentrated to the northwest of the WCF. A finger-like structure is found to the south of the WCF. {$^{13}$CO} emission of GGMC~3 shows a shell-like structure that surrounds the {H{\scriptsize~II}} region Sh~252, and the {$^{13}$CO} emission is enhanced at the wall of the shell structure (Figure~\ref{Fig:figure-GGMC3-B}). To the northwest of the WCF, {C$^{18}$O} emission presents two structures (Figure~\ref{Fig:figure-GGMC3-C}).

The excitation temperature of GGMC~3 is higher than that of GGMC~1 (Figure~\ref{Fig:figure-GGMC3-D}). The maximum temperature of GGMC~3 is 41~K, which is close to that of GGMC~2 (Table~\ref{tbl-GGMC2}, Table~\ref{tbl-GGMC3}). The WCF has higher temperature than the ECF (Figure~\ref{Fig:figure-GGMC3-D}). The molecular gas near {H{\scriptsize~II}} region Sh~252 has the highest temperature, which is above 20~K. We note that the maximum temperatures of all subregions of the ECF are above 20~K.

The GMC includes four compact \Hii\ ({CH{\scriptsize~II}}) regions (\citealt{1977A&A....59...43F}) and five embedded clusters (\citealt{2011MNRAS.414.3769B}) (Figure~\ref{Fig:figure-GGMC3-B}). Most of the objects are located near Sh~252 (Figure~\ref{Fig:figure-GGMC3-B}). \citet{2012MNRAS.424.2486J,2013MNRAS.432.3445J} have identified 577 YSOs in the MCs. This indicates that GGMC~3 has strong star-formation activities.

The total mass of GGMC~3 traced by {$^{12}$CO} is found to be 2.8 $\times$ $10^{5}$ M$_{\odot}$ using the adopted X-factor (Table~\ref{tbl-1}). Under the assumption of LTE, the masses of molecular gases traced by {$^{13}$CO} and {C$^{18}$O} data are calculated to be 1.1 $\times$ $10^{5}$ M$_{\odot}$ and 2.8 $\times$ $10^{4}$ M$_{\odot}$, respectively (Table~\ref{tbl-1}). The mass traced by {C$^{18}$O} of GGMC~3 is about three times higher than that of GGMC~2.

\subsection{GGMC~4}

\begin{figure}
   \centering
   \subfigure[]{
   \label{Fig:figure-GGMC4-A}
   \includegraphics[width=0.55\textwidth, angle=0,clip=true,keepaspectratio=true,trim=0 0 0 0mm]{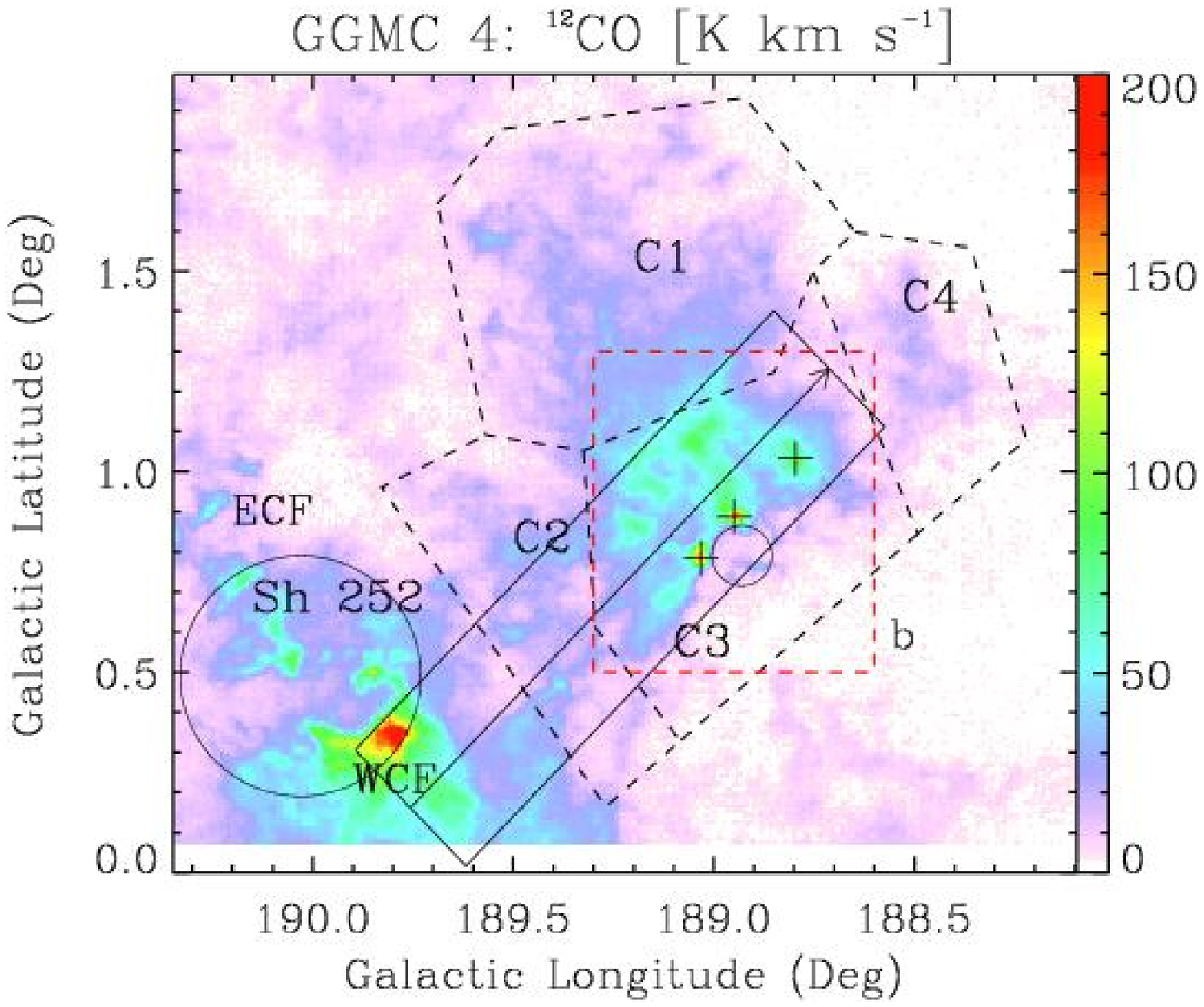}}
   \subfigure[]{
   \label{Fig:figure-GGMC4-B}
   \includegraphics[width=0.42\textwidth, angle=0,clip=true,keepaspectratio=true,trim=0 0 0 0mm]{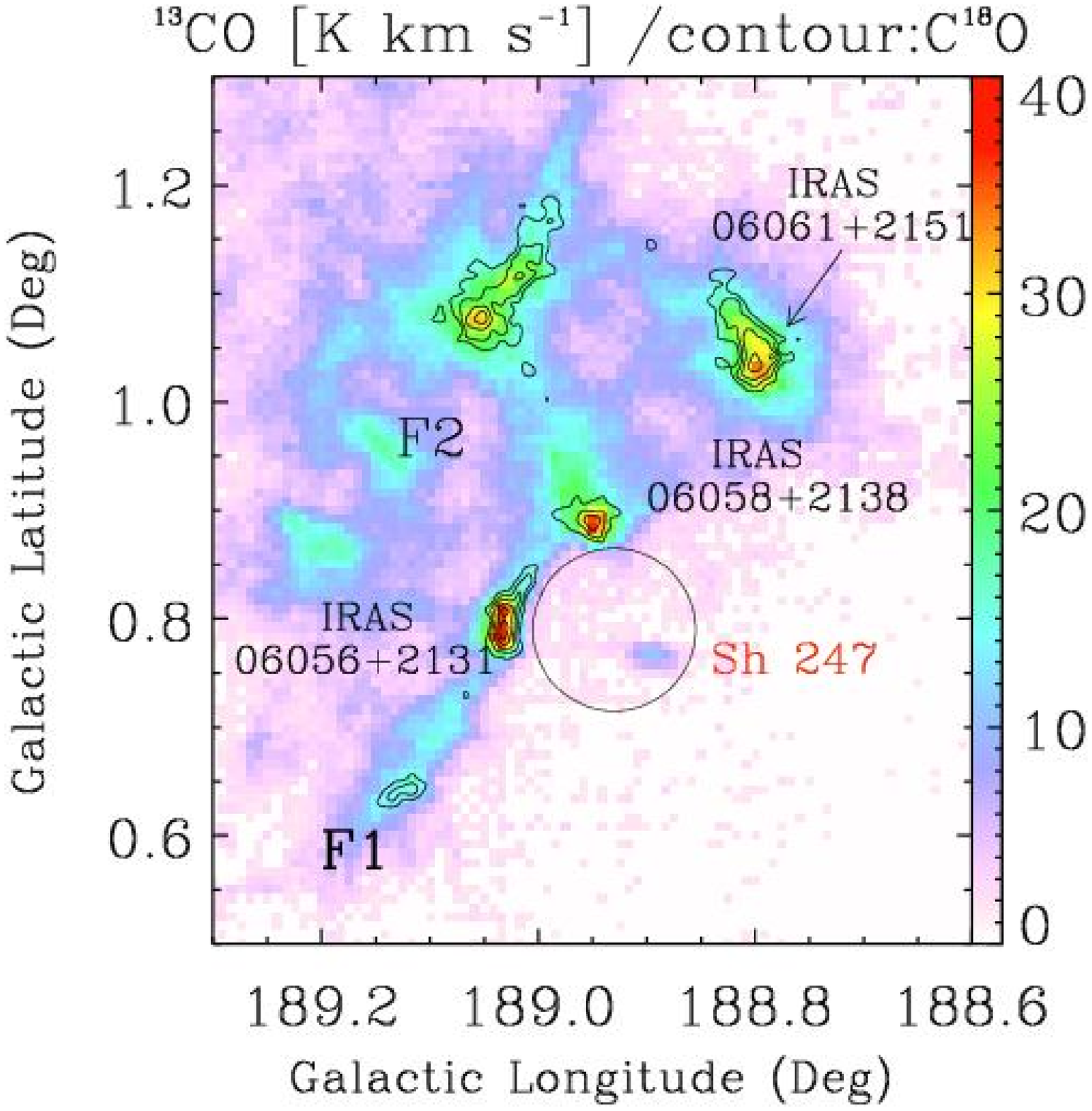}}
   \subfigure[]{
   \label{Fig:figure-GGMC4-C}
   \includegraphics[width=0.5\textwidth, angle=0,clip=true,keepaspectratio=true,trim=0 0 0 0mm]{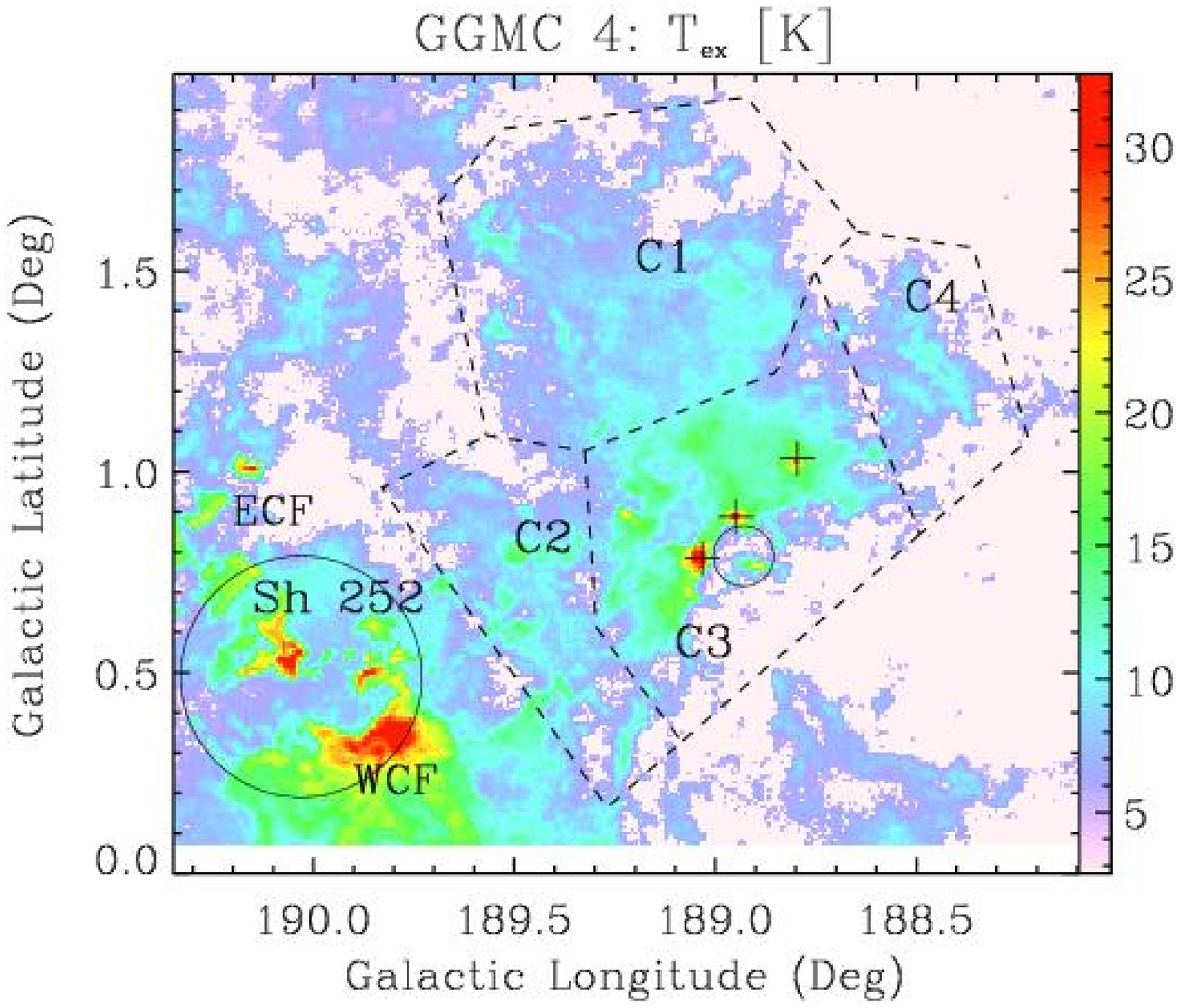}}

   \caption{Gas distribution of GGMC4. (a) Integrated intensity map of {$^{12}$CO} with the range of [-4, 15] {km s$^{-1}$}. The dashed lines show the boundaries of the subregions. The circles represent \Hii\ regions (Sh~247 and Sh~252). The triangles represent compact \Hii\ regions (\citet{2000BASI...28..515G}). The black arrow and rectangle show the direction and width of the P-V diagrams of Figure~\ref{Fig:figure-GGMC4pv}. (b) Integrated intensity map of {$^{13}$CO} over the velocity range of [-4,~14] \kms. The contours indicate the integrated intensity of \coiii\  over the range of [-2.5, 5] \kms, and the steps are 5, 7, 9, 11, 13 $\times$ 0.29 K \kms(1$\sigma$). (c) The excitation temperature map of GGMC4.}
    \label{Fig:figure-GGMC4}
\end{figure}

The last GMC in the complex is GGMC~4, which is located to the west of the complex, and is associated with Sh~247 (Figure~\ref{Fig:figure-3-A}). Figure~\ref{Fig:figure-GGMC4-A} shows the overall molecular gas distribution of GGMC~4, where most of the {$^{12}$CO} emission is located in subregion C3. The size of C1 is larger than that of C3, but the {$^{12}$CO} emission of C1 is more diffuse. A distinct filament, F1, can be identified in subregion C3, and Sh~247 is located southwest of the filament. Three compact \Hii\ regions (\citealt{2000BASI...28..515G}) line up along the filament. The {$^{13}$CO} emission shows another filament, F2, which is parallel to F1 (Figure~\ref{Fig:figure-GGMC4-B}). The {C$^{18}$O} emission in GGMC~4 is also strong (Figure~\ref{Fig:figure-GGMC4-B}), where five clumpy structures are detected. Four of the {C$^{18}$O} MCs are along filament F1, and three of them are associated with compact \Hii\ regions.

Figure~\ref{Fig:figure-GGMC4-C} shows the excitation temperature of GGMC~4. The maximum temperature of GGMC~4 is 33~K, which is higher than that of GGMC~1 but lower than that of GGMC~2 and GGMC~3 (Table~\ref{tbl-GGMC4}). Subregion C3 has a higher temperature than the other subregions.

\begin{figure}
   \centering
   \includegraphics[width=0.55\textwidth, angle=0,clip=true,keepaspectratio=true,trim=0 0 0 0mm]{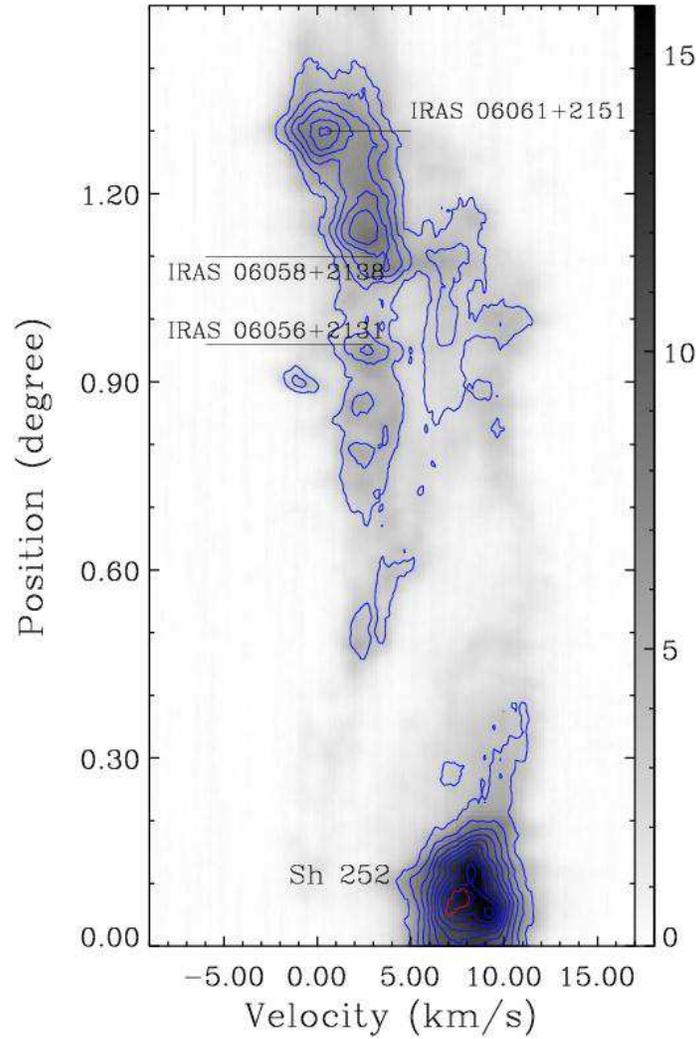}
   \caption{P-V diagram along the arrows in Figure~\ref{Fig:figure-GGMC4-A}; the width of the belt is 48 pixels ($24'$). The background shows the data of \coi. The blue contours indicate {$^{13}$CO} and the contour levels begin at 0.51 K with increments of 0.51 K. The red contours indicate \coiii\ and the step is 0.68 K. There are two velocity components. Figure~\ref{Fig:figure-GGMC4LR} shows the spatial distribution of the components.}
    \label{Fig:figure-GGMC4pv}
\end{figure}

\begin{figure}
   \centering
   \subfigure[]{
   \label{Fig:figure-GGMC4LR-A}
   \includegraphics[width=0.53\textwidth, angle=0,clip=true,keepaspectratio=true,trim=0 0 0 0mm]{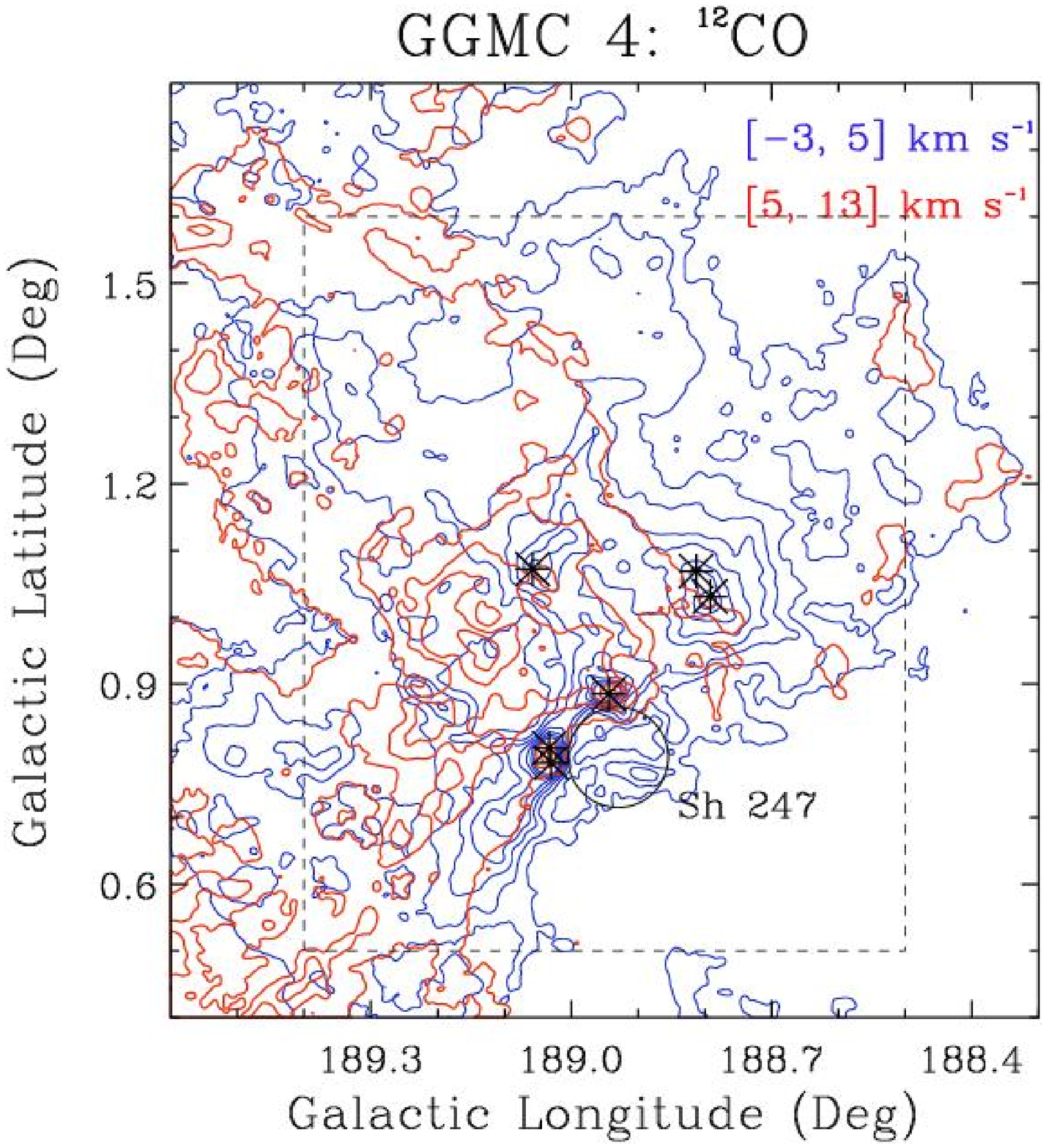}}
   \subfigure[]{
   \label{Fig:figure-GGMC4LR-B}
   \includegraphics[width=0.43\textwidth, angle=0,clip=true,keepaspectratio=true,trim=0 0 0 0mm]{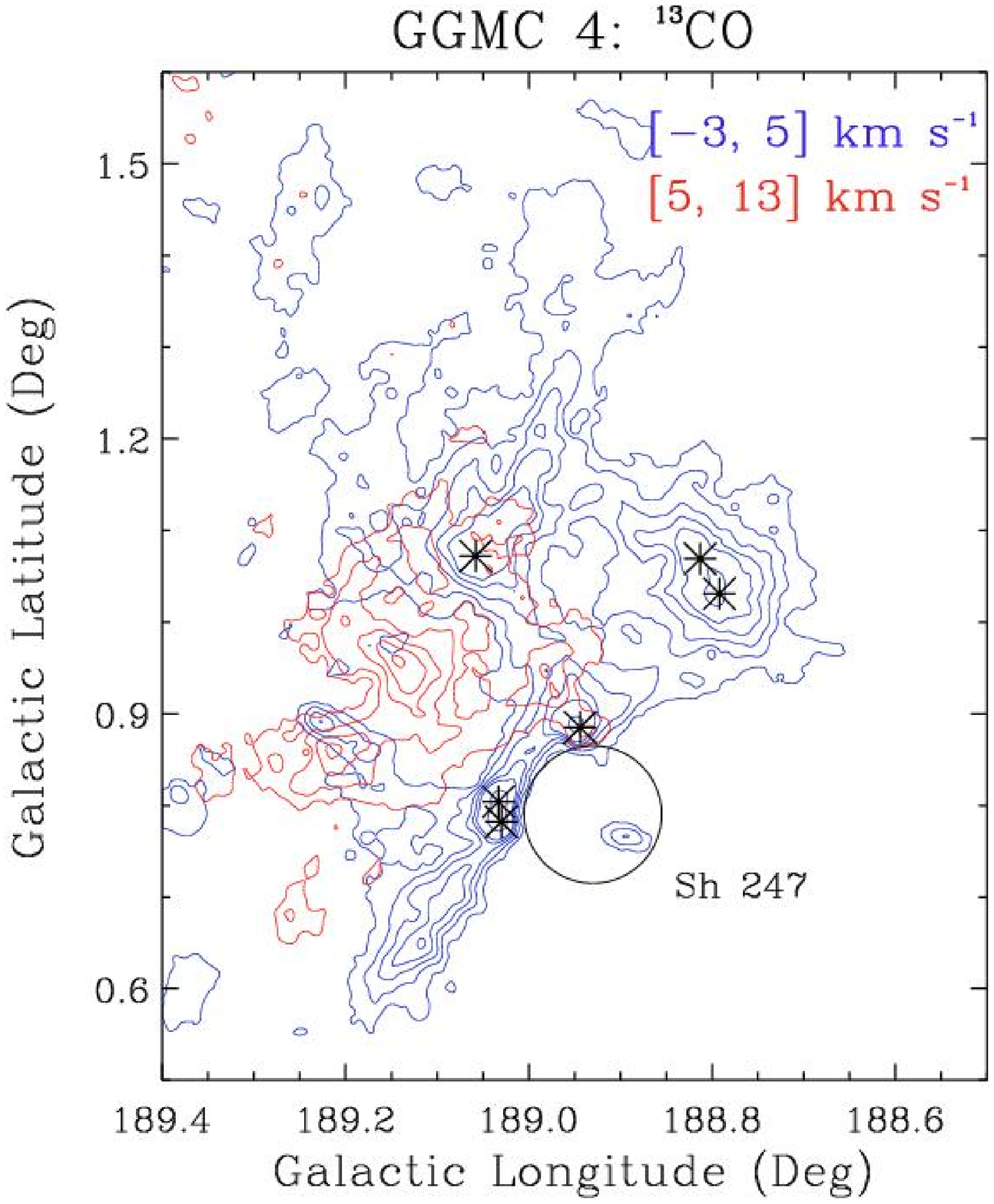}}

   \caption{(a) The blue contours indicate the integrated intensity of {$^{12}$CO} with the range of [-3, 5] \kms. The red contours indicate the integrated intensity of {$^{12}$CO} with the range of [5, 13] \kms; the steps are 1, 3, 5, 7, 9, 20, 25, 30, 35 $\times$ 5 K \kms\ (10$\sigma$). The star symbols represent young embedded clusters ``IR clusters" (\citealt{2013ApJ...768...72S}). (b) The blue contours indicate the integrated intensity of {$^{13}$CO} with the range of [-3, 5] \kms, and the steps are 1, 2, 3, 4, 5, 9 $\times$ 3 K \kms(10$\sigma$). Red contour: the integrated intensity of {$^{13}$CO} with the range of [5,~13] \kms, the steps are 1, 2, 3, 4, 5, 10, 15, 20 $\times$ 3 K \kms\ (10$\sigma$). }
    \label{Fig:figure-GGMC4LR}
\end{figure}

The GGMC~4 includes two major components, according to the spectral analysis. Figure~\ref{Fig:figure-GGMC4pv} shows a P-V diagram along the black arrow in Figure~\ref{Fig:figure-GGMC4-A}. The width of the belt is 24~arcmin. We identify the first component with the velocity interval [-3, 5] km s$^{-1}$ (typical LSR velocity $\sim$2 km s$^{-1}$), while the other is [5, 13] km s$^{-1}$ (typical LSR velocity $\sim$8 km s$^{-1}$). This is consistent with the {$^{12}$CO} (J = 2-1) measurements of \citet{2013ApJ...768...72S}, who identified 11 IR clusters in 2MASSPSC, and concluded that the two components probably collided at some point in the past, which then triggered the observed star formation. Figure~\ref{Fig:figure-GGMC4LR-A} and Figure~\ref{Fig:figure-GGMC4LR-B} show the {$^{12}$CO} and {$^{13}$CO} gas distribution of the two velocity components, respectively. The figures present the spatial coincidence between the 2 km s$^{-1}$ and 8 km s$^{-1}$ components, where it is seen that these two components are mainly detected to the west and east, respectively. Interestingly, the IR clusters and enhanced molecular gas are located at the interface between the two components. We will discuss the possibility of a previous collision in a future paper.

We calculated the mass of GGMC~4 traced by {$^{12}$CO} to be 1.6 $\times$ $10^{5}$ M$_{\odot}$ using the adopted X-factor (Table~\ref{tbl-1}). Next, by assuming LTE, the masses of the molecular gas traced by {$^{13}$CO} and {C$^{18}$O} data are 5.5 $\times$ $10^{4}$ M$_{\odot}$ and 1.1 $\times$ $10^{4}$ M$_{\odot}$, respectively (Table~\ref{tbl-1}).

The four GGMCs in the complex have similar masses, where the mean mass traced by {$^{12}$CO} is 2.1 $\times$ $10^{5}$ M$_{\odot}$ (Table~\ref{tbl-1}). The mean molecular mass traced by {$^{13}$CO} in the GGMCs is 6.9 $\times$ $10^{4}$ M$_{\odot}$ (Table~\ref{tbl-1}), which is similar to that of the W3/4/5 region (\citealt{1978ApJ...226L..39L}) at a similar distance of 2 kpc (\citealt{2006Sci...311...54X}). In total, the GGMCs appear to form a coherent 6.6$^{\circ}$ $\times$ 2.8$^{\circ}$ (230 pc $\times$ 98 pc) complex, which is comparable in size to that of the W3/4/5 region (\citealt{1978ApJ...226L..39L}).

\section{The Local Lynds Dark Clouds}
\label{Sec:lynds}

\begin{figure}
   \centering
   \subfigure[]{
   \label{Fig:figure-Lynds-A}
   \includegraphics[width=0.5\textwidth, angle=0,clip=true,keepaspectratio=true,trim=0 0 0 0mm]{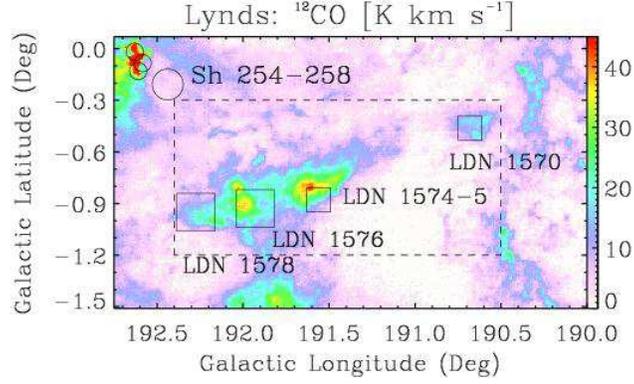}}
   \subfigure[]{
   \label{Fig:figure-Lynds-B}
   \includegraphics[width=0.5\textwidth, angle=0,clip=true,keepaspectratio=true,trim=0 0 0 0mm]{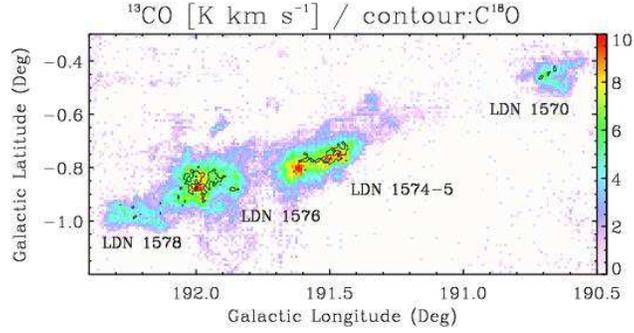}}
   \subfigure[]{
   \label{Fig:figure-Lynds-C}
   \includegraphics[width=0.5\textwidth, angle=0,clip=true,keepaspectratio=true,trim=0 0 0 0mm]{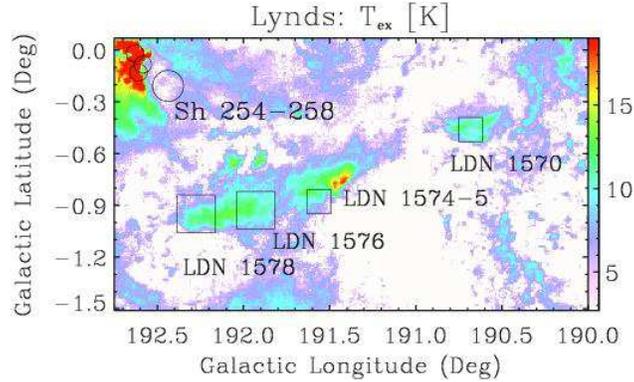}}

   \caption{Gas distribution of Lynds Dark Clouds. (a) Integrated intensity map of {$^{13}$CO} with the range of [-6, 7] {km s$^{-1}$}. The boxes represent regions of Lynds Dark Clouds in optical band, while the circles represent \Hii\ regions. (b) The background is the integrated intensity of {$^{13}$CO} with the range of [-5, 7] {km s$^{-1}$}. The contours indicate the integrated intensity of {C$^{18}$O} with the range of [-3, 2] {km s$^{-1}$}; the steps are 3, 5 $\times$ 0.27 K {km s$^{-1}$} (1$\sigma$). (c) The excitation temperature map of Lynds Dark Clouds.}
    \label{Fig:figure-Lynds}
\end{figure}

In the middle of the observation area, there are a number of Lynds dark clouds (\citealt{1962ApJS....7....1L}), including LDN~1570 (also known as CB~44; \citealt{1988ApJS...68..257C}), LDN~1574-5, LDN~1576, and LDN~1578 (also known as CB~45) (Figure~\ref{Fig:figure-3}).  These dark clouds form a straight chain extending from the northeast by LDN~1570 to the southwest by LDN1578, as shown in Figure~\ref{Fig:figure-Lynds-A}.  All of these dark clouds show prominent {$^{13}$CO} emission, while LDN~ 1574-5, LDN~1576 also show significant {C$^{18}$O} emission (Figure~\ref{Fig:figure-Lynds-B}).

The radial velocities of these dark clouds are within [-6, 7] {km s$^{-1}$}, as shown in (Figure~\ref{Fig:LV}). Considering their proximity both in space and velocity, we regard these clouds to be at the same distance and hence physically associated. As will be shown below, the similarity of their physical properties also supports this suggestion.

The distance of the chain of dark clouds has been estimated by several previous works. \citet{1979PASJ...31..407T} estimated the dark cloud distance to be 300 pc, while \citet{1974AJ.....79...42B} and \citet{2013A&A...556A..65E} obtained distances of 400 pc and 394 pc, respectively. \citet{1995A&AS..113..325H} suggested a distance between 400 and 600 pc. Since more groups found a distance of about 400 pc, including the recent result of \citet{2013A&A...556A..65E}, we adopted a distance of 400 pc in our subsequent analysis. At this distance, these clouds are located in the Local arm, and therefore are not physically related to the GGMCs in Figure~\ref{Fig:figure-3}.

Figure~\ref{Fig:figure-Lynds-C} presents the distribution of the excitation temperature of the chain of local Lynds dark clouds. The excitation temperatures of all the dark clouds are within the range of 4.5-18.7 K (Table~\ref{tbl-Lynds}). Within these dark clouds, LDN~1574-5 shows relatively higher excitation temperatures than the others. Three major enhanced peaks can be identified within the cloud, as defined by their {$^{13}$CO} contours. It is also notable that the {C$^{18}$O} emission is enhanced toward the LDN~1574 area, which will also be discussed in later works.

The size of the dark cloud chain is 2.15$^{\circ}$ $\times$ 0.3$^{\circ}$, which corresponds to a physical size of 15 pc $\times$ 2.1 pc. Using the adopted X-factor between {$^{12}$CO} and molecular hydrogen, we calculated the total mass of all the dark clouds traced by {$^{12}$CO} to be 1750 M$_{\odot}$. The mass of the dark cloud chain is similar to that of L1340 (\citealt{1994A&A...292..249K}; \citealt{1997ApJS..110...21Y}) and L1471 (also known as Barnard 5, \citealt{1985A&A...144..179S}; \citealt{1986ApJ...303L..11G}; \citealt{1989ApJ...337..355L}). Next, and again assuming LTE, the masses of molecular gas traced by {$^{13}$CO} and {C$^{18}$O} are 511 M$_{\odot}$ and 89 M$_{\odot}$, respectively.

In the coldest, densest parts of the clouds, especially in dark clouds and in dense, starless cores, the CO molecules freeze out onto dust grains, and are depleted from the gas (\citealt{2003ApJ...583..789L}). In such cases, the {$^{13}$CO} and {C$^{18}$O} lines do not trace the densest parts of the clouds. We will study this effect in our next paper (C. Wang et al. in preparation).

\section{The West Front}
\label{Sec:west}

\begin{figure}
   \centering
   \subfigure[]{
   \label{Fig:figure-West-A}
   \includegraphics[width=0.67\textwidth, angle=0,clip=true,keepaspectratio=true,trim=0 0 0 0mm]{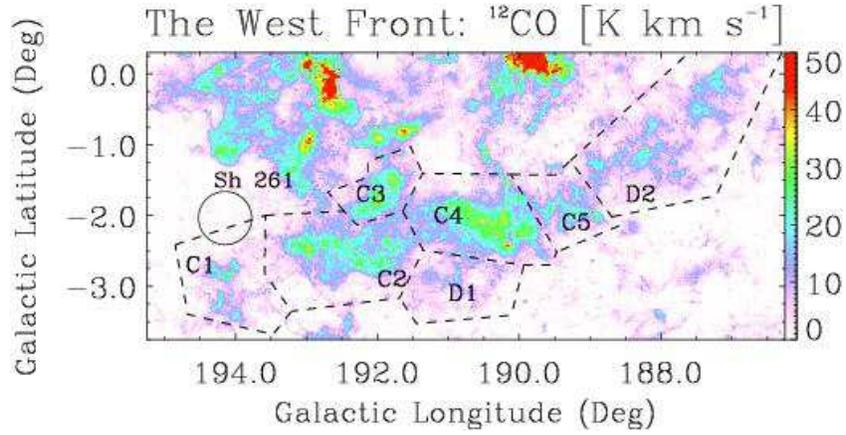}}
   \subfigure[]{
   \label{Fig:figure-West-B}
   \includegraphics[width=0.67\textwidth, angle=0,clip=true,keepaspectratio=true,trim=0 0 0 0mm]{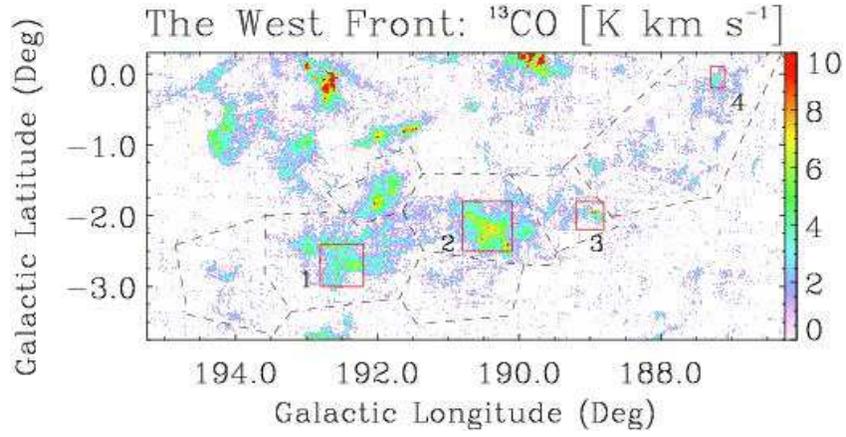}}

   \caption{(a) {$^{12}$CO} integrated intensity map in the range of [-7.5, 10] \kms. The dashed lines show the boundaries of the subregions, and the circles represent \Hii\ regions. (b) {$^{13}$CO} integrated intensity map in the range of [-5, 8] \kms. The four rectangles show the area of Figure~\ref{Fig:figure-West-C1} - Figure~\ref{Fig:figure-West-C4}.}
    \label{Fig:figure-West1}
\end{figure}

\begin{figure}
   \centering
   \subfigure[]{
   \label{Fig:figure-West-C1}
   \includegraphics[width=0.23\textwidth, angle=0,clip=true,keepaspectratio=true,trim=0 0 0 0mm]{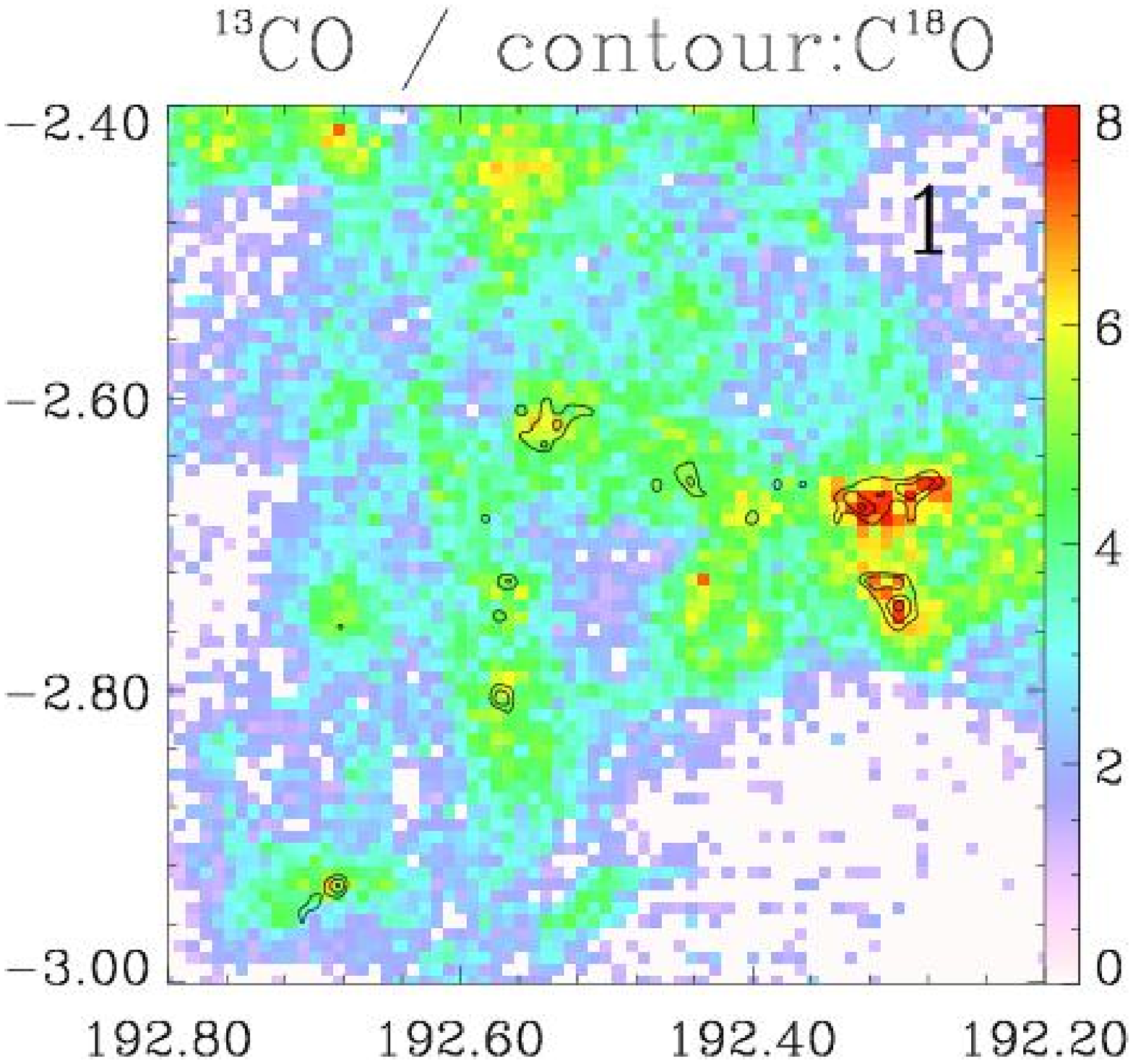}}
   \subfigure[]{
   \label{Fig:figure-West-C2}
   \includegraphics[width=0.23\textwidth, angle=0,clip=true,keepaspectratio=true,trim=0 0 0 0mm]{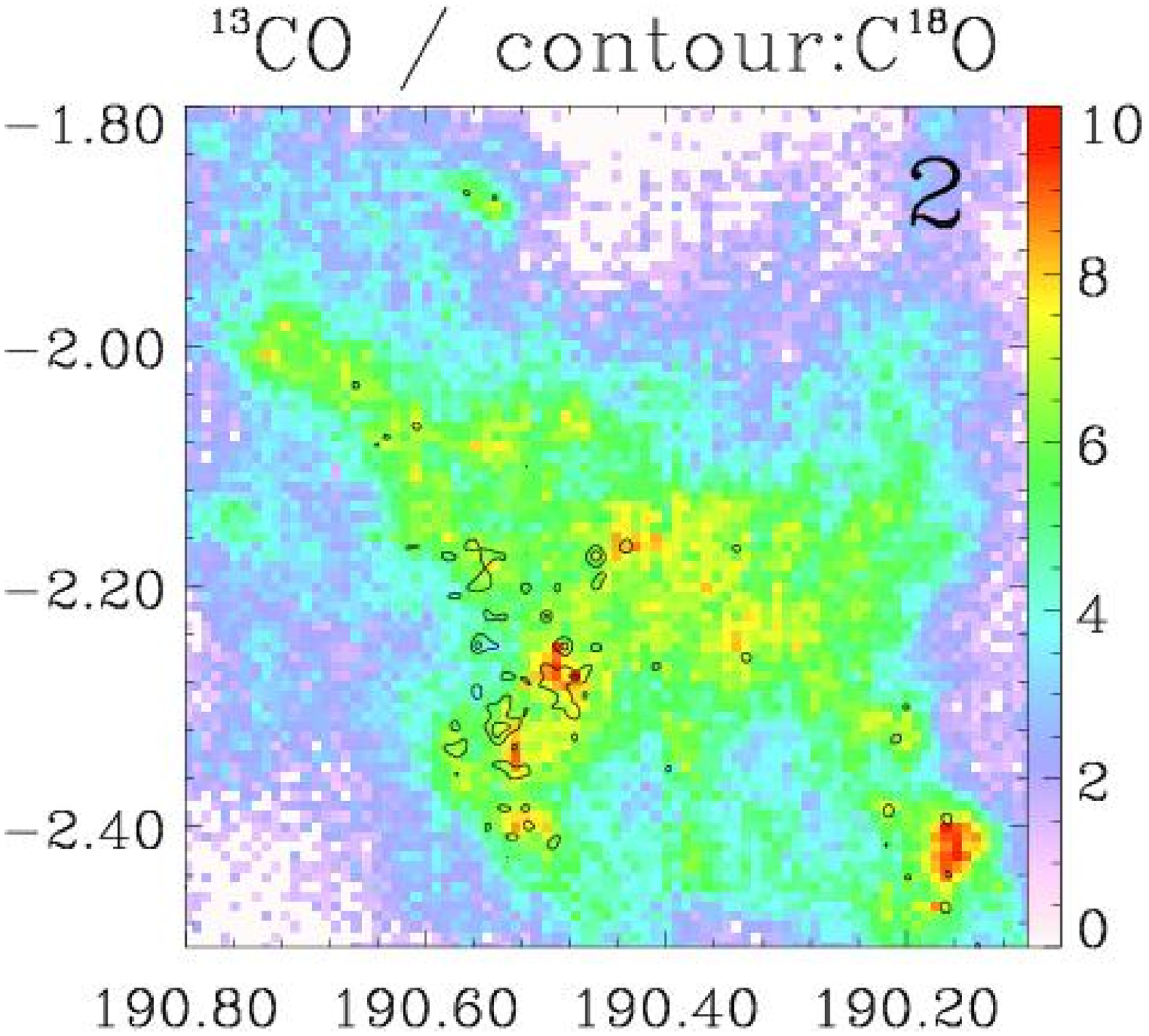}}
   \subfigure[]{
   \label{Fig:figure-West-C3}
   \includegraphics[width=0.23\textwidth, angle=0,clip=true,keepaspectratio=true,trim=0 0 0 0mm]{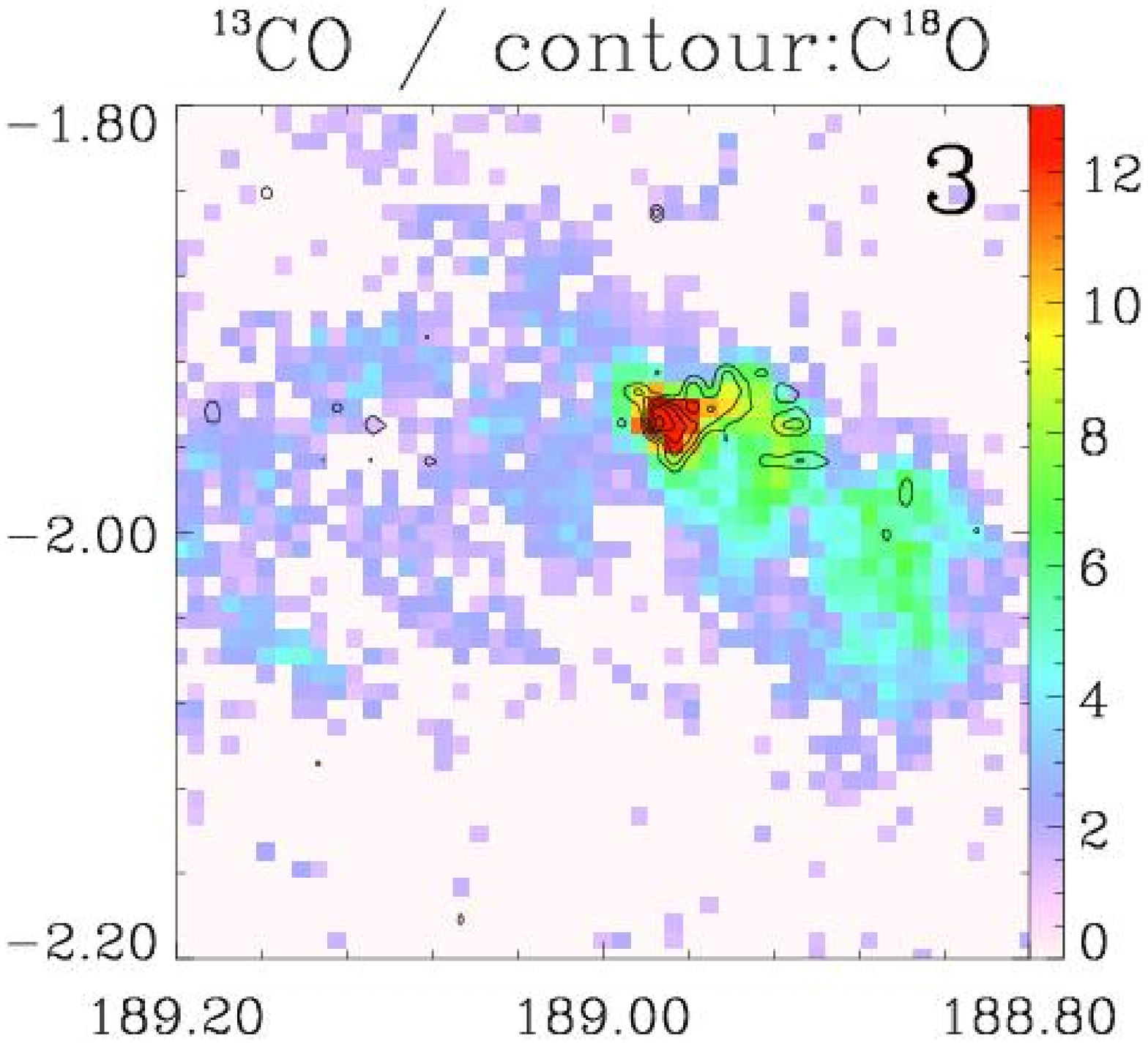}}
   \subfigure[]{
   \label{Fig:figure-West-C4}
   \includegraphics[width=0.18\textwidth, angle=0,clip=true,keepaspectratio=true,trim=0 0 0 0mm]{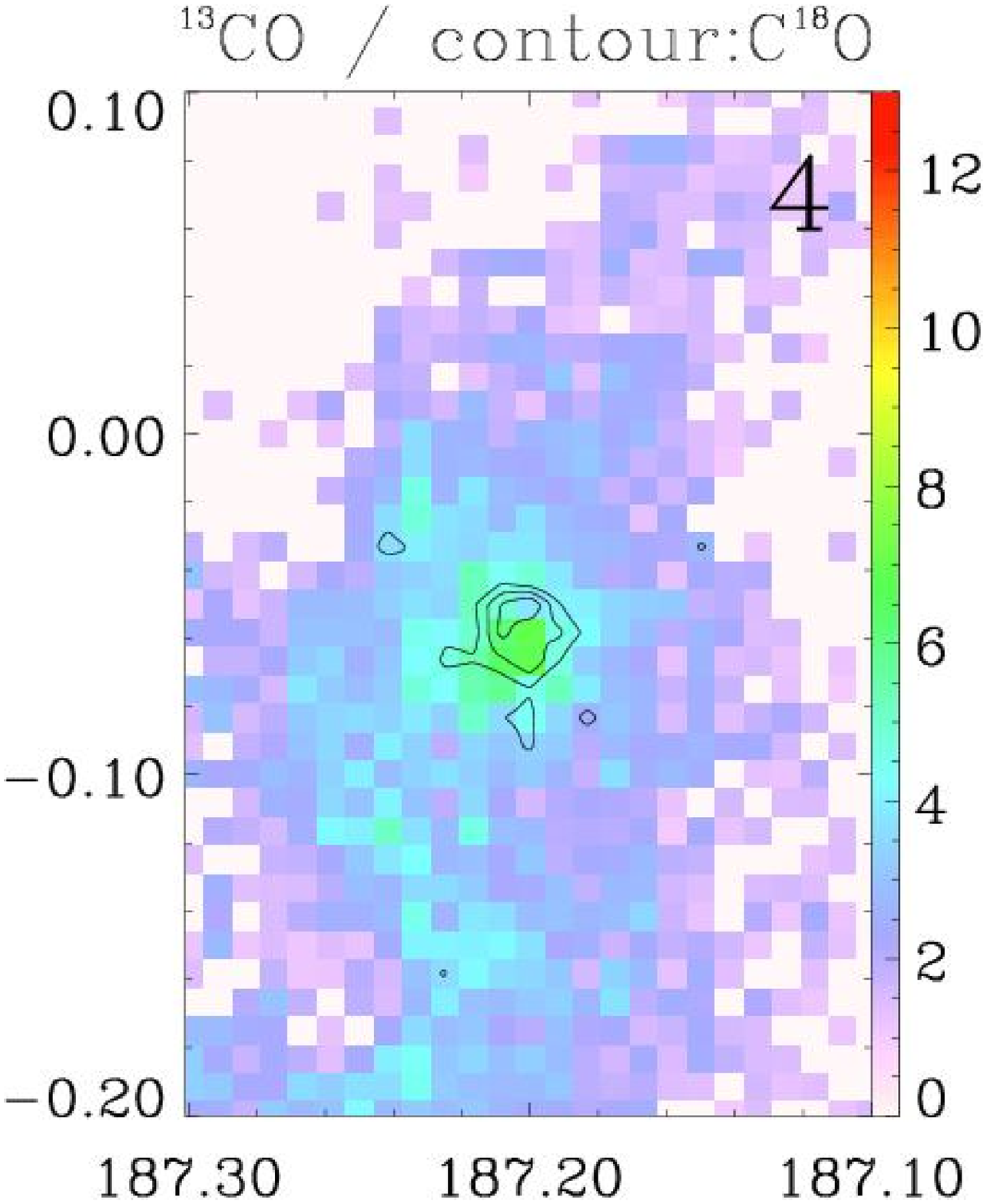}}
   \subfigure[]{
   \label{Fig:figure-West-D}
   \includegraphics[width=0.67\textwidth, angle=0,clip=true,keepaspectratio=true,trim=0 0 0 0mm]{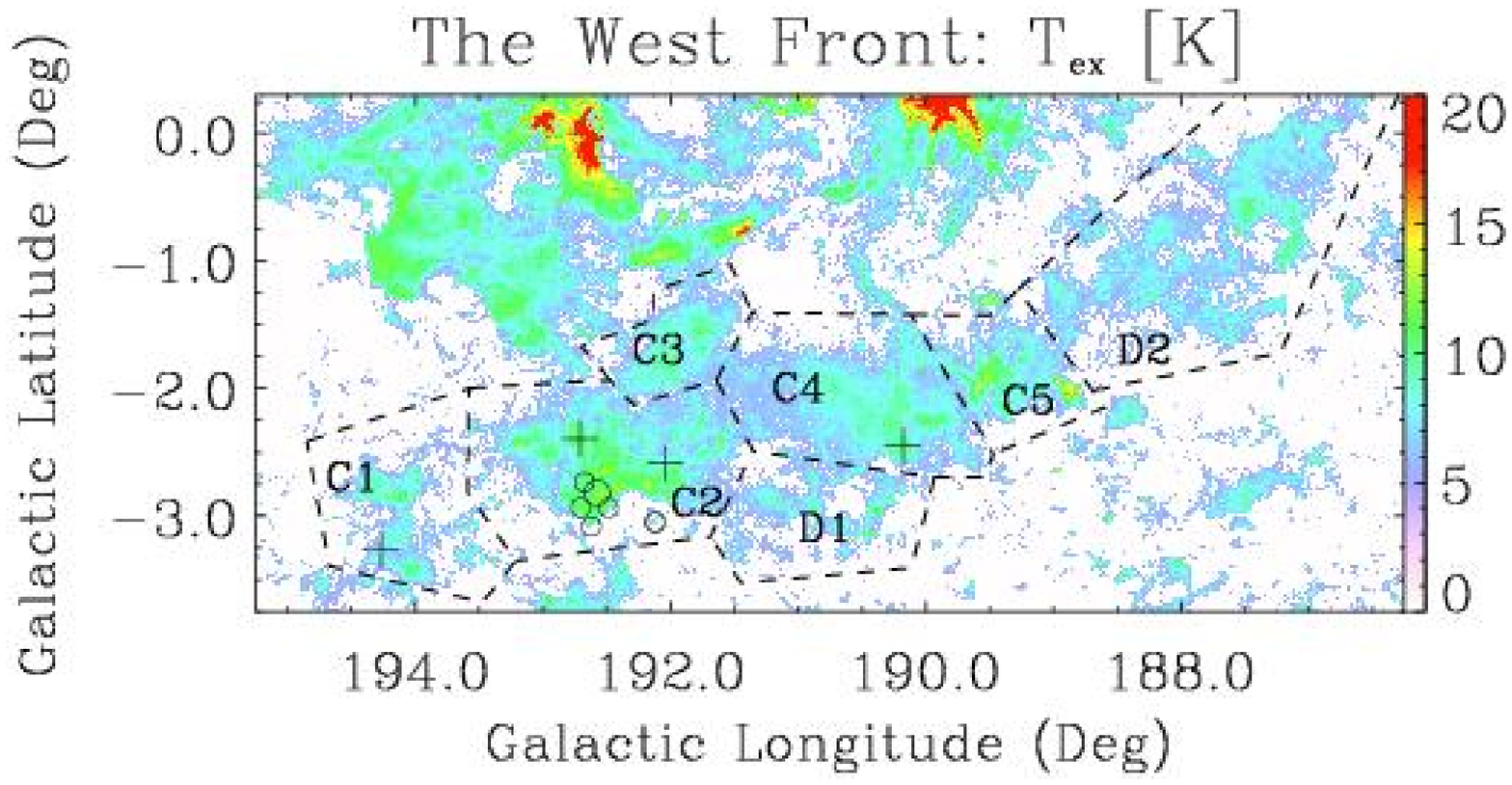}}

   \caption{(a) and (d) The background is {$^{13}$CO} integrated intensity. The contours indicate the {C$^{18}$O} integrated intensity with an interval of [1, 4] {km s$^{-1}$}. The contour levels begin at 0.63 K (3$\sigma$) with increments of 0.21 K. (b) and (c) The background is the {$^{13}$CO} integrated intensity. The contours indicate the {C$^{18}$O} integrated intensity with intervals of [-1, 3] {km s$^{-1}$} and [-3, 1] {km s$^{-1}$}, respectively. The contour levels begin at 0.74 K (3$\sigma$) with increments of 0.25 K. (e) The excitation temperature map of the West Front. The dashed lines show the boundaries of the subregions. The circles show the position of the CB dark clouds (\citealt{1988ApJS...68..257C}) and pluses show the position of the Planck dark clouds (\citealt{2016A&A...594A..28P}).}
    \label{Fig:figure-West-2}
\end{figure}

A very long filamentary cloud is located in the lower part of the observation area, as shown in Figure~\ref{Fig:figure-1}. The cloud extends from the lower left to the right side of the middle of the observation area, and the length of the cloud is about $8.5^{\circ}$ (Figure~\ref{Fig:figure-West-A}). \citet{1995ApJ...445..246C} called this cloud the West Front. {$^{12}$CO} emission is mainly near the middle of the cloud, while gas near both ends of the cloud are fragmented. The {$^{13}$CO} emission is also concentrated in the middle of the cloud, especially in subregions C2, C3, and C4. The gas traced by {$^{13}$CO} at both ends of the cloud is more diffuse. The {C$^{18}$O} emission is relatively weaker than that seen in the GGMCs, and the emission is distributed into four small areas (Figure~\ref{Fig:figure-West-C1} - Figure~\ref{Fig:figure-West-C4}). The radial velocities of the West Front are within [-7, 10] {km s$^{-1}$}, as shown in (Figure~\ref{Fig:LV}). The velocity range is similar to that of the Lynds dark clouds chain. The {H{\scriptsize~II}} region Sh~261 is located above subregion C1 in spatial projection.

The West Front is lacking in systematic research, and information related to its distance is poor. Fortunately, there are four Planck dark clouds in this region which are included in the Planck dark cloud catalog (\citealt{2016A&A...594A..28P}). These dark clouds are PGCCG~190.18-2.45, PGCCG~192.04-2.59, PGCCG~192.71-2.39, and PGCCG~194.27-3.27, which have distances of 0.58~kpc, 0.59~kpc, 0.44~kpc, and 0.63~kpc, respectively (Figure~\ref{Fig:figure-West-D}). Taking distance errors into account, we regard these clouds as being at the same distance. Since the dark clouds overlapped with the West Front in spatial projection, and there are no other velocity components except for the West Front, we suggest that the dark clouds are part of the West front. We adopted the mean distance to the Planck dark clouds of 560 pc as the distance to the West Front in the subsequent analysis. At this distance, the West Front and Lynds dark clouds are locally gas . Additionally, \citet{2015AJ....150..147F} estimated a spectrophotometric distance to Sh~261 of 1.89~kpc, which suggests that Sh~261 is not associated with the West Front.

Figure~\ref{Fig:figure-West-D} shows the distribution of the excitation temperatures in the West Front, which are within the range 4.5-17.5~K (Table~\ref{tbl-west}). The maximum excitation temperature is located at subregion C5, while the excitation temperatures in the other subregions are less than 14~K. Thus, the temperature distribution of the West Front is similar to that of the Lynds dark clouds chain. In addition to the Planck dark clouds, six CB dark clouds (\citealt{1988ApJS...68..257C}) are also associated with the West Front (Figure~\ref{Fig:figure-West-D}). Hence, we regard the West Front as a dark cloud complex.

At a distance of 560~pc, the physical length of the West Front is 83~pc. The total mass of the West Front traced by {$^{12}$CO} is found to be 5.0 $\times10^{4}$~M$_{\odot}$, when using the adopted X-factor. The LTE masses based on {$^{13}$CO} and {C$^{18}$O} are 8.4 $\times10^{3}$~M$_{\odot}$ and 1.9 $\times10^{2}$~M$_{\odot}$, respectively. The mass of the West Front is similar to several giant molecular filaments reported by \citet{2014A&A...568A..73R}. Since the West Front is local, it is a good sample with which to study the physical properties and structures of elongated clouds.

\section{The Other Clouds}
\label{Sec:others}

In addition to the clouds that were discussed above, there are three other MCs presented here: Swallow, Horn, and the Remote Clouds. The typical velocities of these three clouds are higher than those of the clouds formerly discussed. Their details are introduced in the following.

\subsection{Swallow}

\begin{figure}
   \centering
   \subfigure[]{
   \label{Fig:figure-Swallow-A}
   \includegraphics[width=0.5\textwidth, angle=0,clip=true,keepaspectratio=true,trim=0 0 0 0mm]{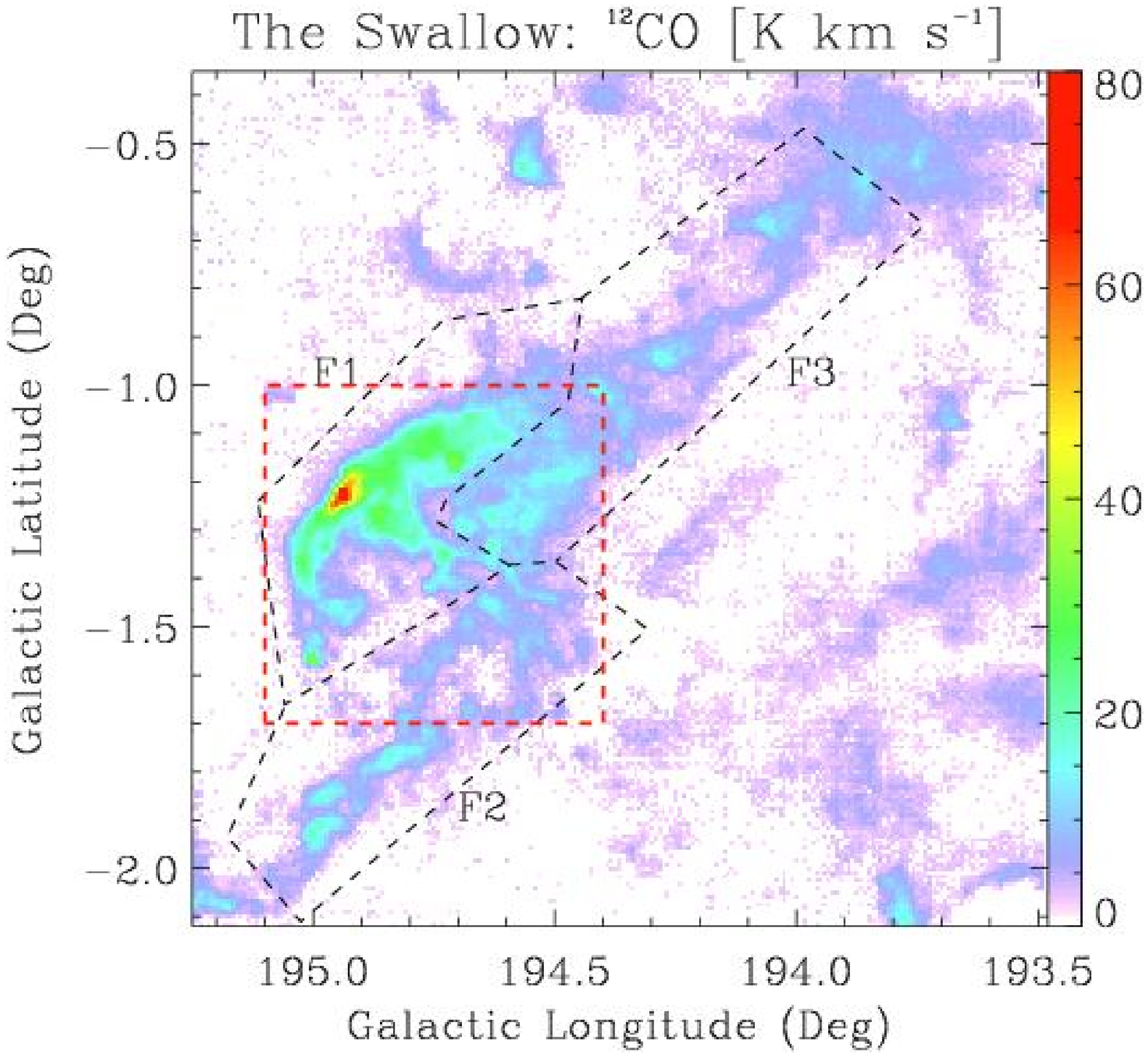}}
   \subfigure[]{
   \label{Fig:figure-Swallow-B}
   \includegraphics[width=0.4\textwidth, angle=0,clip=true,keepaspectratio=true,trim=0 0 0 0mm]{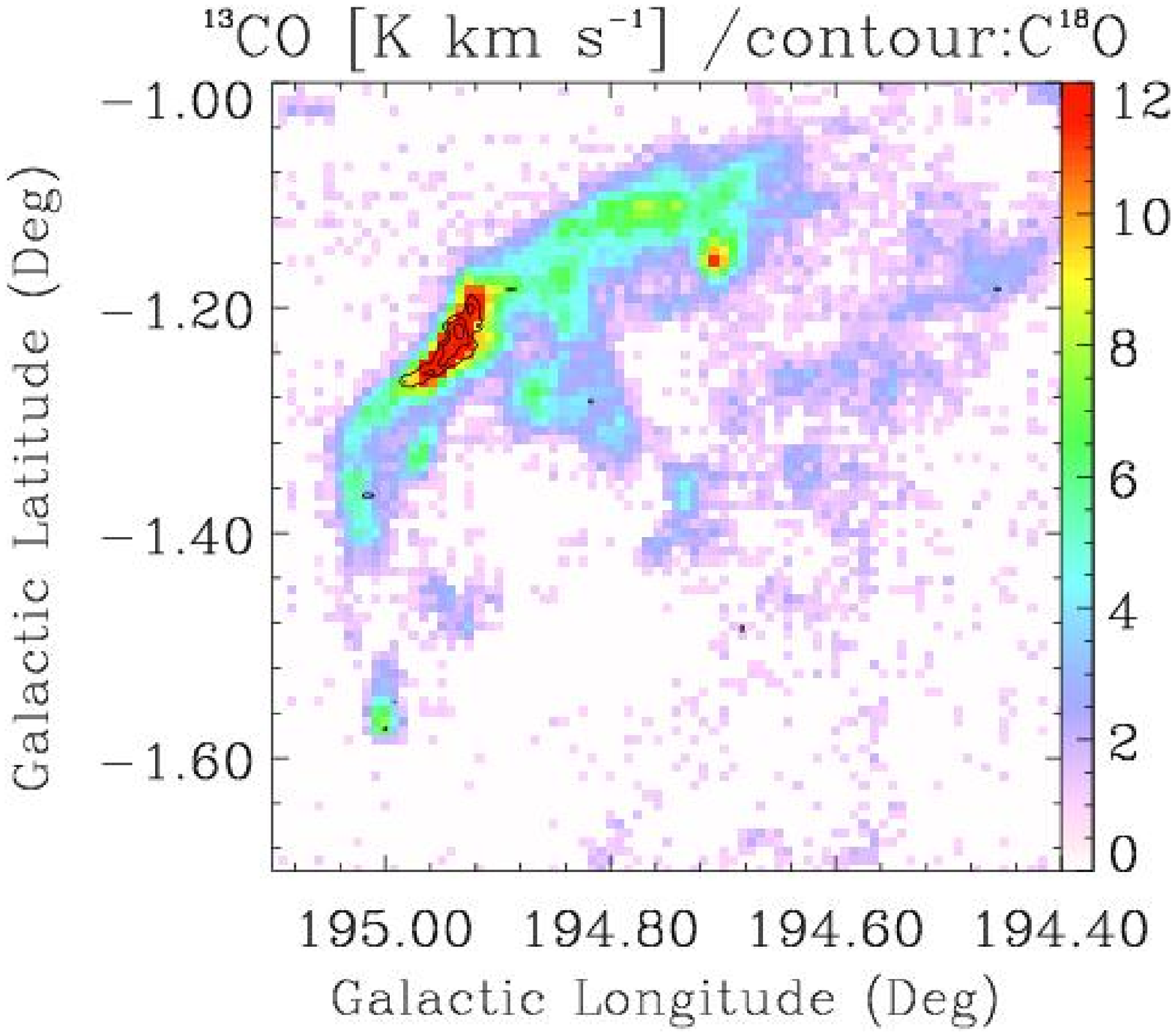}}
   \subfigure[]{
   \label{Fig:figure-Swallow-C}
   \includegraphics[width=0.5\textwidth, angle=0,clip=true,keepaspectratio=true,trim=0 0 0 0mm]{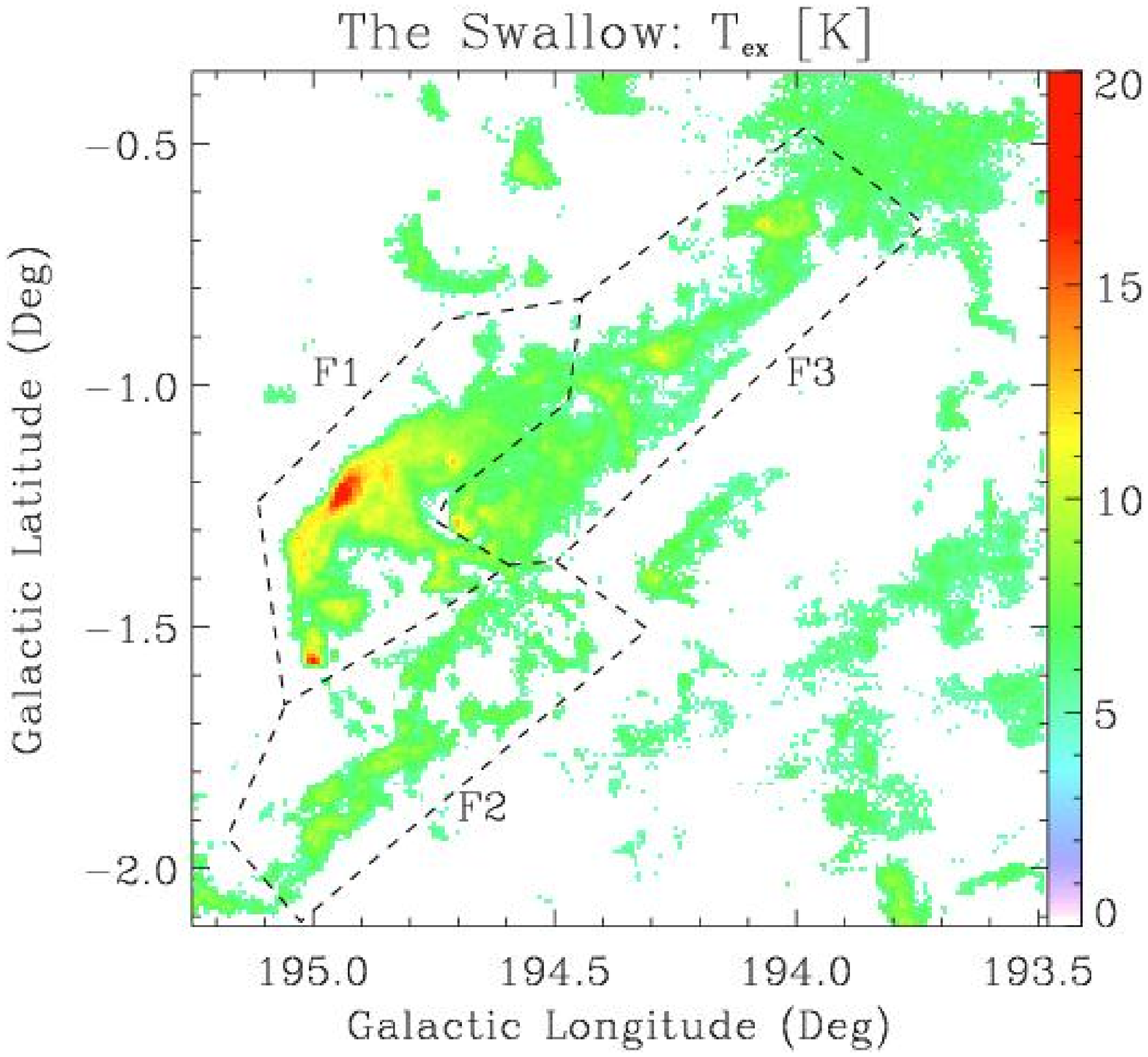}}

   \caption{Gas distribution of Swallow. (a) {$^{12}$CO} integrated intensity map in the range of [11, 20] \kms. The dashed lines show the boundaries of the subregions. The red rectangle shows the area of Figure~\ref{Fig:figure-Swallow-B}. (b): The background is the {$^{13}$CO} integrated intensity in the range of [11, 19] \kms. The contours indicate the integrated intensity of \coiii\  with the range of [12, 18] \kms, the steps are 4, 6, 8 $\times$ 0.25 K \kms (1$\sigma$). (c) The excitation temperature map of Swallow MCs.}
    \label{Fig:figure-Swallow}
\end{figure}

The Swallow cloud is to the southeast of GGMC~1, and the clouds appear close together, as projected on the plane of the sky (Figure~\ref{Fig:figure-3}). The {$^{12}$CO} emission of Swallow shows three filaments (Figure~\ref{Fig:figure-Swallow-A}). Filament F1 is an arc-like structure and is coherent in space, while the others are more diffuse. The {$^{13}$CO} emission is concentrated to F1 (Figure~\ref{Fig:figure-Swallow-B}), while the {C$^{18}$O} emission is only located at the peak emission of F1 (Figure~\ref{Fig:figure-Swallow-B}). {C$^{18}$O} shows a crescent-shaped structure in the integrated intensity map (Figure~\ref{Fig:figure-Swallow-B}).

The radial velocities of Swallow are within [10, 20] km s$^{-1}$, which is different from that of GGMC~1 (Figure~\ref{Fig:LV}). Hence, they are probably independent clouds. The excitation temperature distribution of Swallow is shown in Figure~\ref{Fig:figure-Swallow-C}, where the maximum temperature is 20~K (Table~\ref{tbl-swallow}), which is located at the center of filament F1. The excitation temperatures of F2 and F3 are below 14~K. We note that the southernmost part of filament F1 has a maximum temperature of 17~K. Its distance will be discussed in Sect~\ref{Sec:discussion}.

Interestingly, there is a cloud (194.9-01.2) in the catalog of \citet{1998ApJS..117..387K} that is located in the Swallow area. By comparing 194.9-01.2 with our {$^{12}$CO}, {$^{13}$CO} emission map (Figure~\ref{Fig:figure-Swallow-A},Figure~\ref{Fig:figure-Swallow-B}), we find that it is located in subregion F1.

\subsection{Horn}

\begin{figure}
   \centering
   \subfigure[]{
   \label{Fig:figure-Horn-A}
   \includegraphics[width=0.46\textwidth, angle=0,clip=true,keepaspectratio=true,trim=0 0 0 0mm]{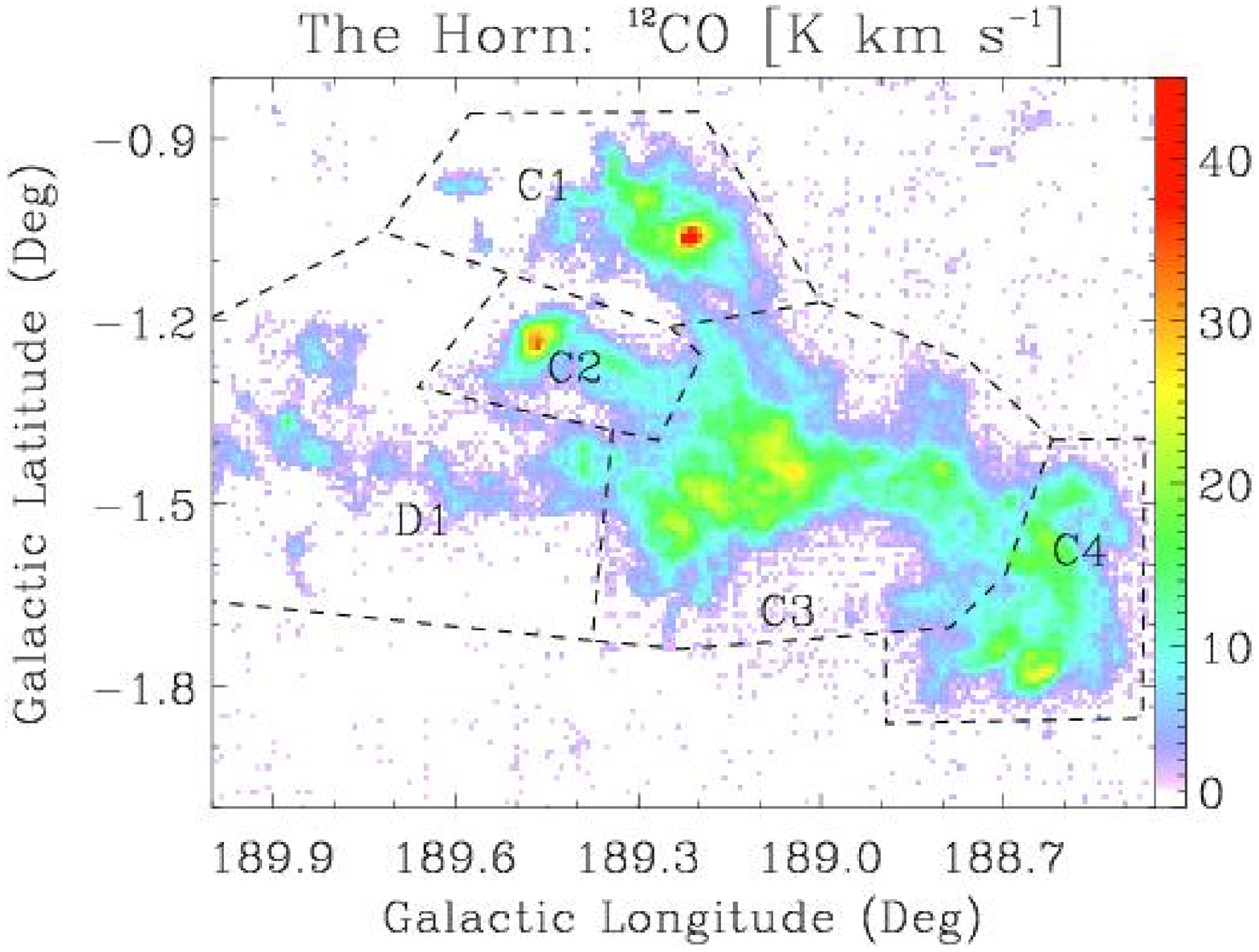}}
   \subfigure[]{
   \label{Fig:figure-Horn-B}
   \includegraphics[width=0.46\textwidth, angle=0,clip=true,keepaspectratio=true,trim=0 0 0 0mm]{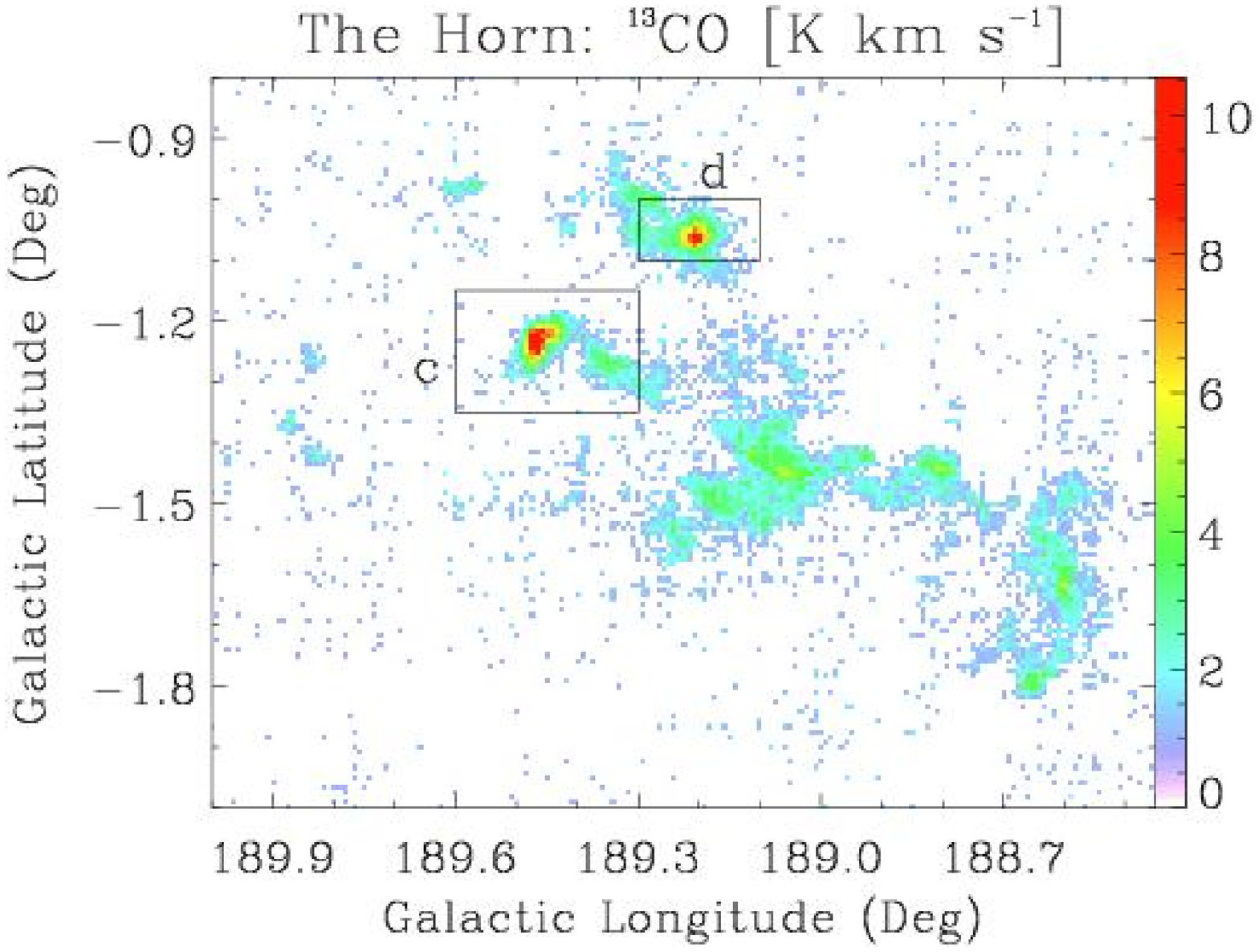}}
   \subfigure[]{
   \label{Fig:figure-Horn-C}
   \includegraphics[width=0.4\textwidth, angle=0,clip=true,keepaspectratio=true,trim=0 0 0 0mm]{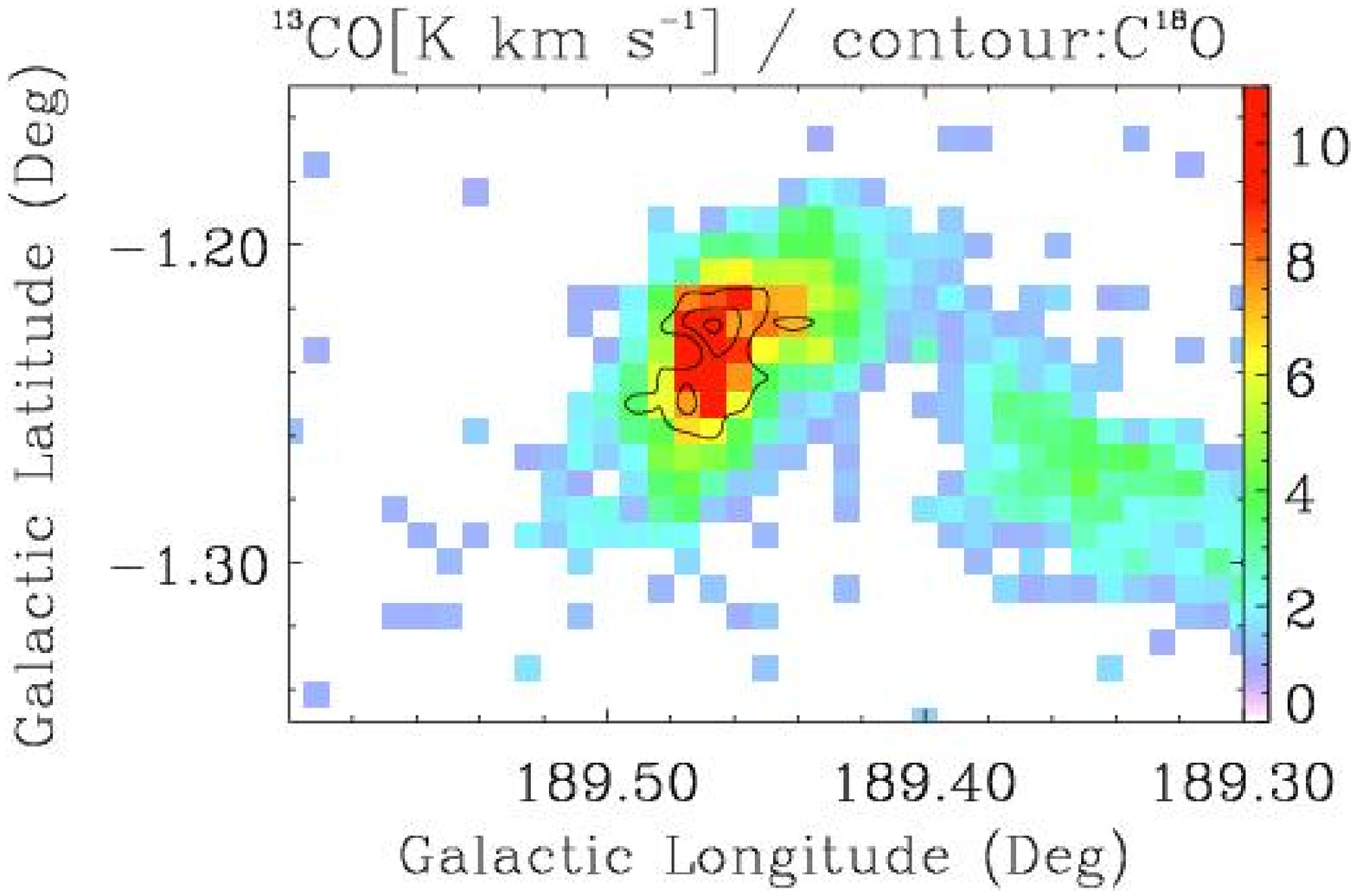}}
   \subfigure[]{
   \label{Fig:figure-Horn-D}
   \includegraphics[width=0.4\textwidth, angle=0,clip=true,keepaspectratio=true,trim=0 0 0 0mm]{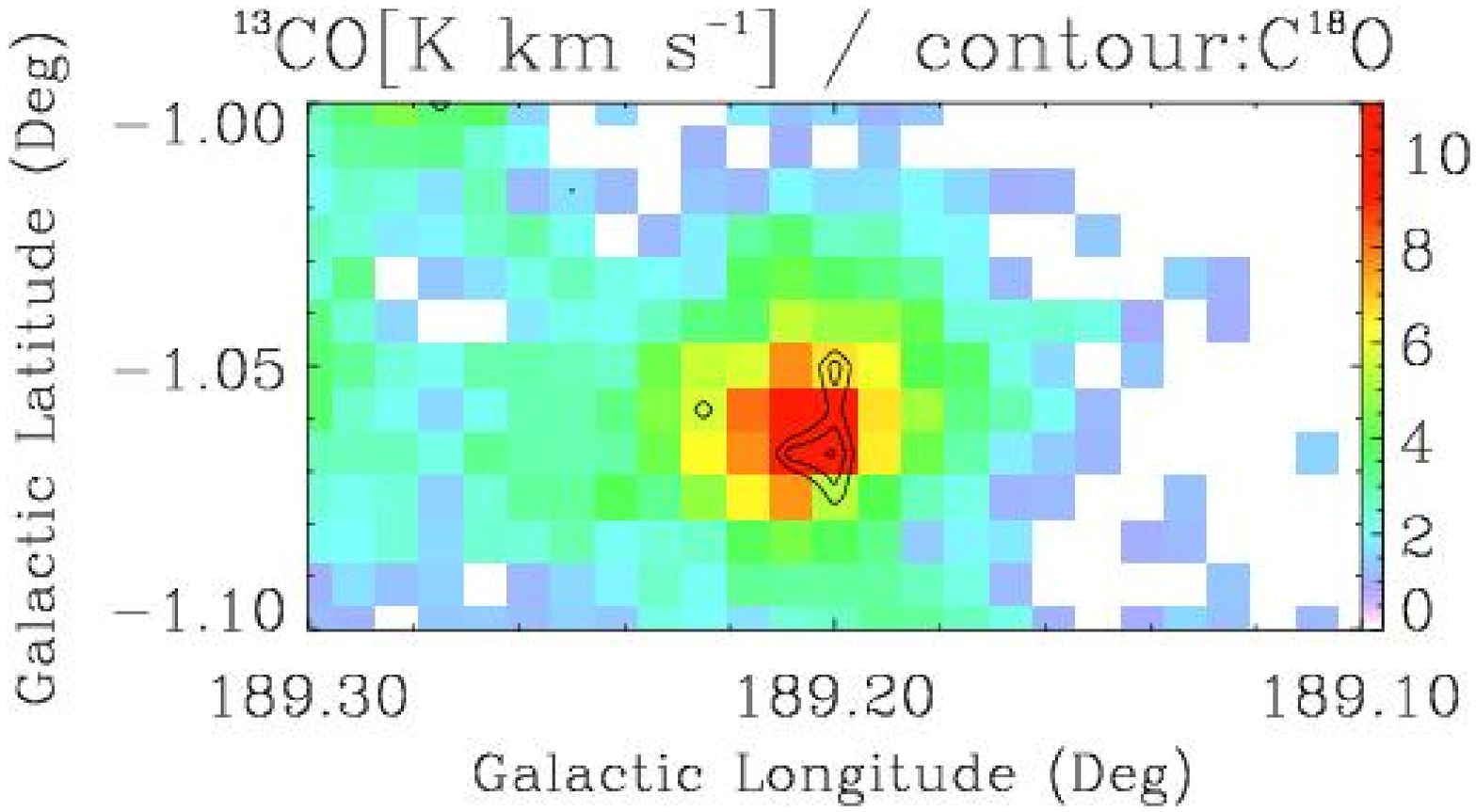}}

   \caption{Gas distribution of Horn. (a) {$^{12}$CO} integrated intensity map in the range of [11, 21] \kms. The dashed lines show the boundaries of the subregions. (b) Map of the {$^{12}$CO} integrated intensity in the range of [11, 20] \kms. The rectangles show the areas of Figure~\ref{Fig:figure-Horn-C} and Figure~\ref{Fig:figure-Horn-D}. (c) and (d) The background is a map of the {$^{13}$CO} integrated intensity. The contours indicate the integrated intensity of \coiii\  with the range of [12, 18] \kms; the steps are 4, 6, 8 $\times$ 0.25 K \kms(1$\sigma$).}
    \label{Fig:figure-Horn}
\end{figure}

The Horn is located to the southwest of GGMC~3, and it overlaps with the West Front in spatial distribution (Figure~\ref{Fig:figure-3-C}), but with a different velocity interval of [11, 21] km s$^{-1}$ (Figure~\ref{Fig:LV}). Thus, the Horn and the West Front are probably independent clouds. The cloud is filamentary, as shown in Figure~\ref{Fig:figure-Horn-A}. The {$^{13}$CO emission} shows two structures in subregions C1 and C2 (Figure~\ref{Fig:figure-Horn-B}). The {C$^{18}$O} emission is weak, and it is located in an area of enhanced {$^{13}$CO} emission (Figure~\ref{Fig:figure-Horn-C}, Figure~\ref{Fig:figure-Horn-D}).

\begin{figure}
   \centering
   \includegraphics[width=\textwidth, angle=0,clip=true,keepaspectratio=true,trim=0 0 0 0mm]{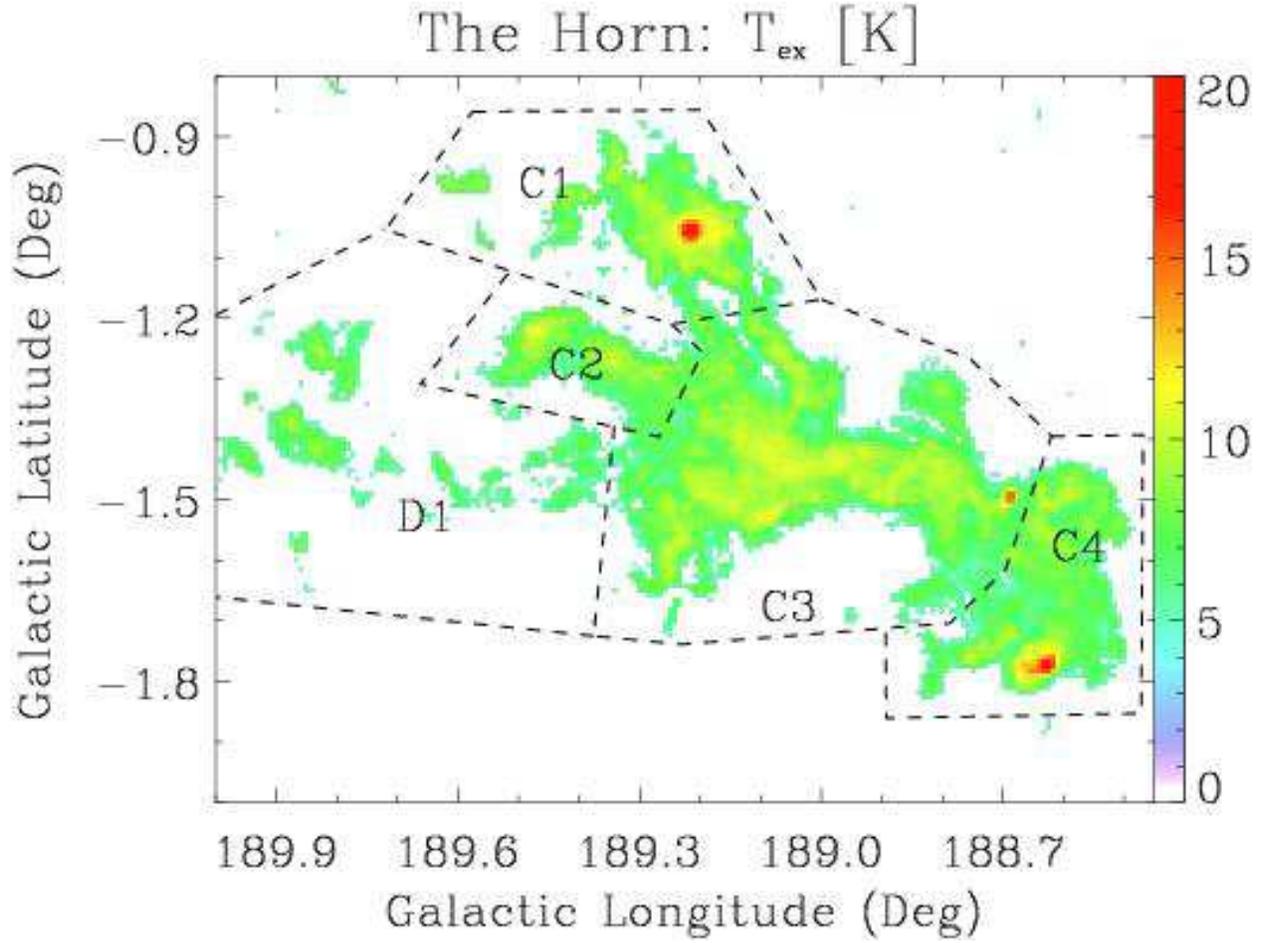}
   \caption{ Excitation temperature map of the MC Horn. The dashed lines show the boundaries of the subregions.}
    \label{Fig:figure-Horn-tex}
\end{figure}

Figure~\ref{Fig:figure-Horn-tex} shows the excitation temperature distribution of Horn. The Horn has two temperature maxima: one is 20~K, which is located in subregion C1, which has strong {$^{13}$CO} emission and weak {C$^{18}$O} emission, while the other is 18~K, which is located in the southernmost part of Horn that has weak {$^{13}$CO} emission and no {C$^{18}$O} emission (Table~\ref{tbl-horn}, Figure~\ref{Fig:figure-Horn-tex}). Most of the molecular gas temperatures in the cloud are below 14~K. The distance information of Horn will be discussed in Sects~\ref{Sec:discussion}.

In a past study, \citet{1998ApJS..117..387K} found two clouds in this area, which they named 188.8-01.5 and 189.2-01.1. By comparing these clouds with our {$^{12}$CO} map, we find that the {$^{13}$CO} emission map (Figure~\ref{Fig:figure-Horn-A}, Figure~\ref{Fig:figure-Horn-B}), corresponds to the middle of the C3 and C1 subregions.

\subsection{Remote Clouds}

\begin{figure}
   \centering
   \subfigure[]{
   \label{Fig:figure-Remote-A}
   \includegraphics[width=0.45\textwidth, angle=0,clip=true,keepaspectratio=true,trim=0 0 0 0mm]{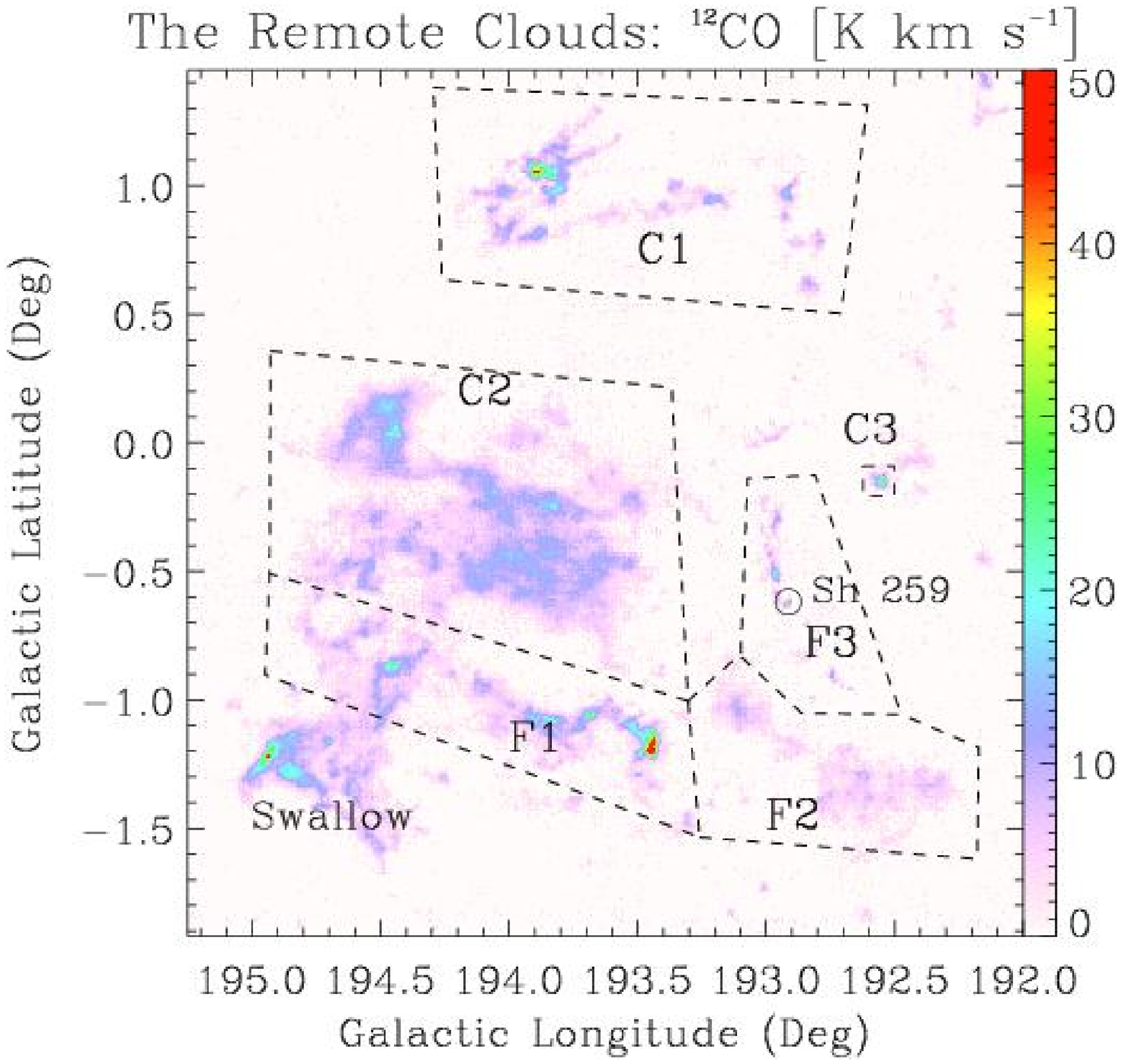}}
   \subfigure[]{
   \label{Fig:figure-Remote-B}
   \includegraphics[width=0.45\textwidth, angle=0,clip=true,keepaspectratio=true,trim=0 0 0 0mm]{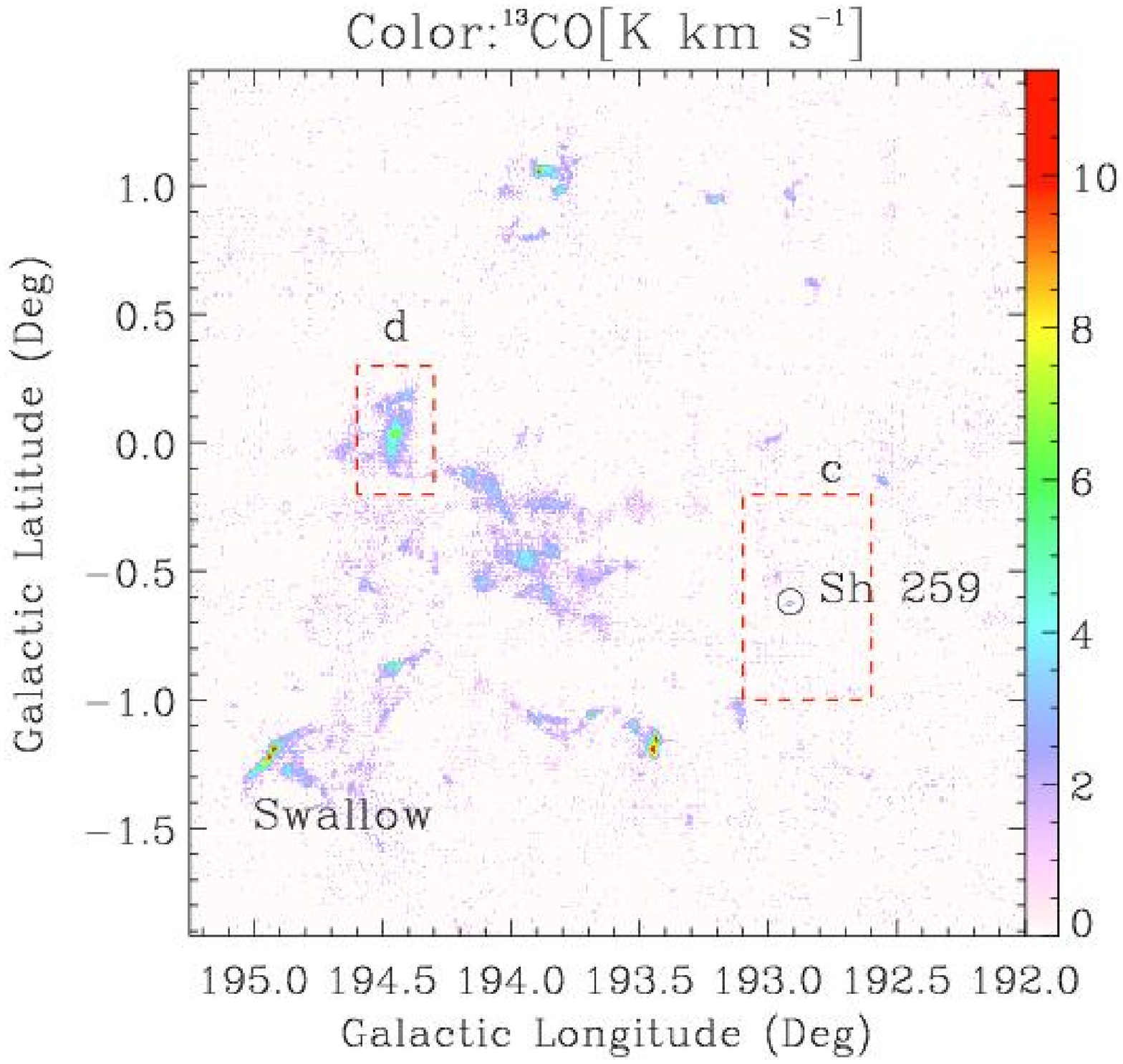}}
   \subfigure[]{
   \label{Fig:figure-Remote-C}
   \includegraphics[width=0.4\textwidth, angle=0,clip=true,keepaspectratio=true,trim=0 0 0 0mm]{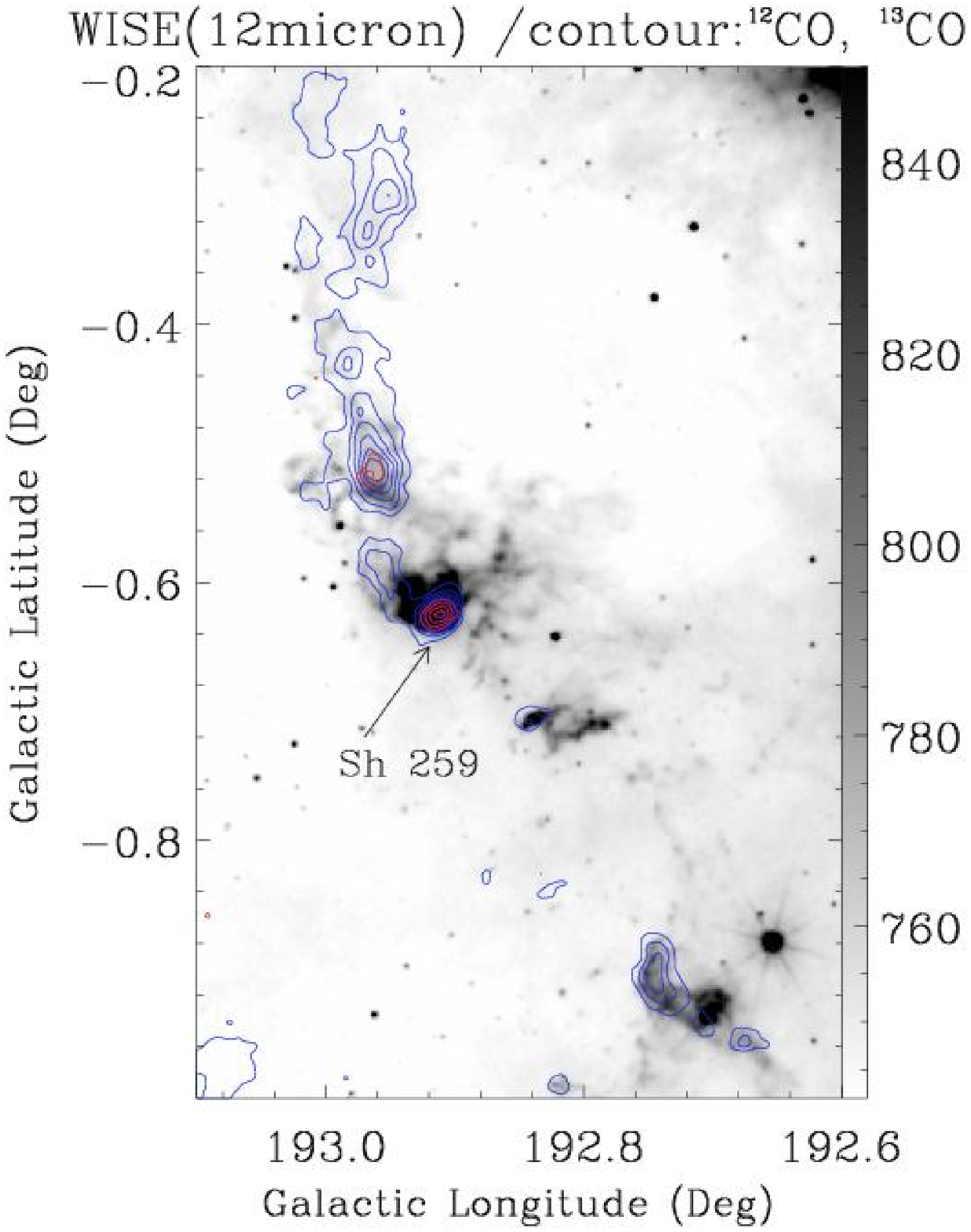}}
   \subfigure[]{
   \label{Fig:figure-Remote-D}
   \includegraphics[width=0.4\textwidth, angle=0,clip=true,keepaspectratio=true,trim=0 0 0 0mm]{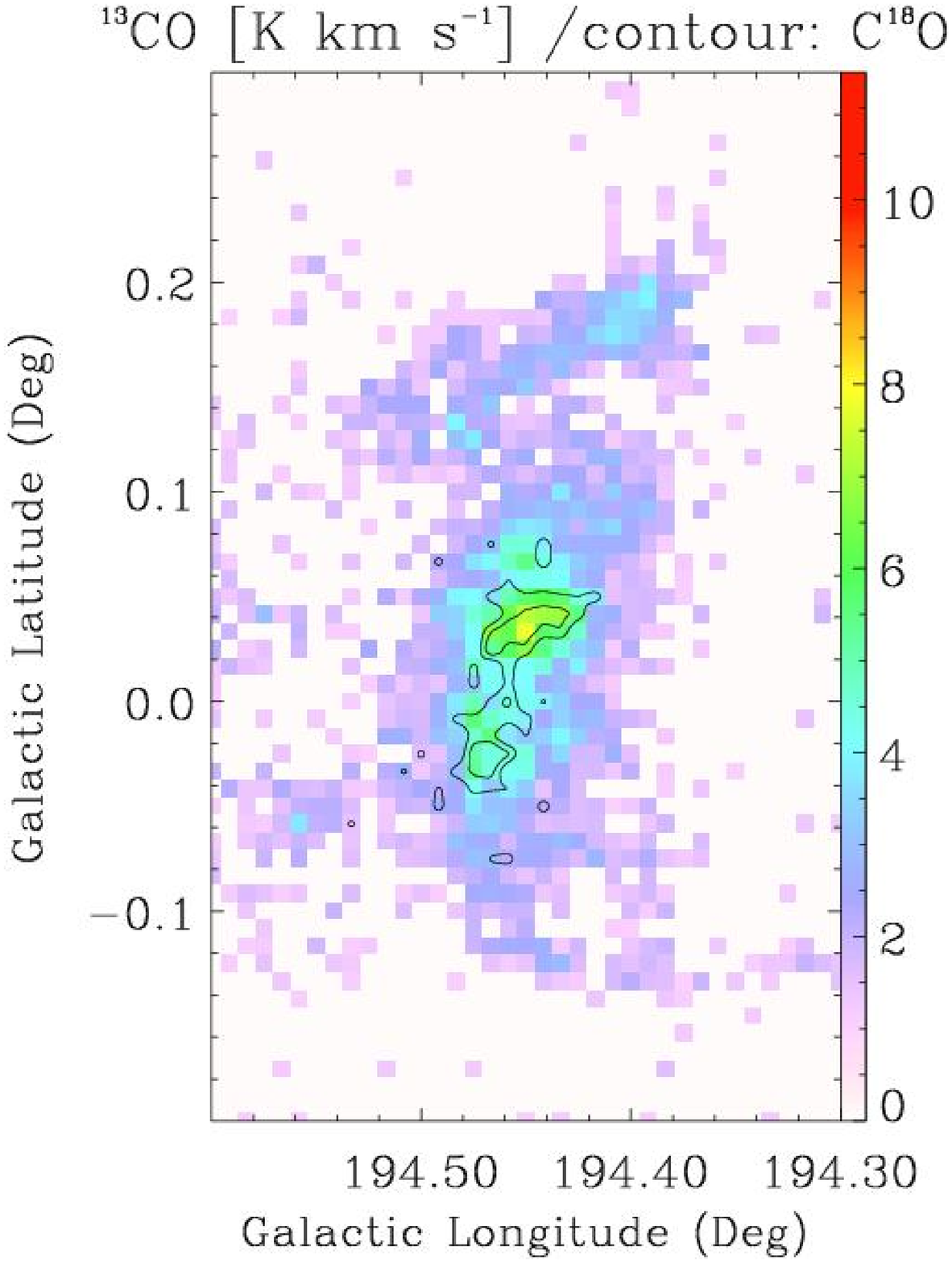}}

   \caption{Gas distribution of the Remote Clouds. (a) {$^{12}$CO} integrated intensity map in the range of [16,~32] {km s$^{-1}$}. The circle represents the \Hii\ region. The dashed lines show the boundaries of the subregions. (b) Map of the {$^{13}$CO} integrated intensity in the range of [18,~28] {km s$^{-1}$}. (c) Image of a WISE 12 micron map. The blue contours indicate the integrated intensity of {$^{12}$CO}, and the contour levels begin at 3.1 K {km s$^{-1}$} with increments of 3.1 K {km s$^{-1}$} (5$\sigma$). The red contour indicates the integrated intensity of {$^{13}$CO}, and the contour levels begin at 1.5 K {km s$^{-1}$} with increments of 0.78 K {km s$^{-1}$} (2$\sigma$). (d) Image of the integrated intensity map of {$^{13}$CO}. The contours indicate the integrated intensity of {C$^{18}$O}  with the range of [18, 20] {km s$^{-1}$}; the steps are 3, 5 $\times$ 0.25 K \kms (1$\sigma$).}
    \label{Fig:figure-Remote}
\end{figure}

The Remote Clouds are located to the east of the observation area, with a maximum velocity of [16, 28] km s$^{-1}$ (Figure~\ref{Fig:figure-3-C}). These clouds are probably the most distant clouds according to Galactic rotation curves. The {$^{12}$CO} emission of the Remote Clouds shows six distinct subregions (Figure~\ref{Fig:figure-Remote-A}). The cloud area is about 2.7$^{\circ}$ $\times$ 2.9$^{\circ}$ in Galactic coordinates. Subregion C2 is the largest cloud in the area, and there are three filaments to the south and west of this area. C3 is located to the northwest of F3, which is a cloud with a small size but relatively strong {$^{12}$CO} emission. The mean {$^{12}$CO} integrated intensity of the Remote Clouds is less than that of the other clouds studied in this paper. The {$^{13}$CO} integrated intensity of the Remote Clouds shows a clumpy distribution (Figure~\ref{Fig:figure-Remote-B}) and the mean intensity is weaker than that of the GGMC complex and local clouds presented above. {C$^{18}$O} emission is absent in most subregions except for C2 (Figure~\ref{Fig:figure-Remote-D}).

Sh~259 is a small \Hii\ region (\citet{1959ApJS....4..257S}) located in the peak emission of the subregion F3 area (Figure~\ref{Fig:figure-Remote-C}), thus indicating they are physically associated (\citealt{1995ApJ...445..246C}). Moreover, the WISE band 3 (12 micrometers) map also shows a filamentary structure near Sh~259, which also suggests a physical correlation with our CO filamentary structure.

\begin{figure}
   \centering
   \includegraphics[width=\textwidth, angle=0,clip=true,keepaspectratio=true,trim=0 0 0 0mm]{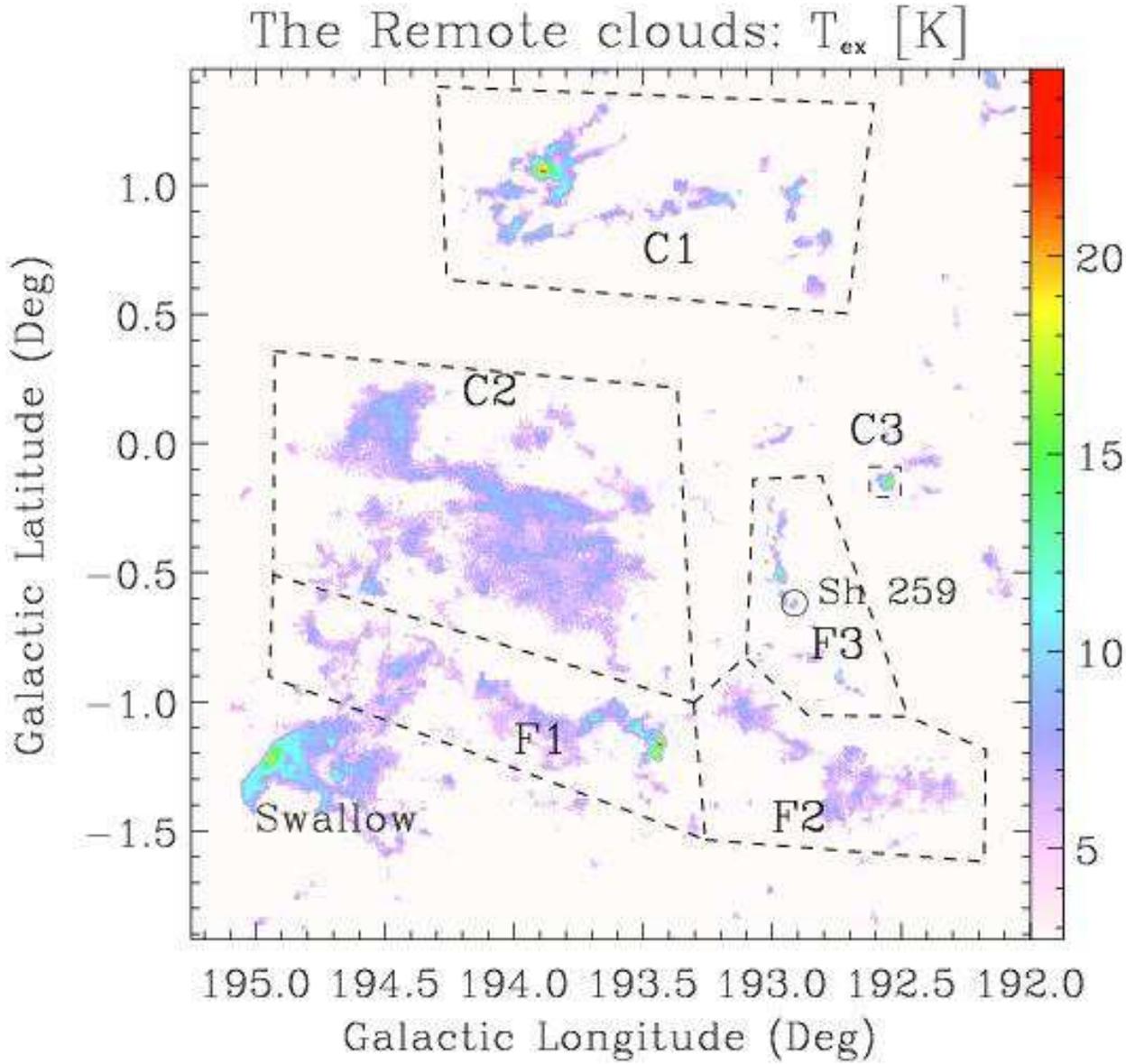}
   \caption{ The excitation temperature map of the Remote Clouds. The dashed lines show the boundaries of the subregions. }
    \label{Fig:figure-Remote-tex}
\end{figure}

The excitation temperature distribution of the Remote Clouds is presented in Figure~\ref{Fig:figure-Remote-tex}. The maximum temperatures of C1 and F1 are 23~K and 25~K, respectively (Table~\ref{tbl-remote}). The small cloud C3 is within the range of 4.5-19~K, and the other three cloud temperatures are below 20~K.  We note that cloud C2, which is the only one that has {C$^{18}$O} emission in the Remote Clouds along with relatively strong {$^{13}$CO} emission, is very cold, with a maximum temperature of 10~K. The distance to the Remote Clouds will be discussed in Sects~\ref{Sec:discussion}.

\section{Discussion}
\label{Sec:discussion}

Due to the degeneracy of the radial velocity in the Galactic anti-center, the kinematic distance uncertainties of the MCs are large. Some groups have searched for celestial bodies that are associated with the cloud, such as \Hii\ regions, massive stars, and maser sources, to determine a distance to a given cloud (\citealt{1995ApJ...445..246C}, \citealt{2009ApJ...693..397R}, \citealt{2010A&A...511A...2R}, \citealt{2011PASJ...63....9N}). Other groups obtained cloud distances via extinction measurements (e.g., \citealt{2013A&A...556A..65E}). In this work, the distances to the GGMCs, Lynds dark clouds and West Front Clouds are discussed in Sects~\ref{Sec:ggmc} - Sect~\ref{Sec:west}.

So far there is no distance information for Swallow and Horn, and there are no other sources nearby that can be traced to determine their distances. There is a dark cloud that just overlaps with Horn, which is listed in the Planck dark clouds catalog as PGCCG~188.82-1.46 at a distance of 3.78 kpc (\citealt{2016A&A...594A..28P}). We note that the dark cloud has strong CO emission in subregion C3 in the Horn, but there is some weak {$^{12}$CO} emission from the West Front along the LoS to the Planck dark cloud. As the West Front is local, while the distance of the Planck dark cloud is 3.78 kpc, we regard the dark cloud as being associated with the Horn and it is in the background of the West Front. It is consistent with the velocity of Horn, which is higher than that of the West Front. Since the influence of weak {$^{12}$CO} emission is from the foreground , 3.78 kpc should be considered as an upper limit to the distance to the Horn.

The Swallow and Horn MCs have the same velocity interval, [11, 21] km s$^{-1}$, but their spatial separation is quite large ($\sim$$4^{\circ}$). They have a similar spatial size (0.6 and 0.8 deg$^{2}$), and the mean column densities derived from {$^{12}$CO} and {$^{13}$CO} are also similar (Table~\ref{tbl-1}). They are probably at the same distance from Earth on the basis of the galactic rotation curve. If we assume that the distance of the clouds is 3.78 kpc, the masses of the Swallow and Horn respectively are $9.2\times10^{4}$~M$_{\odot}$ and $7.9\times10^{4}$~M$_{\odot}$, as derived from {$^{12}$CO}. If their distance is 2~kpc, their respective masses are $2.6\times10^{4}$~M$_{\odot}$ and $2.2\times10^{4}$~M$_{\odot}$, as derived from {$^{12}$CO}.

The Remote Clouds include six subregions (Figure~\ref{Fig:figure-Remote-A}). Those subregions are close to each other in space, and they have the same velocity interval. However, the sizes and the shapes of the clouds are different. Since Sh~259 is physically associated with subregion F3, the distance of F3 can be determined. \citet{1979A&AS...38..197M} suggested that Sh~259 is 8.3 kpc~distant, based on spectroscopy of the excited star, while \citet{2015AJ....150..147F} found a distance of 8.71~kpc. In this paper, we suggest that the distance to F3 is 8.71~kpc, which is consistent with the small size and lower intensity of the CO emission detected in F3. Hence the masses of subregion F3 are $2.1\times10^{4}$~M$_{\odot}$ and $4.7\times10^{3}$~M$_{\odot}$, as derived from {$^{12}$CO} and {$^{13}$CO}, respectively.

As there is no distance information of the other subregions, we are not sure if the clouds are located at the same distance. Assuming the clouds are at the same distance, we calculated the mass of the clouds, which are presented in Table~\ref{tbl-remote}. Under this assumption, the masses of the Remote Clouds are $1\times10^{6}$~M$_{\odot}$, $2.3\times10^{5}$~M$_{\odot}$, and $3.1\times10^{4}~$ M$_{\odot}$, as derived from {$^{12}$CO}, {$^{13}$CO}, and {C$^{18}$O}, respectively. Next, the size of the complex is 410~pc $\times$ 440~pc. These inferred sizes and the masses suggest that the Remote Clouds are larger if located at a distance of $\sim$17 kpc from the Galaxy Center. We thus conclude that the clouds probably are not at the same distance as F3.

\begin{figure}
   \centering
   \subfigure[]{
   \label{Fig:figure-statistics-A}
   \includegraphics[width=0.45\textwidth, angle=0,clip=true,keepaspectratio=true,trim=0 0 0 0mm]{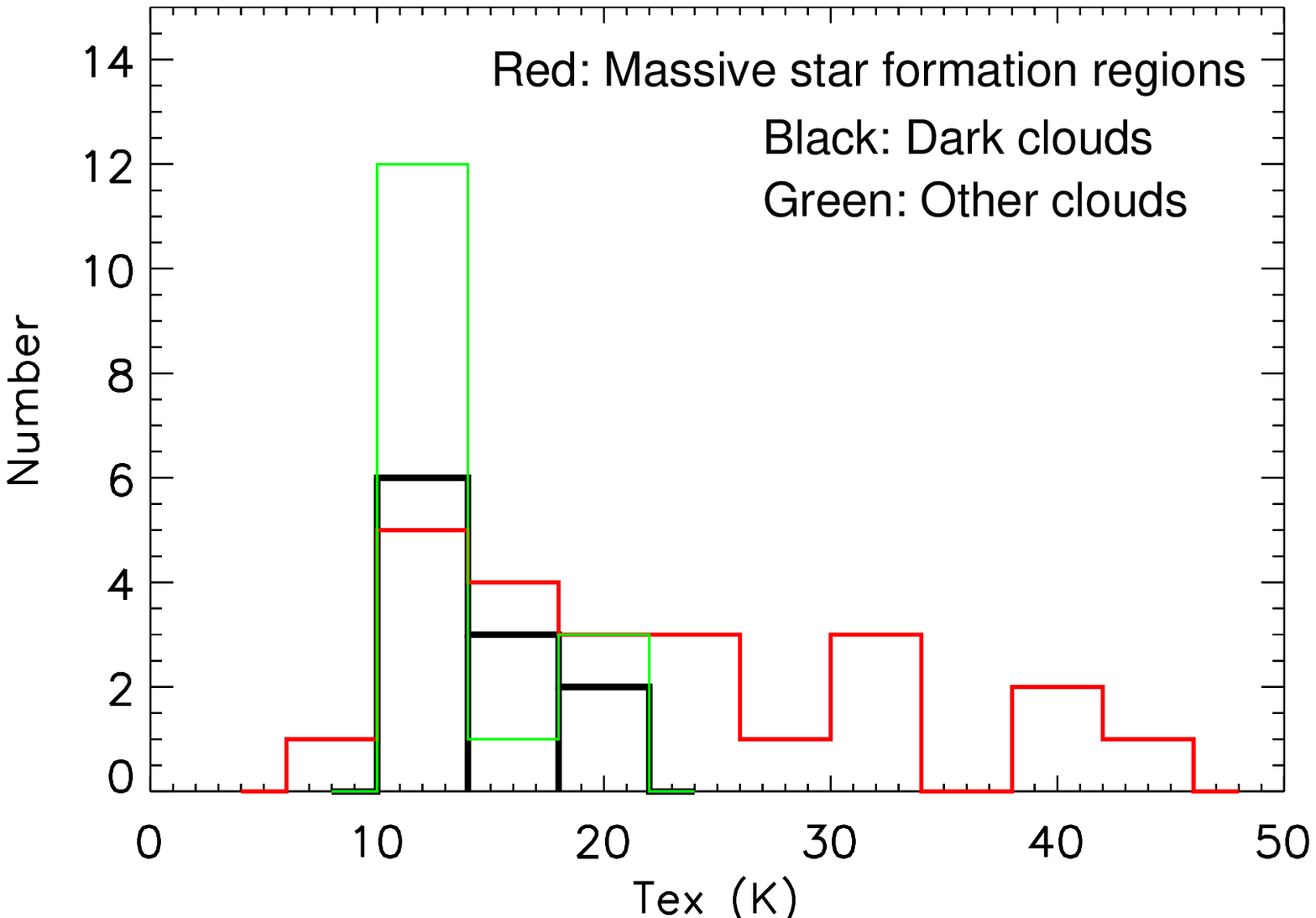}}
   \subfigure[]{
   \label{Fig:figure-statistics-B}
   \includegraphics[width=0.45\textwidth, angle=0,clip=true,keepaspectratio=true,trim=0 0 0 0mm]{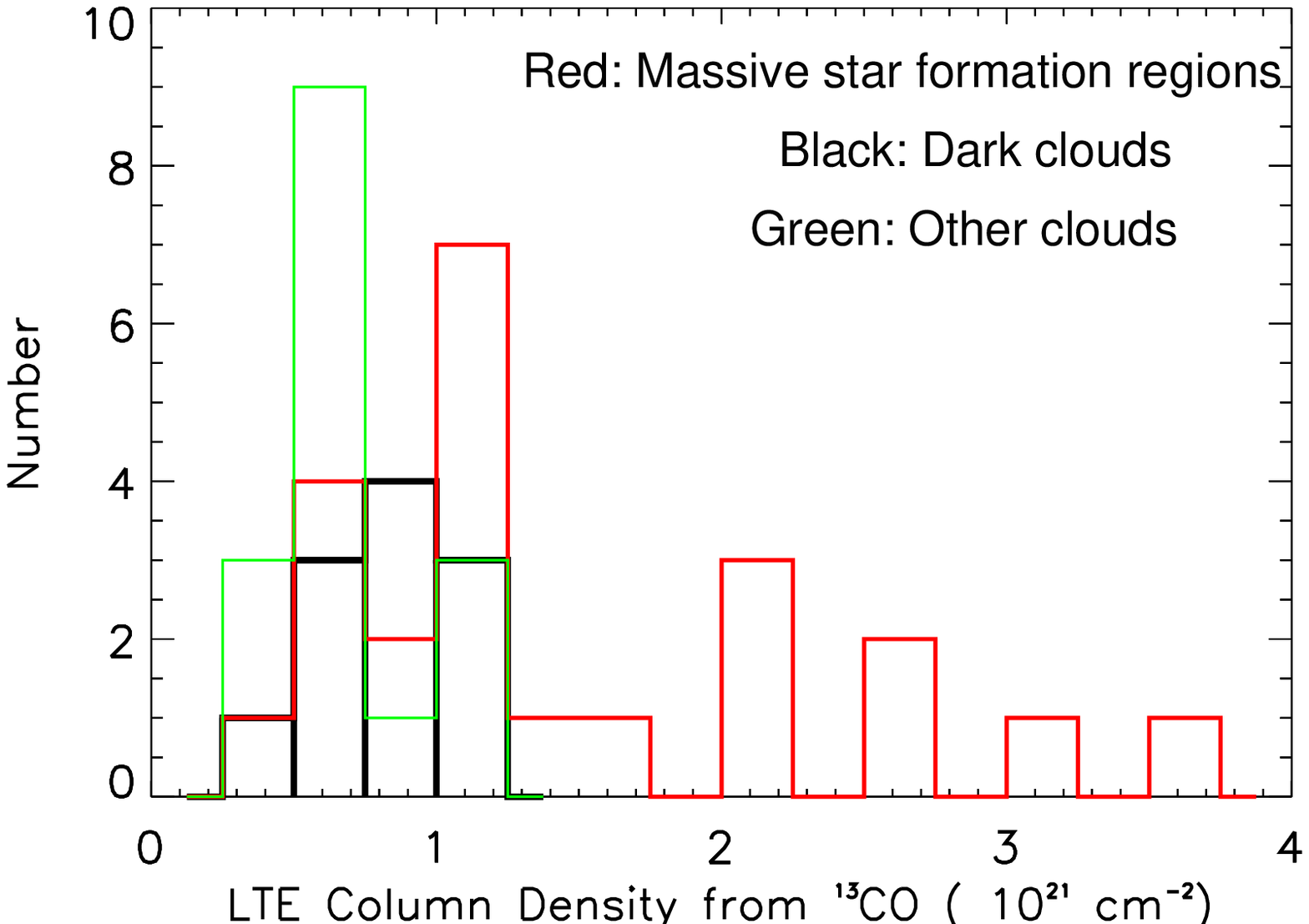}}
   \subfigure[]{
   \label{Fig:figure-statistics-C}
   \includegraphics[width=0.45\textwidth, angle=0,clip=true,keepaspectratio=true,trim=0 0 0 0mm]{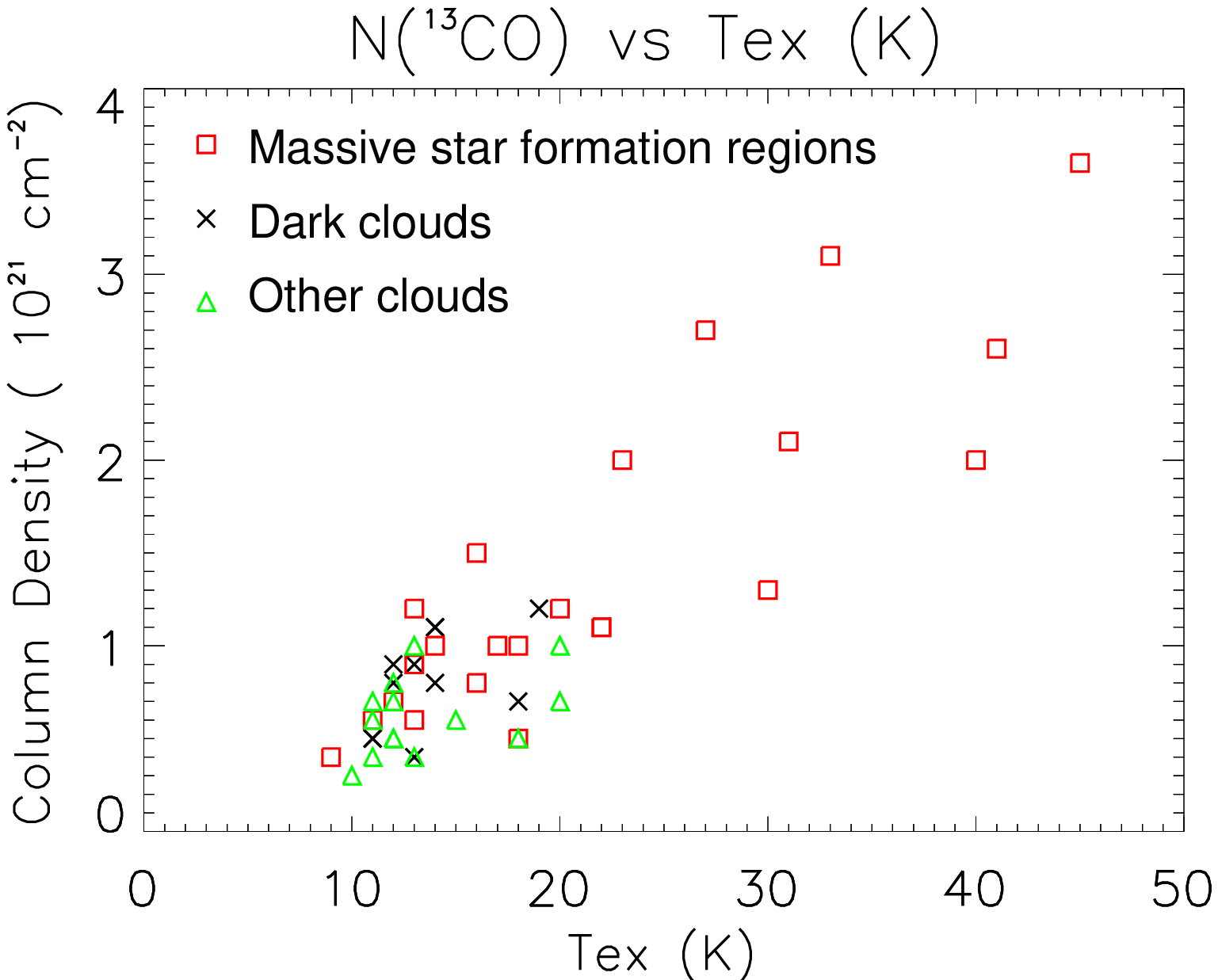}}

   \caption{(a): Histogram of the excitation temperature maximum from different subregions. (b) Histogram of the mean column density of gas derived from \coii. (c) $N$($^{13}$CO) vs. $T_\mathrm{ex}$, The relation between the mean column density of gas derived from {$^{13}$CO} and excitation temperature maximum. The red histograms (in (a) and (b)) and squares (in (c)) represent data from massive star formation region, black histograms (in (a) and (b)) and crosses (in (c)) represent data from dark clouds; and the green histograms (in (a) and (b)) and the triangles (in (c)) represent data from other clouds except the Remote Clouds.}
    \label{Fig:figure-statistics}
\end{figure}

Our survey includes many clouds with different types, for example: massive star formation regions (GGMC~2, GGMC~3, GGMC~4), dark clouds (Lynds Dark Clouds, the West Front), and other types (GGMC~1, Swallow, Horn). The molecular gas associated with massive star formation regions has higher maximum excitation temperatures than the other regions (Figure~\ref{Fig:figure-statistics-A}). Massive star formation regions have higher molecular gas column densities traced by {$^{13}$CO} than others (Figure~\ref{Fig:figure-statistics-B}). There is a relationship between the maximum excitation temperature and molecular gas column density (Figure~\ref{Fig:figure-statistics-C}). Dark clouds and other cloud types have lower excitation temperatures and column densities. Because CO molecules freeze out onto dust grains in the coldest, densest parts of the clouds, the column densities of the dark clouds (Lynds Dark Clouds and the West Front) estimated above are hence lower limits. In massive star formation regions, the gas column densities are positively correlated with excitation temperature, as seen in Figure~\ref{Fig:figure-statistics-C}. We note that the subregions with higher excitation temperatures are associated with \Hii\ regions, except GGMC~2 C5; thus, the molecular gas in these subregions may be heated by the \Hii\ regions and the density of the subregions may be enhanced. The relation between the column density and excitation temperature in massive star formation regions is also present in the Rosette Nebula (C. Li et al. in preparation). We will study the physical properties and chemical abundances of the clouds pixel-by-pixel in our next paper (C. Wang et al. in preparation).

\section{Summary}

We have presented large-scale observations in the \coi, {$^{13}$CO}, and \coiii\ $(1-0)$ transitions of the GEM OB1 MC complex. The observed region covers a total area of 58.5 deg$^2$, with ranges of $186.25^{\circ} < l < 195.25^{\circ}$ and $-3.75^{\circ} < b < 2.75^{\circ}$. We presented the large-scale distribution and physical properties of the MCs studied in this paper. Our main results are summarized as follows.

1. We divided the CO emission region into four MC components based on their spatial distribution and velocity ranges: the GGMC complex, the Lynds Dark Clouds and the West Front Clouds, the Swallow and the Horn, and the Remote Clouds. The GGMC Complex includes four GMCs and is located in the Perseus arm at a distance of 2~kpc. The Lynds Dark Clouds and the West Front Clouds are local clouds, with distances of 400~pc and 560~pc, respectively. Swallow and Horn are presented for the first time in this paper. They have velocity intervals of [11, 21] km s$^{-1}$; however, their distances could not determined. The velocity of the Remote Clouds is the highest, and we suggest that one cloud in the Remote Clouds, F3, is at a distance of 8.71 kpc.

2. The physical size of the GGMC complex is about 230~pc $\times$ 98~pc. A special filament was discovered in GGMC~1, which includes four components. One is located in the middle of the filament, while the others line up along the axis of the filament. GGMC~4 contains two components with velocity intervals of [-3, 5] and [5, 13] {km s$^{-1}$}. There are enhanced gas and IR clusters at the interface between the two components.

3. Swallow and Horn have similar angular sizes (0.8 and 0.6 deg$^{2}$) and mean column densities determined from {$^{12}$CO} and {$^{13}$CO}. The {C$^{18}$O} emission in Swallow shows a crescent-shaped structure.

4. Two elongated structures were discovered in an $L-V$ map within the velocity interval [11,~30] km s$^{-1}$. One component, [10,~18] km s$^{-1}$, connects Swallow and Horn, while the other [15,~30] km s$^{-1}$, connects the Remote Clouds and Horn.

5. The {C$^{18}$O} emission in this survey was mainly detected from the GGMC Complex. The intensity of the {C$^{18}$O} emission in all the other clouds was relatively weak.

\begin{acknowledgements}
This work is part of the Milky Way Image Scroll Painting (MWISP) project, which is based on observations made with the PMO 13.7 m telescope at Delingha. We would like to thank all the staff members of Qinghai Radio Observing Station at Delingha for their help during the observations. We also thank the anonymous referee for his careful reading and valuable comments that improved this paper. This work also used the data from (1) WISE, and (2) the Palomar Digital Sky Survey. We acknowledge the use of NASA's SkyView facility (http://skyview.gsfc.nasa.gov) hosted at NASA Goddard Space Flight Center. This work is funded by the National Natural Science Foundation of China (NSFC) grants No.11233007 and the Key Laboratory for Radio Astronomy, CAS. The MWISP is supported by the Strategic Priority Research Program, the Emergence of Cosmological Structures of the Chinese Academy of Sciences, grant No. XDB09000000. This work is also partially supported by Millimeter Wave Radio Astronomy Database (http://www.radioast.csdb.cn/), and the Millimeter \& Sub-Millimeter Wave Laboratory of Purple Mountain Observatory (http://mwlab.pmo.ac.cn/).

\end{acknowledgements}

\bibliographystyle{aa}
\bibliography{bibtex}

\clearpage

\clearpage

\clearpage

\clearpage

\clearpage

\clearpage

\begin{deluxetable}{cccccccccccccccccccccccccccccccccccc}
\tabletypesize{\tiny}
\rotate
\tablecaption{Properties of the MCs\label{tbl-1}}
\tablewidth{0pt}
\tabcolsep = 2 pt
\tablehead{
Cloud & Distance & \multicolumn{5}{c}{\coi}  &  \hspace{0.2em} & \multicolumn{5}{c}{\coii}&  \hspace{0.1em} & \multicolumn{5}{c}{{C$^{18}$O}} \\   \cline{3-7}  \cline{9-13} \cline{15-19} \noalign{\smallskip}
 && $V_\mathrm{range}$ & Size & $T_\mathrm{peak}$ & $N_\mathrm{H_{2}}$ & Mass && $V_\mathrm{range}$ & Size & $T_\mathrm{peak}$ & $N_\mathrm{H_{2}}$ & Mass && $V_\mathrm{range}$ & Size & $T_\mathrm{peak}$ & $N_\mathrm{H_{2}}$ & Mass \\
 & (kpc) & (\kms) & (deg$^{2}$) & (K) & ($10^{21}$cm$^{-2}$) & (M$_{\odot}$) &&  (\kms) & (deg$^{2}$) & (K) & ($10^{21}$cm$^{-2}$) & (M$_{\odot}$) && (\kms) & (arcmin$^{2}$) & (K) & ($10^{21}$cm$^{-2}$) & (M$_{\odot}$) \\}
\startdata
GGMC1 & 2\tablenotemark{[1]} & [-3, 10] & 3.7 & 9.5 & 2.0 & $2.0\times10^{5}$ && [-0.3, 10] & 1.48 & 5.8 & 0.9 & $3.6\times10^{4}$ && [2, 9] & 25 & 2.2 & 4.4 & $2.0\times10^{3}$ & \\
GGMC2 & 2\tablenotemark{[1]}\tablenotemark{[2]} & [-5, 16] & 2.1 & 41.7 & 3.1 & $1.8\times10^{5}$ && [-1.5, 16] & 1.15 & 17.2 & 1.9 & $7.4\times10^{4}$ && [4, 11] & 109 & 2.9 & 11.6 & $1.0\times10^{4}$ \\
GGMC3 & 2\tablenotemark{[1]}\tablenotemark{[2]} & [-5, 16] & 3.8 & 37.9 & 3.1 & $2.8\times10^{5}$ && [-1.5, 16] & 1.93 & 16.8 & 2.3 & $1.1\times10^{5}$ && [4, 11] & 219 & 3.8 & 17.5 & $2.8\times10^{4}$ &\\
GGMC4 & 2\tablenotemark{[1]}\tablenotemark{[2]} & [-4, 15] & 2.3 & 29.2 & 2.6 & $1.6\times10^{5}$ && [-4, 14] & 0.98 & 13.2 & 2.0 & $5.5\times10^{4}$ && [-2.5, 5] & 87 & 2.2 & 8.3 & $1.1\times10^{4}$ & \\
L1570 & 0.4\tablenotemark{[3]} & [-6, 7] & 0.1 & 9.3 & 1.2 & $1.7\times10^{2}$ && [-5, 6.8] & 0.04 & 5.8 & 1.2 & $4.6\times10^{1}$ && [-3, 2] & 4 & 1.8 & 4.2 & $4.9\times10^{0}$ \\
L1574-5 & 0.4\tablenotemark{[3]} & [-6, 7] & 0.3 & 15.2 & 2.2 & $7.3\times10^{2}$ && [-5, 6.8] & 0.13 & 5.9 & 1.5 & $2.1\times10^{2}$ && [-3, 2] & 22 & 2.3 & 5.4 & $3.6\times10^{1}$ \\
L1576 & 0.4\tablenotemark{[3]} & [-6, 7] & 0.2 & 11.0 & 2.6 & $6.3\times10^{2}$ && [-5, 6.8] & 0.13 & 5.3 & 1.5 & $2.0\times10^{2}$ && [-3, 2] & 31 & 1.8 & 4.8 & $4.5\times10^{1}$ \\
L1578 & 0.4\tablenotemark{[3]} & [-6, 7] & 0.1 & 10.2 & 1.8 & $2.2\times10^{2}$ && [-5, 6.8] & 0.04 & 4.6 & 1.3 & $5.5\times10^{1}$ && [-3, 2] & 2.3 & 1.4 & 4.0 & $3.0\times10^{0}$ \\
West Front & 0.56\tablenotemark{[4]} & [-7, 10] & 13.9 & 14.1 & 1.7 & $5.0\times10^{4}$ && [-5, 8] & 4.85 & 8.5 & 0.8 & $8.4\times10^{3}$ && [-1, 4] & 22 & 2.1 & 3.5 & $1.9\times10^{2}$ \\
Swallow\tablenotemark{a} & $\cdot\cdot\cdot$ & [11, 20] & 0.8 & 16.5 & 1.2	 & $\cdot\cdot\cdot$ && [11, 19] & 0.31 & 7.6 & 0.8	 & $\cdot\cdot\cdot$ && [12, 18] & 6 & 1.3 & 4.8 & $\cdot\cdot\cdot$ \\		
Horn\tablenotemark{a} & $\cdot\cdot\cdot$ & [11, 21] & 0.6 & 16.1 & 1.3	 & $\cdot\cdot\cdot$ && [11, 20] & 0.21 & 4.9 & 0.6	 & $\cdot\cdot\cdot$ && [12, 18] & 5 & 1.1 & 4.4 & $\cdot\cdot\cdot$ \\		
Remote Clouds\tablenotemark{a} & $\cdot\cdot\cdot$ & [16, 32] & 2.2 & 21.3 & 0.6 & $\cdot\cdot\cdot$ && [16, 32] & 0.49 & 5.5 & 0.5 & $\cdot\cdot\cdot$ && [18, 20] & 10.3 & 1.2 & 1.6 & $\cdot\cdot\cdot$ \\
\enddata

\tablerefs{[1] \citet{1995ApJ...445..246C}; [2] \citet{2009ApJ...693..397R}, \citet{2011PASJ...63....9N}, \citet{2010A&A...511A...2R}; [3] \citet{1974AJ.....79...42B}, \citet{2013A&A...556A..65E}; [4] \citet{2016A&A...594A..28P}.}
\tablecomments{Columns are cloud names, distances, and the properties derived from {$^{12}$CO}, {$^{13}$CO}, and {C$^{18}$O} respectively, including velocity ranges, sizes with pixels great than 3$\sigma$ of integrated intensity, the maxima of spectra peak, mean column densities of H$_{2}$, and mass. The column density and mass for {$^{12}$CO} are derived with the X-factor, and those for {$^{13}$CO} and {C$^{18}$O} are derived based on the LTE assumption. The masses of Swallow, Horn, and Remote Clouds are absent.}
\tablenotetext{a}{The masses of Swallow, Horn, and Remote Clouds are absent because the distances are undetermined; see the discussion of Sect~\ref{Sec:discussion}. }

\end{deluxetable}

\clearpage

\begin{deluxetable}{cccccccccccccccccccccccccccccccccccc}
\tabletypesize{\scriptsize}
\rotate
\tablecaption{Properties of the subregions in the GGMC 1\label{tbl-GGMC1}}
\tablewidth{0pt}
\tabcolsep = 2 pt
\tablehead{
Subregion & & \hspace{0.1em} & \multicolumn{4}{c}{\coi}  &  \hspace{0.1em} & \multicolumn{4}{c}{\coii}&  \hspace{0.1em} & \multicolumn{4}{c}{{C$^{18}$O}} \\   \cline{4-7}  \cline{9-12} \cline{14-17} \noalign{\smallskip}
 & $T_\mathrm{ex}$ && Size & $\sigma$ & $N_\mathrm{H_{2}}$ & Mass && Size & $\sigma$ & $N_\mathrm{H_{2}}$ & Mass && Size & $\sigma$ & $N_\mathrm{H_{2}}$ & Mass \\
 & (K) && (arcmin$^{-2}$) & (\kms) & ($10^{21}$cm$^{-2}$) & (M$_{\odot}$) && (arcmin$^{-2}$) & (\kms) & ($10^{21}$cm$^{-2}$) & (M$_{\odot}$) && (arcmin$^{-2}$) & (\kms) & ($10^{21}$cm$^{-2}$) & (M$_{\odot}$) \\}
\startdata
A1 & 13 && 1297 & 2.1 & 2.0 & $1.9\times10^{4}$ && 592 & 1.3 & 1.3 & $5.8\times10^{3}$ && 20 & 0.5 & 4.4  & $1.7\times10^{3}$ \\

C1 & 12 && 996 & 2.0 & 1.2 & $5.2\times10^{3}$ && 167 & 1.2 & 0.9 & $1.2\times10^{3}$ && 5 & 0.6 & 4.4 & $3.3\times10^{2}$ \\

C2 & 12 && 1640 & 1.8 & 1.3 & $1.4\times10^{4}$ && 252 & 0.7 & 0.6 & $1.1\times10^{3}$ && $\cdot\cdot\cdot$ & $\cdot\cdot\cdot$ & $\cdot\cdot\cdot$ & $\cdot\cdot\cdot$ \\

C3 & 12 && 1637 & 2.4 & 2.2 & $2.4\times10^{4}$ && 462 & 0.9 & 0.8 & $3.3\times10^{3}$ && $\cdot\cdot\cdot$ & $\cdot\cdot\cdot$ & $\cdot\cdot\cdot$ & $\cdot\cdot\cdot$ \\

C4 & 11 && 1511 & 2.4 & 2.4 & $2.7\times10^{4}$ && 800 & 1.0 & 0.7 & $4.5\times10^{3}$ && $\cdot\cdot\cdot$ & $\cdot\cdot\cdot$ & $\cdot\cdot\cdot$ & $\cdot\cdot\cdot$ \\

C5 & 13 && 1072 & 2.7 & 2.8 & $2.1\times10^{4}$ && 686 & 1.3 & 1.2 & $6.4\times10^{3}$ && $\cdot\cdot\cdot$ & $\cdot\cdot\cdot$ & $\cdot\cdot\cdot$ & $\cdot\cdot\cdot$ \\

C6 & 12 && 4235 & 2.9 & 2.4 & $6.8\times10^{4}$ && 1744 & 1.3 & 0.9 & $1.2\times10^{4}$ && $\cdot\cdot\cdot$ & $\cdot\cdot\cdot$ & $\cdot\cdot\cdot$ & $\cdot\cdot\cdot$ \\

F1 & 12 && 1637 & 2.6 & 2.0 & $2.2\times10^{4}$ && 520 & 1.1 & 0.8 & $3.3\times10^{3}$ && $\cdot\cdot\cdot$ & $\cdot\cdot\cdot$ & $\cdot\cdot\cdot$ & $\cdot\cdot\cdot$ \\

\enddata
\tablecomments{Columns are subregion names, maximas of excitation temperature, and the properties derived from {$^{12}$CO}, {$^{13}$CO}, and {C$^{18}$O} respectively, including size with pixels greater than 3$\sigma$ of integrated intensity, line widths of averaged spectra, mean column density of H$_{2}$, and mass.}
\end{deluxetable}

\begin{deluxetable}{cccccccccccccccccccccccccccccccccccc}
\tabletypesize{\scriptsize}
\rotate
\tablecaption{Properties of the subregions in the GGMC 2\label{tbl-GGMC2}}
\tablewidth{0pt}
\tabcolsep = 2 pt
\tablehead{
Subregion & & \hspace{0.1em} & \multicolumn{4}{c}{\coi}  &  \hspace{0.1em} & \multicolumn{4}{c}{\coii}&  \hspace{0.1em} & \multicolumn{4}{c}{{C$^{18}$O}} \\   \cline{4-7}  \cline{9-12} \cline{14-17} \noalign{\smallskip}
 & $T_{ex}$ && Size & $\sigma$ & $N_\mathrm{H_{2}}$ & Mass && Size & $\sigma$ & $N_\mathrm{H_{2}}$ & Mass && Size & $\sigma$ & $N_\mathrm{H_{2}}$ & Mass \\
 & (K) && (arcmin$^{-2}$) & (\kms) & ($10^{21}$cm$^{-2}$) & (M$_{\odot}$) && (arcmin$^{-2}$) & (\kms) & ($10^{21}$cm$^{-2}$) & (M$_{\odot}$) && (arcmin$^{-2}$) & (\kms) & ($10^{21}$cm$^{-2}$) & (M$_{\odot}$) \\}
\startdata
A1 & 18 && 997 & 2.5 & 1.1 & $1.5\times10^{4}$ && 330 & 1.7 & 0.7 & $1.6\times10^{3}$ && $\cdot\cdot\cdot$ & $\cdot\cdot\cdot$ & $\cdot\cdot\cdot$ & $\cdot\cdot\cdot$ \\

C1 & 9 && 316 & 3.3 & 1.5 & $5.6\times10^{3}$ && 116 & 2.1 & 0.5 & $4.2\times10^{2}$ && $\cdot\cdot\cdot$ & $\cdot\cdot\cdot$ & $\cdot\cdot\cdot$ & $\cdot\cdot\cdot$ \\					

C2 & 13 && 327 & 3.8 & 3.8 & $9.6\times10^{3}$ && 278 & 2.1 & 1.5 & $3.1\times10^{3}$ && 2 & 1.2 & 5.0 & $8.4\times10^{1}$ \\

C3 & 13 && 323 & 4.3 & 3.9 & $9.5\times10^{3}$ && 293 & 2.2 & 1.1 & $2.4\times10^{3}$ && $\cdot\cdot\cdot$ & $\cdot\cdot\cdot$ & $\cdot\cdot\cdot$ & $\cdot\cdot\cdot$ \\
			
C4 & 16 && 657 & 3 & 2.9 & $1.5\times10^{4}$ && 468 & 1.8 & 1 & $3.4\times10^{3}$ && $\cdot\cdot\cdot$ & $\cdot\cdot\cdot$ & $\cdot\cdot\cdot$ & $\cdot\cdot\cdot$ \\					

C5 & 27 && 351 & 3.1 & 5.1 & $1.4\times10^{4}$ && 311 & 2 & 3.3 & $7.9\times10^{3}$ && 20 & 1.5 & 10.6 & $1.6\times10^{3}$ \\

C6 & 45 && 1010 & 3.5 & 5.9 & $4.5\times10^{4}$ && 905 & 2.3 & 4.3 & $3.0\times10^{4}$ && 74 & 1.7 & 13.2 & $7.4\times10^{3}$ \\

C7 & 16 && 946 & 4.1 & 4.7 & $3.4\times10^{4}$ && 798 & 2.5 & 1.8 & $1.1\times10^{4}$ && 12 & 1.5 & 5.8 & $5.4\times10^{2}$ \\

F1 & 11 && 901 & 2.9 & 1.9 & $1.4\times10^{4}$ && 399 & 1.9 & 0.7 & $2.0\times10^{3}$ && $\cdot\cdot\cdot$ & $\cdot\cdot\cdot$ & $\cdot\cdot\cdot$ & $\cdot\cdot\cdot$ \\		

F2 & 13 && 490 & 2.1 & 1.2 & $5.9\times10^{3}$ && 70 & 1.6 & 0.7 & $3.4\times10^{2}$ && $\cdot\cdot\cdot$ & $\cdot\cdot\cdot$ & $\cdot\cdot\cdot$ & $\cdot\cdot\cdot$ \\

F3 & 18 && 556 & 2.7 & 2.2 & $1.1\times10^{4}$ && 372 & 1.7 & 1.1 & $3.3\times10^{3}$ && 2 & 1.3 & 4.3 & $7.2\times10^{1}$ \\

F4 & 31 && 269 & 2.3 & 2.5 & $6.3\times10^{3}$ && 173 & 1.7 & 2.5 & $3.3\times10^{3}$ && 8 & 1.4 & 12.9 & $7.5\times10^{2}$ \\

\enddata
\tablecomments{Columns are the same as Table~\ref{tbl-GGMC1}}
\end{deluxetable}

\begin{deluxetable}{cccccccccccccccccccccccccccccccccccc}
\tabletypesize{\scriptsize}
\rotate
\tablecaption{Properties of the subregions in the GGMC 3\label{tbl-GGMC3}}
\tablewidth{0pt}
\tabcolsep = 2 pt
\tablehead{
Subregion & & \hspace{0.1em} & \multicolumn{4}{c}{\coi}  &  \hspace{0.1em} & \multicolumn{4}{c}{\coii}&  \hspace{0.1em} & \multicolumn{4}{c}{{C$^{18}$O}} \\   \cline{4-7}  \cline{9-12} \cline{14-17} \noalign{\smallskip}
 & $T_{ex}$ && Size & $\sigma$ & $N_\mathrm{H_{2}}$ & Mass && Size & $\sigma$ & $N_\mathrm{H_{2}}$ & Mass && Size & $\sigma$ & $N_\mathrm{H_{2}}$ & Mass \\
 & (K) && (arcmin$^{-2}$) & (\kms) & ($10^{21}$cm$^{-2}$) & (M$_{\odot}$) && (arcmin$^{-2}$) & (\kms) & ($10^{21}$cm$^{-2}$) & (M$_{\odot}$) && (arcmin$^{-2}$) & (\kms) & ($10^{21}$cm$^{-2}$) & (M$_{\odot}$) \\}
\startdata

ECF1 & 40 && 444 & 3.5 & 4.6 & $1.5\times10^{4}$ && 376 & 2.1 & 2.4 & $6.7\times10^{3}$ && 18 & 1.2 & 6.6 & $8.7\times10^{2}$ \\

ECF2 & 23 && 181 & 3.3 & 3.2 & $6.0\times10^{3}$ && 135 & 2.1 & 2.4 & $2.4\times10^{3}$ && 8 & 1.7 & 11.2 & $7.0\times10^{2}$ \\

ECF3 & 30 && 302 & 3 & 2.6 & $6.9\times10^{3}$ && 156 & 1.7 & 1.5 & $1.8\times10^{3}$ && 1 & 1 & 7.9 & $5.9\times10^{1}$ \\

ECF4 & 22 && 834 & 3 & 2.1 & $1.6\times10^{4}$ && 461 & 2 & 1.3 & $4.7\times10^{3}$ && 3.5 & 1.4 & 5.8 & $1.5\times10^{2}$ \\

ECF5 & 20 && 1308 & 3.2 & 2.5 & $2.9\times10^{4}$ && 773 & 2.2 & 1.4 & $7.9\times10^{3}$ && 3.5 & 1.2 & 4.0 & $1.0\times10^{2}$ \\

ECF6 & 22 && 837 & 2.3 & 2.2 & $1.6\times10^{4}$ && 487 & 1.5 & 1.3 & $4.8\times10^{3}$ && $\cdot\cdot\cdot$ & $\cdot\cdot\cdot$ & $\cdot\cdot\cdot$ & $\cdot\cdot\cdot$ \\

WCF & 41 && 4117 & 4.3 & 4.7 & $1.6\times10^{5}$ && 3140 & 3 & 3.2 & $7.5\times10^{4}$ && 175 & 2.1 & 19.4 & $2.6\times10^{4}$ \\

\enddata
\tablecomments{Columns are the same as Table~\ref{tbl-GGMC1}}
\end{deluxetable}

\begin{deluxetable}{cccccccccccccccccccccccccccccccccccc}
\tabletypesize{\scriptsize}
\rotate
\tablecaption{Properties of the subregions in the GGMC 4\label{tbl-GGMC4}}
\tablewidth{0pt}
\tabcolsep = 2 pt
\tablehead{
Subregion & & \hspace{0.1em} & \multicolumn{4}{c}{\coi}  &  \hspace{0.1em} & \multicolumn{4}{c}{\coii}&  \hspace{0.1em} & \multicolumn{4}{c}{{C$^{18}$O}} \\   \cline{4-7}  \cline{9-12} \cline{14-17} \noalign{\smallskip}
 & $T_{ex}$ && Size & $\sigma$ & $N_\mathrm{H_{2}}$ & Mass && Size & $\sigma$ & $N_\mathrm{H_{2}}$ & Mass && Size & $\sigma$ & $N_\mathrm{H_{2}}$ & Mass \\
 & (K) && (arcmin$^{-2}$) & (\kms) & ($10^{21}$cm$^{-2}$) & (M$_{\odot}$) && (arcmin$^{-2}$) & (\kms) & ($10^{21}$cm$^{-2}$) & (M$_{\odot}$) && (arcmin$^{-2}$) & (\kms) & ($10^{21}$cm$^{-2}$) & (M$_{\odot}$) \\}
\startdata

C1 & 14 && 2385 & 3.2 & 2.7 & $4.8\times10^{4}$ && 1304 & 1.1 & 1.2 & $1.2\times10^{4}$ && $\cdot\cdot\cdot$ & $\cdot\cdot\cdot$ & $\cdot\cdot\cdot$ & $\cdot\cdot\cdot$ \\

C2& 17 && 1099 & 2.5 & 2.6 & $2.2\times10^{4}$ && 606 & 0.8 & 1.2 & $5.5\times10^{3}$ && $\cdot\cdot\cdot$ & $\cdot\cdot\cdot$ & $\cdot\cdot\cdot$ & $\cdot\cdot\cdot$ \\			

C3 & 33 && 1913 & 4.5 & 4.5 & $6.4\times10^{4}$ && 1160 & 2.4 & 3.9 & $3.4\times10^{4}$ && 87 & 0.4 & 8.3 & $1.1\times10^{3}$ \\

C4 & 12 && 878 & 1.7 & 1.2 & $7.7\times10^{3}$ && 151 & 0.3 & 0.8 & $9.3\times10^{2}$ && $\cdot\cdot\cdot$ & $\cdot\cdot\cdot$ & $\cdot\cdot\cdot$ & $\cdot\cdot\cdot$ \\					

\enddata
\tablecomments{Columns are the same as Table~\ref{tbl-GGMC1}}
\end{deluxetable}

\begin{deluxetable}{cccccccccccccccccccccccccccccccccccc}
\tabletypesize{\scriptsize}
\rotate
\tablecaption{Properties of the subregions in the Lynds Dark Clouds\label{tbl-Lynds}}
\tablewidth{0pt}
\tabcolsep = 2 pt
\tablehead{
Subregion & & \hspace{0.1em} & \multicolumn{4}{c}{\coi}  &  \hspace{0.1em} & \multicolumn{4}{c}{\coii}&  \hspace{0.1em} & \multicolumn{4}{c}{{C$^{18}$O}} \\   \cline{4-7}  \cline{9-12} \cline{14-17} \noalign{\smallskip}
 & $T_{ex}$ && Size & $\sigma$ & $N_\mathrm{H_{2}}$ & Mass && Size & $\sigma$ & $N_\mathrm{H_{2}}$ & Mass && Size & $\sigma$ & $N_\mathrm{H_{2}}$ & Mass \\
 & (K) && (arcmin$^{-2}$) & (\kms) & ($10^{21}$cm$^{-2}$) & (M$_{\odot}$) && (arcmin$^{-2}$) & (\kms) & ($10^{21}$cm$^{-2}$) & (M$_{\odot}$) && (arcmin$^{-2}$) & (\kms) & ($10^{21}$cm$^{-2}$) & (M$_{\odot}$) \\}
\startdata

L1570 & 13 && 563 & 1.5 & 1.2 & $1.7\times10^{2}$ && 132 & 0.9 & 1.2 & $4.6\times10^{1}$ && 4 & 0.6 & 4.2 & $4.9\times10^{0}$ \\	

L1574 & 19 && 1220 & 2.7 & 2.2 & $7.3\times10^{2}$ && 473 & 1.2 & 1.5 & $2.1\times10^{2}$ && 22 & 0.6 & 5.4 & $3.6\times10^{1}$ \\	

L1576 & 14 && 821 & 2.7 & 2.6 & $6.3\times10^{2}$ && 452 & 1.2 & 1.5 & $2.0\times10^{2}$ && 31 & 0.5 & 4.8 & $4.5\times10^{1}$ \\

L1578 & 14 && 429 & 2 & 1.8 & $2.2\times10^{2}$ && 140 & 1.1 & 1.3 & $5.5\times10^{1}$ && 2 & $\cdot\cdot\cdot$\tablenotemark{a} & 4.0 & $3.0\times10^{0}$ \\			
\enddata
\tablecomments{Columns are the same as Table~\ref{tbl-GGMC1}}
\tablenotetext{a}{The line width of C1 derived from {C$^{18}$O} is not listed here because the emissions are so weak that the value is uncalculable. }
\end{deluxetable}

\begin{deluxetable}{cccccccccccccccccccccccccccccccccccc}
\tabletypesize{\scriptsize}
\rotate
\tablecaption{Properties of the subregions in the West Front\label{tbl-west}}
\tablewidth{0pt}
\tabcolsep = 2 pt
\tablehead{
Subregion & & \hspace{0.1em} & \multicolumn{4}{c}{\coi}  &  \hspace{0.1em} & \multicolumn{4}{c}{\coii}&  \hspace{0.1em} & \multicolumn{4}{c}{{C$^{18}$O}} \\   \cline{4-7}  \cline{9-12} \cline{14-17} \noalign{\smallskip}
 & $T_{ex}$ && Size & $\sigma$ & $N_\mathrm{H_{2}}$ & Mass && Size & $\sigma$ & $N_\mathrm{H_{2}}$ & Mass && Size & $\sigma$ & $N_\mathrm{H_{2}}$ & Mass \\
 & (K) && (arcmin$^{-2}$) & (\kms) & ($10^{21}$cm$^{-2}$) & (M$_{\odot}$) && (arcmin$^{-2}$) & (\kms) & ($10^{21}$cm$^{-2}$) & (M$_{\odot}$) && (arcmin$^{-2}$) & (\kms) & ($10^{21}$cm$^{-2}$) & (M$_{\odot}$) \\}
\startdata

C1 & 11 && 2516 & 2.1 & 1.3 & $2.9\times10^{3}$ && 746 & 0.9 & 0.5 & $2.4\times10^{2}$ && $\cdot\cdot\cdot$ & $\cdot\cdot\cdot$ & $\cdot\cdot\cdot$ & $\cdot\cdot\cdot$ \\		

C2 & 14 && 6257 & 2.7 & 2.2 & $1.0\times10^{4}$ && 4198 & 1 & 0.9 & $2.3\times10^{3}$ && 16 & 0.5 & 3.6 & $1.6\times10^{2}$ \\

C3 & 12 && 2342 & 3.3 & 2.5 & $4.2\times10^{3}$ && 1564 & 1.7 & 1.0 & $1.0\times10^{3}$ && $\cdot\cdot\cdot$ & $\cdot\cdot\cdot$ & $\cdot\cdot\cdot$ & $\cdot\cdot\cdot$ \\

C4 & 12 && 6073 & 3.8 & 2.7 & $1.2\times10^{4}$ && 4003 & 1.8 & 1.0 & $2.3\times10^{3}$ && 2 & 0.7 & 3.2 & $2.1\times10^{1}$ \\

C5 & 18 && 2491 & 2.4 & 1.9 & $3.5\times10^{3}$ && 1040 & 1 & 0.8 & $4.9\times10^{2}$ && 1 & $\cdot\cdot\cdot$\tablenotemark{a} & 3.3 & $4.3\times10^{0}$ \\		

D1 & 13 && 2313 & 1.8 & 1.2 & $2.6\times10^{3}$ && 871 & 0.8 & 0.5 & $2.5\times10^{2}$ && $\cdot\cdot\cdot$ & $\cdot\cdot\cdot$ & $\cdot\cdot\cdot$ & $\cdot\cdot\cdot$ \\

D2 & 11 && 8246 & 2.2 & 1.4 & $9.8\times10^{3}$ && 2714 & 0.9 & 0.6 & $1.0\times10^{3}$ && 3 & 0.6 & 3.0 & $8.4\times10^{0}$ \\
		
\enddata
\tablecomments{Columns are the same as Table~\ref{tbl-GGMC1}}
\tablenotetext{a}{The line width of C1 derived from {C$^{18}$O} is not listed here because the emissions are so weak that the value is uncalculable. }
\end{deluxetable}

\begin{deluxetable}{cccccccccccccccccccccccccccccccccccc}
\tabletypesize{\scriptsize}
\rotate
\tablecaption{Properties of the subregions in the Swallow \label{tbl-swallow}}
\tablewidth{0pt}
\tabcolsep = 2 pt
\tablehead{
Subregion & & \hspace{0.1em} & \multicolumn{4}{c}{\coi}  &  \hspace{0.1em} & \multicolumn{4}{c}{\coii}&  \hspace{0.1em} & \multicolumn{4}{c}{{C$^{18}$O}} \\   \cline{4-7}  \cline{9-12} \cline{14-17} \noalign{\smallskip}
 & $T_{ex}$ && Size & $\sigma$ & $N_\mathrm{H_{2}}$ & Mass && Size & $\sigma$ & $N_\mathrm{H_{2}}$ & Mass && Size & $\sigma$ & $N_\mathrm{H_{2}}$ & Mass \\
 & (K) && (arcmin$^{-2}$) & (\kms) & ($10^{21}$cm$^{-2}$) & (M$_{\odot}$) && (arcmin$^{-2}$) & (\kms) & ($10^{21}$cm$^{-2}$) & (M$_{\odot}$) && (arcmin$^{-2}$) & (\kms) & ($10^{21}$cm$^{-2}$) & (M$_{\odot}$) \\}
\startdata

F1 & 20 && 596 & 2.7 & 1.9 & $\cdot\cdot\cdot$ && 466 & 1.7 & 1.3 & $\cdot\cdot\cdot$ && 6 & 0.5 & 4.8 & $\cdot\cdot\cdot$ \\

F2 & 11 && 391 & 1.7 & 1 & $\cdot\cdot\cdot$ && 202 & 1.1 & 0.5 & $\cdot\cdot\cdot$ && $\cdot\cdot\cdot$ & $\cdot\cdot\cdot$ & $\cdot\cdot\cdot$ & $\cdot\cdot\cdot$ \\

F3 & 13 && 752 & 1.8 & 1.1 & $\cdot\cdot\cdot$ && 369 & 0.9 & 0.5 & $\cdot\cdot\cdot$  && $\cdot\cdot\cdot$ & $\cdot\cdot\cdot$ & $\cdot\cdot\cdot$ & $\cdot\cdot\cdot$ \\

\enddata
\tablecomments{Columns are the same as Table~\ref{tbl-GGMC1}}
\end{deluxetable}

\begin{deluxetable}{cccccccccccccccccccccccccccccccccccc}
\tabletypesize{\scriptsize}
\rotate
\tablecaption{Properties of the subregions in the Horn \label{tbl-horn}}
\tablewidth{0pt}
\tabcolsep = 2 pt
\tablehead{
Subregion & & \hspace{0.1em} & \multicolumn{4}{c}{\coi}  &  \hspace{0.1em} & \multicolumn{4}{c}{\coii}&  \hspace{0.1em} & \multicolumn{4}{c}{{C$^{18}$O}} \\   \cline{4-7}  \cline{9-12} \cline{14-17} \noalign{\smallskip}
 & $T_{ex}$ && Size & $\sigma$ & $N_\mathrm{H_{2}}$ & Mass && Size & $\sigma$ & $N_\mathrm{H_{2}}$ & Mass && Size & $\sigma$ & $N_\mathrm{H_{2}}$ & Mass \\
 & (K) && (arcmin$^{-2}$) & (\kms) & ($10^{21}$cm$^{-2}$) & (M$_{\odot}$) && (arcmin$^{-2}$) & (\kms) & ($10^{21}$cm$^{-2}$) & (M$_{\odot}$) && (arcmin$^{-2}$) & (\kms) & ($10^{21}$cm$^{-2}$) & (M$_{\odot}$) \\}
\startdata

C1 & 20 && 228 & 1.9 & 1.3 & $\cdot\cdot\cdot$ && 136 & 1.1 & 0.8 & $\cdot\cdot\cdot$  && 1 & $\cdot\cdot\cdot$\tablenotemark{a} & 4.0 & $\cdot\cdot\cdot$ \\

C2 & 11 && 146 & 2.1 & 1.3 & $\cdot\cdot\cdot$ && 95 & 1.4 & 1 & $\cdot\cdot\cdot$ && 4 & 0.9 & 4.5 & $\cdot\cdot\cdot$ \\

C3 & 15 && 697 & 2.1 & 1.5 & $\cdot\cdot\cdot$ && 371 & 1 & 0.7 & $\cdot\cdot\cdot$  && $\cdot\cdot\cdot$ & $\cdot\cdot\cdot$ & $\cdot\cdot\cdot$ & $\cdot\cdot\cdot$ \\

C4 & 18 && 280 & 2.1 & 1.5 & $\cdot\cdot\cdot$ && 169 & 1 & 0.6 & $\cdot\cdot\cdot$  && $\cdot\cdot\cdot$ & $\cdot\cdot\cdot$ & $\cdot\cdot\cdot$ & $\cdot\cdot\cdot$ \\	

D1 & 10 && 165 & 1.3 & 0.7 & $\cdot\cdot\cdot$ && 82 & 0.7 & 0.4 & $\cdot\cdot\cdot$  && $\cdot\cdot\cdot$ & $\cdot\cdot\cdot$ & $\cdot\cdot\cdot$ & $\cdot\cdot\cdot$ \\

\enddata
\tablecomments{Columns are the same as Table~\ref{tbl-GGMC1}}
\tablenotetext{a}{The line width of C1 derived from {C$^{18}$O} is not listed here because the emission is so weak that the value is uncalculable. }
\end{deluxetable}

\begin{deluxetable}{cccccccccccccccccccccccccccccccccccc}
\tabletypesize{\scriptsize}
\rotate
\tablecaption{Properties of the subregions in the Remote Clouds \label{tbl-remote}}
\tablewidth{0pt}
\tabcolsep = 2 pt
\tablehead{
Subregion & & \hspace{0.1em} & \multicolumn{4}{c}{\coi}  &  \hspace{0.1em} & \multicolumn{4}{c}{\coii}&  \hspace{0.1em} & \multicolumn{4}{c}{{C$^{18}$O}} \\   \cline{4-7}  \cline{9-12} \cline{14-17} \noalign{\smallskip}
 & $T_{ex}$ && Size & $\sigma$ & $N_\mathrm{H_{2}}$ & Mass && Size & $\sigma$ & $N_\mathrm{H_{2}}$ & Mass && Size & $\sigma$ & $N_\mathrm{H_{2}}$ & Mass \\
 & (K) && (arcmin$^{-2}$) & (\kms) & ($10^{21}$cm$^{-2}$) & (M$_{\odot}$) && (arcmin$^{-2}$) & (\kms) & ($10^{21}$cm$^{-2}$) & (M$_{\odot}$) && (arcmin$^{-2}$) & (\kms) & ($10^{21}$cm$^{-2}$) & (M$_{\odot}$) \\}
\startdata

C1 & 23 && 540 & 1.5 & 0.9 & $1.1\times10^{5}$ && 261 & 1.1 & 0.7 & $2.3\times10^{4}$ && $\cdot\cdot\cdot$ & $\cdot\cdot\cdot$ & $\cdot\cdot\cdot$ & $\cdot\cdot\cdot$ \\

C2 & 10 && 2082 & 1.8 & 1.1 & $5.1\times10^{5}$ && 1106 & 1 & 0.6 & $8.9\times10^{4}$ && 10.3 & 0.4 & 1.6 & $3.1\times10^{4}$ \\

C3 & 19 && 20 & 1.7 & 1.2 & $5.8\times10^{3}$ && 12 & 1 & 0.8 & $1.2\times10^{3}$ && $\cdot\cdot\cdot$ & $\cdot\cdot\cdot$ & $\cdot\cdot\cdot$ & $\cdot\cdot\cdot$ \\

F1 & 25 && 803 & 1.9 & 1 & $2.0\times10^{5}$ && 385 & 1.6 & 0.6 & $3.2\times10^{4}$ && $\cdot\cdot\cdot$ & $\cdot\cdot\cdot$ & $\cdot\cdot\cdot$ & $\cdot\cdot\cdot$ \\

F2 & 9 && 448 & 1.2 & 0.7 & $7.1\times10^{4}$ && 112 & 0.8 & 0.5 & $4.7\times10^{3}$ && $\cdot\cdot\cdot$ & $\cdot\cdot\cdot$ & $\cdot\cdot\cdot$ & $\cdot\cdot\cdot$ \\

F3 & 17 && 91 & 1.3 & 0.7 & $2.1\times10^{4}$ && 81 & 1 & 0.5 & $4.7\times10^{3}$ && $\cdot\cdot\cdot$ & $\cdot\cdot\cdot$ & $\cdot\cdot\cdot$ & $\cdot\cdot\cdot$ \\

\enddata
\tablecomments{Columns are the same as Table~\ref{tbl-GGMC1}. These clouds are probably associated because of spatial vicinities and consistent velocity ranges. Here we set the distance of the other clouds to be the same as the cloud F3, for which the distance is determined to be 8.71 kpc (\citealt{2015AJ....150..147F}). However, the mass of the clouds calculated through the assumed distance seems to be higher. So, the masses are listed here for reference.}
\end{deluxetable}

\end{document}